\documentclass[a4paper,11pt]{article}
\pdfoutput=1
\usepackage{jheppub}
\usepackage{adjustbox}
\usepackage{amsmath,amssymb,bm,graphicx,bbold,epsf,colordvi}
\usepackage{slashed}
\usepackage{diagbox}
\def\bea#1\eea{\begin{align}#1\end{align}} 
\usepackage{diagbox}
\usepackage{lipsum}
\usepackage{soul}
\usepackage{braket}
\usepackage{bm}
\allowdisplaybreaks 
\usepackage{color}
\usepackage[dvipsnames]{xcolor}

\usepackage{mathrsfs}
\usepackage{appendix}
\usepackage{bm}
\usepackage{lipsum}

\newcommand{\bef}{\begin{figure}[hbt]\centering}
\newcommand{\eef}{\end{figure}}

\usepackage{mathtools}
\usepackage{subfigure}
\usepackage{booktabs}
\usepackage{graphicx}
\graphicspath{ {./figure/} }
\usepackage{caption}
\usepackage{lipsum}

\newcommand{\nnu}{\nonumber\\}

\newcommand{\beq}{\begin{equation}}
\newcommand{\eeq}{\end{equation}}
\def\bea#1\eea{\begin{align}#1\end{align}}
\newcommand{\sla}[1]{{#1}\!\!\!\slash}

\def \be  {\begin{equation}}
\def \ee  {\end{equation}}
\def \ba  {\begin{eqnarray}}
\def \ea  {\end{eqnarray}}
\newcommand\as{\alpha_s}

\allowdisplaybreaks

\usepackage{graphicx}

\makeatletter
\def\@fpheader{~}
\makeatother

\usepackage{stackengine}
\newcommand\xrowht[2][0]{\addstackgap[.5\dimexpr#2\relax]{\vphantom{#1}}}

\title{Spin asymmetries in electron-jet production at the future electron ion collider}

\author[a,b,c]{Zhong-Bo Kang}
\author[d,e]{, Kyle Lee}
\author[f,g]{, Ding Yu Shao}
\author[a,b]{, Fanyi Zhao}

\affiliation[a]{Department of Physics and Astronomy, University of California, Los Angeles, CA 90095, USA}
\affiliation[b]{Mani L. Bhaumik Institute for Theoretical Physics, University of California, Los Angeles, CA 90095, USA}
\affiliation[c]{Center for Frontiers in Nuclear Science, Stony Brook University, Stony Brook, NY 11794, USA}
\affiliation[d]{Nuclear Science Division, Lawrence Berkeley National Laboratory, Berkeley, CA 94720, USA}
\affiliation[e]{Physics Department, University of California, Berkeley, CA 94720, USA}
\affiliation[f]{Department of Physics and Center for Field Theory and Particle Physics, Fudan University, Shanghai, 200433, China}
\affiliation[g]{Key Laboratory of Nuclear Physics and Ion-beam Application (MOE), Fudan University, Shanghai, 200433, China}

\emailAdd{zkang@g.ucla.edu}
\emailAdd{kylelee@lbl.gov}
\emailAdd{dingyu.shao@cern.ch}
\emailAdd{fanyizhao@physics.ucla.edu}

\abstract{We study all the possible spin asymmetries that can arise in back-to-back electron-jet production, $ep\rightarrow e+\text{jet}+X$, as well as the associated jet fragmentation process, $ep\rightarrow e+ \text{jet} (h)+X$, in electron-proton collisions. We derive the factorization formalism for these spin asymmetries and perform the corresponding phenomenology for the kinematics relevant to the future electron ion collider. In the case of unpolarized electron-proton scattering, we also give predictions for azimuthal asymmetries for the HERA experiment. This demonstrates that electron-jet production is an outstanding process for probing unpolarized and polarized transverse momentum dependent parton distribution functions and fragmentation functions.}

\begin{document}

\maketitle

\section{Introduction}
In recent years, studies of jets and their substructures have been used as important probes to test the fundamental and emergent properties of Quantum Chromodynamics (QCD) and strong interactions~\cite{Sterman:1977wj,Sapeta:2015gee,Altheimer:2012mn,Abdesselam:2010pt,Ali:2010tw,Salam:2009jx,Buttar:2008jx,Ellis:2007ib,Larkoski:2017jix}, as well as for searching for beyond-standard-model signals~\cite{Asquith:2018igt,Marzani:2019hun}. For example, jets and their various substructures have served as major tools for mapping out the partonic structure of a nucleon~\cite{Martin:2001es,Lai:1996mg}, and for unveiling the basic properties of quark-gluon plasma (QGP) produced in heavy-ion collisions~\cite{Connors:2017ptx}.  

The advent of the Electron-Ion Collider (EIC) with polarized beams would unlock the full potential of jets as novel probes for the three-dimensional (3D) structure of the nucleon and nuclei~\cite{AbdulKhalek:2021gbh}. Such 3D structure in the momentum space is encoded in the so-called transverse momentum dependent parton distribution functions (TMDPDFs), whose study is one of the scientific pillars at the EIC~\cite{Accardi:2012qut,Boer:2011fh,Proceedings:2020eah}. Besides jet production, the study of jet substructure, in particular, hadron distribution within jets has received growing attention in the last several years as an efficient instrument to understand the process of fragmentation, explaining how the color-carrying partons turn into color-neutral particles such as hadrons~\cite{Larkoski:2017jix}. It is essential to understand such a fragmentation process as it will provide us with an insight into the  hadronization process. The jet fragmentation functions (JFFs) describe the momentum distribution of hadrons within a fully reconstructed jet. In comparison with inclusive hadron production in $pp$~\cite{Jager:2002xm,Ellis:1985er} and $ep$~\cite{Gamberg:2014eia,Hinderer:2015hra} collisions, 
by studying the longitudinal momentum distribution of the hadron in the jet rather than the hadron itself, one is able to obtain differential information on the fraction of jet momentum taken by the hadron. This process is described by collinear JFFs which are closely related to the standard collinear fragmentation functions (FFs), allowing us to more explicitly extract the universal FFs by tracking the differential momentum fraction dependency. In the sense of exclusive jet production, the theoretical advances in the collinear JFFs were first studied~\cite{Procura:2009vm,Jain:2011xz,Jain:2011iu,Chien:2015ctp,Kang:2019ahe}, then later studied in the context of inclusive jet production~\cite{Arleo:2013tya,Kaufmann:2015hma,Kang:2016ehg,Dai:2016hzf,Kang:2017yde,Bain:2017wvk,Wang:2020kar}. Besides longitudinal momentum fraction, one can study the transverse momentum distribution of the hadrons in the jet with respect to the jet axis by including transverse momentum dependence to the JFFs, namely transverse momentum dependent jet fragmentation functions (TMDJFFs)~\cite{Kang:2017glf,Kang:2020xyq,Bain:2016rrv,Makris:2017arq}, which have close connections with the standard transverse-momentum-dependent fragmentation functions (TMDFFs)~\cite{Bacchetta:2000jk,Mulders:2000sh,Metz:2016swz}.

Recently, production of back-to-back electron+jet and the corresponding jet substructure at the EIC has been proposed as novel probes of both TMDPDFs and TMDFFs~\cite{Liu:2018trl,Arratia:2020nxw,Liu:2020dct}. In this work, we present the general theoretical framework for the hadron distribution in a jet in back-to-back electron-jet production from electron-proton~($ep$) colliders, 
\bea
e^- + p \rightarrow e^- + \left(\text{jet}\left({\bm q}_T\right) h\left(z_h, {\bm j}_\perp\right)\right)+X\,,
\eea
where both incoming particles (an electron and a proton) and outgoing hadrons inside the jet have general polarizations. Here, ${\bm q}_T$ is the imbalance of the transverse momentum of the final-state electron and the jet, which is measured with respect to the beam direction (of the electron and the proton).\footnote{Alternatively, TMDPDFs were studied using the Breit frame in~\cite{Gutierrez-Reyes:2018qez,Gutierrez-Reyes:2019vbx}.} On the other hand, $z_h$ is the momentum fraction of the jet carried by the hadron, and ${\bm j}_\perp$ is the transverse momentum of the hadron inside the jet with respect to the jet axis. As we will demonstrate below, using the simultaneous differential information on ${\bm q}_T$ and ${\bm j}_\perp$, we are able to separately constrain TMDPDFs and TMDFFs. In particular, ${\bm q}_T$ is only sensitive to TMDPDFs, while the ${\bm j}_\perp$-dependence is sensitive to TMDFFs alone. On the other hand, TMDPDFs and TMDFFs are always convolved for the analogous semi-inclusive deep inelastic scattering (SIDIS) process without a jet~\cite{Bacchetta:2006tn}, and thus additional processes such as Drell-Yan production in  $pp$ collisions and dihadron production in $e^+e^-$ scatterings are required to constrain TMDPDFs and TMDFFs separately. 

We derive all the correlations which come from different combinations of TMDPDFs and TMDFFs. These correlations manifest themselves via characteristic azimuthal asymmetries with respect to the scattering plane. We find that the distribution of hadrons in the jet in back-to-back electron-jet production is sensitive to {\it all} the leading-twist TMDPDFs and TMDFFs, making this an extremely promising process to study all different flavors of standard~\footnote{By standard, we mean the TMDPDFs and TMDFFs that can be probed in the conventional SIDIS, Drell-Yan, and dihdron in $e^+e^-$ collisions~\cite{Collins:2011zzd,Ji:2004wu,Bacchetta:2006tn}. In recent years, there has been new TMD functions describing hadron distribution with respect to the winner-take-all or groomed jet axis~\cite{Neill:2016vbi,Gutierrez-Reyes:2019msa,Chien:2020hzh,Liu:2021ewb}.} TMDs found in the literature. It is instructive to compare this back-to-back electron+jet production with the production of the single inclusive jet in $ep$ collisions, which was studied in our previous paper~\cite{Kang:2020xyq}. In the case of single inclusive jet production, $e+p\to \left(\text{jet}\, h\left(z_h, {\bm j}_\perp\right)\right)+X$, the electron is not observed~\cite{Kang:2011jw,Hinderer:2015hra,Abelof:2016pby} in the final state. As shown in~\cite{Kang:2020xyq,Kang:2017glf}, single inclusive jets probe only collinear PDFs. Consequently, being differential in ${\bm j}_\perp$ would enable one to probe a single TMDFF alone (on top of a collinear PDF). On the other hand, back-to-back electron-jet production with the imbalance ${\bm q}_T$ measurements provides sensitivity to TMDPDFs and further sensitivity to TMDFFs when hadrons in the jet are measured with their transverse momentum ${\bm j}_\perp$ relative to the jet axis. With the era of the EIC looming in the near future, we select a few azimuthal asymmetries to carry out phenomenological analysis using the EIC kinematics. In the case of unpolarized electron-proton collisions, we also provide predictions for the HERA kinematics~\cite{Newman:2013ada}. 

In the rest of the paper, we develop the theoretical framework and study the phenomenological relevance of back-to-back electron-jet production from $ep$ colliders, with general polarization for the incoming electron and proton. We develop the paper sequentially by increasing the complexity of the final states produced. In section~\ref{sec:jet}, we begin by studying the back-to-back electron-jet production without observation of a hadron in the jet. In order to illustrate its relevance, we study transverse-longitudinal spin asymmetry, which allows us to access the transversal helicity TMDPDF $g_{1T}$. In section~\ref{sec:un_h}, we generalize to a case when an unpolarized hadron in a jet is observed. As a phenomenological application, we consider the $\cos(\phi_{q}-\hat{\phi}_h)$ azimuthal asymmetry sensitive to the Boer-Mulders TMDPDF $h_1^\perp$ and the Collins TMDFF $H_1^\perp$. Finally, in section~\ref{sec:pol_h}, we further generalize the framework to the case where a polarized spin-$\frac{1}{2}$ hadron, a $\Lambda$ baryon, is produced inside the jet. With this most general framework, we present $\Lambda$ transverse polarization in the jet to study the polarizing TMDFF  $D_{1T}^\perp$ using the future EIC kinematics. We conclude our paper in section~\ref{sec:summary}.

\section{Electron-jet production}
\label{sec:jet}

In this section, we study the back-to-back electron-jet production in $e+p$ collisions. By measuring fully reconstructed jets instead of hadrons, this process will be sensitive to a single TMDPDF, making it particularly useful to constrain TMDPDFs. This is to be compared with the standard TMD processes, namely SIDIS,  Drell-Yan, and $e^+e^-\to$ dihadrons, where two TMD functions appear in convolution, and thus require extra efforts to decouple them. For example, Drell-Yan process probes two TMDPDFs~\cite{Collins:1984kg,Ji:2004xq}, SIDIS is sensitive to a convolution of a TMDPDF and a TMDFF~\cite{Ji:2004wu,Bacchetta:2006tn}, while $e^+e^-\to$ dihadrons probes two TMDFFs~\cite{Collins:1981uk,Boer:1997mf}. As we will demonstrate below, the electron-jet correlation would allow us to probe all the chiral-even TMDPDFs, without any additional convolution of another TMDPDFs or TMDFFs. In this case, the chiral-odd ones cannot be studied as they require two chiral-odd functions to be convolved with each other.\footnote{We always use standard jet axis to define the jet in this paper. In a recent study~\cite{Liu:2021ewb} when a winner-take-all (WTA) jet axis is used, the chiral-odd component of jets can arise and thus would allow probing the chiral-odd TMDPDFs. How sensitive such a method is to the TMDPDFs remains to be investigated.}

The back-to-back electron-jet production has been previously proposed in context to study quark Sivers functions~\cite{Liu:2018trl,Arratia:2020nxw}. After presenting the theoretical framework to study electron-jet production with general polarizations, as a new example, we carry out a phenomenological study of transversal helicity TMDPDF, $g_{1T}$. This TMDPDF describes longitudinally polarized quarks in a transversely polarized proton, often referred to as a ``worm-gear'' function or Kotzinian-Mulders~\cite{Tangerman:1994eh,Kotzinian:1995cz} function.

\subsection{Theoretical framework}
\label{sec2:theoretical}
\begin{figure}
\centering
\includegraphics[width=3.4in]{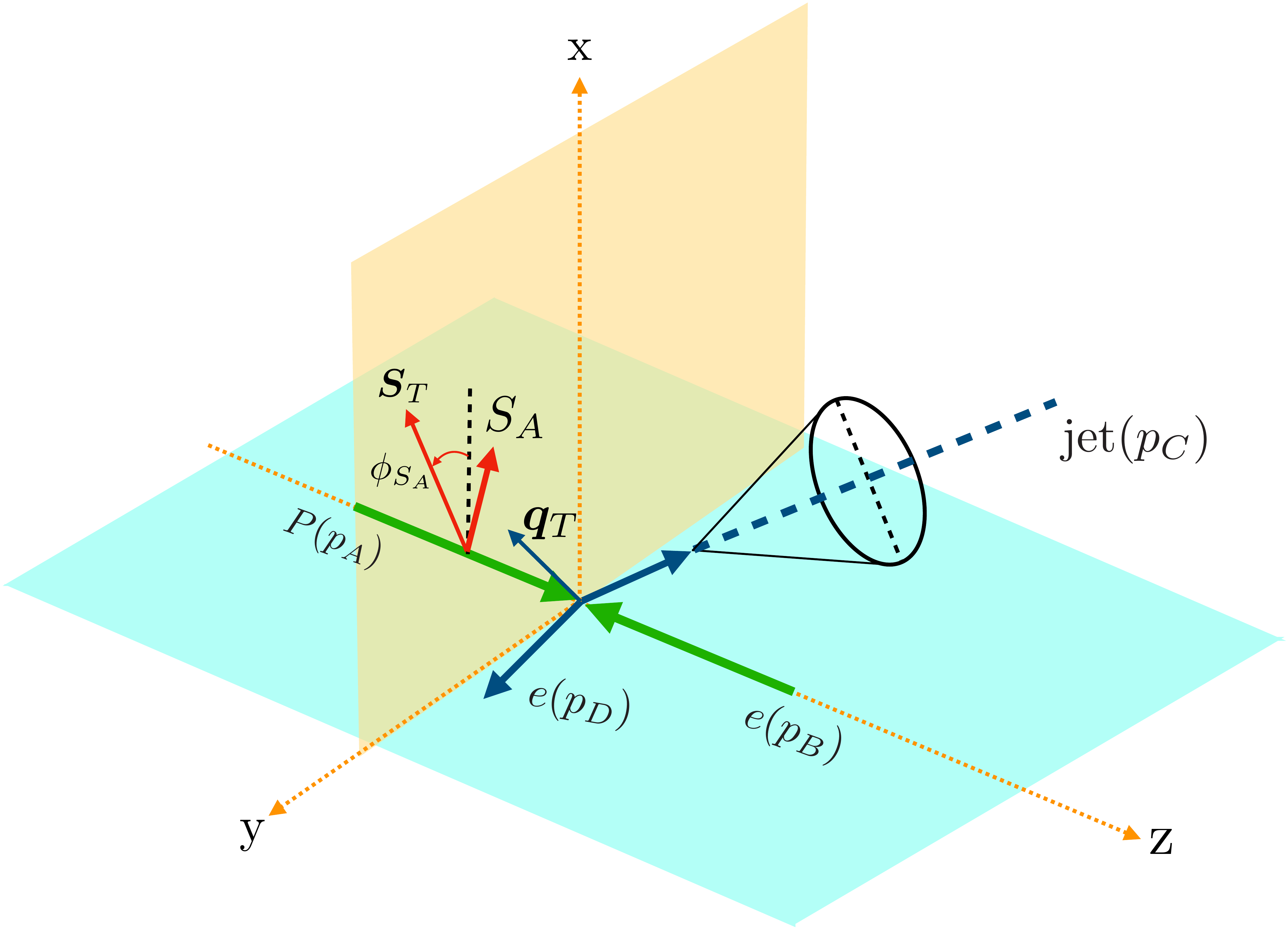}  \caption{Back-to-back electron and jet production in electron-proton collision, where $S_A$ indicates the spin of the incoming proton and $\bm q_T$ is the imbalance momentum between the outgoing electron and the final-state jet.}
\label{fig:qT}
\end{figure}

As shown in Fig. \ref{fig:qT}, we label back-to-back production of electron and jet from electron-proton collision process as
\bea
p({p}_A,{S}_{A})+e({p}_B, \lambda_e)\rightarrow \text{jet}({p}_C)+e({p}_D)+X\,,
\eea
where an electron with momentum ${p}_B$ (unpolarized or longitudinally polarized with helicity $\lambda_e$, moving along ``$-z$'' direction) scatters with a polarized proton with momentum ${p}_A$ and polarization $S_A$ (moving along ``$+z$'' direction), and such a scattering produces a jet with momentum ${p}_C$ and an electron with momentum ${p}_D$ in the final state.
To conveniently parametrize the momentum and the spin vectors involved, we first define a light-cone vector $n_a^\mu=\frac{1}{\sqrt{2}}(1,0,0,1)$ and its conjugate vector ${n}_b^\mu=\frac{1}{\sqrt{2}}(1,0,0,-1)$, such that $n_a^2 = n_b^2 = 0$ and $n_a\cdot{n_b} = 1$. Using the defined light-cone vectors, we can decompose any four-vector $p^\mu$ as $p^\mu = [p^+, p^-,p_T]$. That is, 
\bea
p^\mu =p^+ n_a^{\mu}+ p^- n_b^{\mu} + p_T^\mu\,,
\eea
where $p^+ = n_b\cdot p = \frac{1}{\sqrt{2}}\left(p^0 + p^z\right)$ and $p^- = {n_a}\cdot p = \frac{1}{\sqrt{2}}\left(p^0 - p^z\right)$. We study our process using the center-of-mass (CM) frame of the $ep$ collision, where the incoming momenta $p_{A,B}$ and the proton spin vector $S_A$ can be written as  
\bea
{p}_A^\mu=&\sqrt{\frac{s}{2}} n_a^\mu + \mathcal{O}(M)\,,\\ {p}_B^\mu=&\sqrt{\frac{s}{2}} n_b^\mu + \mathcal{O}(m_e)\,,\\
S_A^\mu =&\left[\lambda_p \frac{p_A^+}{M},-\lambda_p \frac{M}{2p_A^+},{\bm S}_T\right]\,.
\label{eq:light-cone}
\eea
Here $s=\left(p_A+p_B\right)^2$ is the CM energy,  $\lambda_p$ and $\bm S_T$ are the helicity and transverse spin vector of the incoming proton, and $M$ ($m_e$) are the mass of the proton (electron). We define the usual $Q^2 = -(p_B-p_D)^2$, the virtuality of the exchanged photon, and the event inelasticity $y= Q^2/\left(x_B s\right)$ with $x_B$ the standard Bjorken-$x$~\cite{Bacchetta:2006tn}. We set the final observed jet to be produced in the $xz$-plane as shown in Fig.~\ref{fig:qT}, with the four-momentum 
\bea
{p}_C^\mu=&E_J(1,\sin\theta_J,0,\cos\theta_J)\,,
\label{eq:jetmom}
\eea
where the angle $\theta_J$ is measured with respect to the beam direction. We refer such a $xz$-plane as the scattering plane, which is formed by the jet momentum and incoming electron-proton beam directions. Note that we use $(t,x,y,z)$ momentum representation for $p_C^\mu$, to be distinguished from the light-cone component representation in the form of brackets, e.g. in Eq.~\eqref{eq:light-cone}. 

We consider the kinematic region where the electron and the jet are produced back-to-back with the  transverse momentum imbalance $q_T = |\bm q_T|$ much smaller than the average transverse momentum $p_T = |\bm p_T|$ of the electron and the jet: $q_T \ll p_T$. Here $\bm q_T$ and $\bm p_T$ are given by
\bea
\bm{q}_T &= \bm{p}_{C,T} + \bm{p}_{D,T}\,,
\\
\bm{p}_T &=  \left(\bm{p}_{C,T} - \bm{p}_{D,T}\right)/2\,.
\label{eq:pT}
\eea
We parametrize $\bm{q}_T$ and transverse spin vector $\bm{S}_T$ in terms of their azimuthal angles as
\bea
\label{eq:qT}
\bm{q}_T  &= q_T(\cos\phi_{q},\sin\phi_{q})\,,\\
\label{eq:sT}
\bm{S}_T &= S_T(\cos\phi_{S_A},\sin\phi_{S_A})\,,
\eea
where $T$ subscript denotes that the vector is transverse with respect to the beam direction. The azimuthal angles are measured in the frame where the incoming beams and outgoing jet defines the $xz$-plane, as shown in Fig.\ \ref{fig:qT}. 
Note that we use a slight abuse of notation, $S_T = |\bm{S}_T|$ and $q_T=|\bm{q}_T|$, to denote the magnitude of the transverse vectors in Eqs.~\eqref{eq:qT} and\ \eqref{eq:sT}. This notation needs to be used with caution, since representing the four-vector as $q_T^\mu = (0,0,\bm{q}_T)$ (similarly for $S_T^\mu$) would lead to a contradiction $q_T^2 = - \bm{q}_T^2$ if one interprets $q_T^2 = q_T^\mu \,q_{T,\mu}$ . We always use $q_T$ and $S_T$ to denote the magnitude of the transverse momentum and spin, and explicitly write indices, for instance $q_T^\mu \,q_{T,\mu}$, to represent the invariant mass of a four-momentum. We also use unbolded text with Latin indices, for example ${k}_T^i$ or ${S}_T^i$, to denote the components of transverse vectors. 

Working in the one-photon exchange approximation and neglecting the electron mass, the cross section of $e+$jet production from $ep$ collision can then be expressed in terms of structure functions as
\bea
\label{eq:unpjet}
\frac{d\sigma^{p(S_A)+e(\lambda_e)\to e+\text{jet}+X}}{d{p}^2_T dy_J d^2{\bm q}_T}=&F_{UU}+\lambda_p\lambda_e F_{LL}\nnu
&\hspace{-2cm}+S_T\bigg\{\sin(\phi_q-{\phi}_{S_A})F^{\sin(\phi_{q}-{\phi}_{S_A})}_{TU}+\lambda_e\cos(\phi_{q}-{\phi}_{S_A})F^{\cos(\phi_q-{\phi}_{S_A})}_{TL}\bigg\}\,,
\eea
where $y_J$ is the rapidity of the jet, and the subscripts $A$ and $B$ in the structure functions $F_{AB}$ indicate the polarization of the incoming proton and electron. Specifically, we use $U,~L,~T$ to represent the unpolarized, longitudinal polarized, and transversely polarized particles, respectively. In the back-to-back limit where $q_T \ll p_T$, these structure functions can then be analyzed within the TMD factorization formalism, and thus allow us to study the relevant TMDPDFs. 

Before we write down the explicit TMD factorization formalism, let us first provide a short review for the definitions of the TMDPDFs for later convenience. TMDPDFs are defined through the so-called quark-quark correlation function~\cite{Mulders:1995dh}, $\Phi(x,\bm{k}_T;S)$, 
\bea
\Phi(x, \bm{k}_T; S) = \int \frac{d\xi^- d^2 \bm \xi_T}{(2\pi)^3} e^{i k\cdot \xi}
\left.\langle PS| \bar\psi(0) \psi(\xi)|PS\rangle\right|_{\xi^+ = 0}\,,
\eea
where $k^ += xp^+$ with $p^+$ is the large light-cone component of the proton, and $\bm{k}_T$ is the quark transverse momentum with respect to the parent proton. Here we have suppressed the relevant gauge link for our process, which is the same as that for SIDIS process and renders the expression on the right-hand side gauge invariant. In different processes, the structure of the gauge link can change which leads to the important and nontrivial process-dependence of the TMDPDFs~\cite{Collins:2002kn,Boer:2003cm,Bomhof:2004aw,Bacchetta:2005rm,Collins:2011zzd,Kang:2020xez,Buffing:2018ggv}. The correlation function
$\Phi(x, \bm{k}_T; S)$ can be parametrized by TMDPDFs at leading twist accuracy~\cite{Mulders:1995dh,Goeke:2005hb,Bacchetta:2006tn} as
\bea
\label{eq:TMDPDFs}
\Phi(x,\bm{k}_T;S)=&\frac{1}{2}\Bigg[\left(f_{1} - \frac{\epsilon_{T}^{ij} k_{T}^i S_{T}^j}{M}f_{1 T}^{\perp}\right) \slashed{n}_a+\left(\lambda_p g_{1L} + \frac{{\bm k}_T\cdot{\bm S}_T}{M}g_{1T}\right)\gamma_{5} \slashed{n}_a\\
& -i\sigma_{i\mu}n_a^\mu \left(h_1{S_{T}^i}\gamma_5 -ih_1^{\perp}\frac{k_{T}^i}{M}+  h_{1L}^\perp\frac{\lambda_p k_{T}^i}{M} \gamma_5+h_{1T}^\perp\frac{{\bm k}_T\cdot{\bm S}_T k_{T}^i - \frac{1}{2}k_T^2S_{T}^i}{M^2}\gamma_5\right)\Bigg]\,, \notag
\eea
where $\sigma_{\mu\nu}=\frac{i}{2}\left[\gamma_\mu,\gamma_\nu\right]$. We have eight quark TMDPDFs $f_1(x, k_T^2)$, $f_{1T}^{\perp}(x, k_T^2)$, $g_{1L}(x, k_T^2)$, $g_{1T}(x, k_T^2)$, $h_1(x, k_T^2)$, $h_1^\perp(x, k_T^2)$, $h_{1L}^\perp(x, k_T^2)$, and $h_{1T}^\perp(x, k_T^2)$, and their physical interpretations are summarized in Table.~\ref{intpdf}. For details, see~\cite{Mulders:1995dh,Goeke:2005hb,Bacchetta:2006tn,Bacchetta:2004jz,Boer:2011fh,Accardi:2012qut}.

As usual, we find it convenient to work in the Fourier or $\bm{b}$-space. Taking the Fourier transformation of the correlation function, we have 
\bea
\tilde{\Phi}\left(x, \bm{b}; S\right)=\int d^2\bm{k}_T\,e^{-i\bm{k}_T\cdot \bm{b}}\, \Phi(x,\boldsymbol{k}_T;S)\,,
\eea
and the $\bm{b}$-space correlation function $\tilde{\Phi}(x, \bm{b}; S)$ at leading twist is given by~\cite{Boer:2011xd}
\bea
\label{eq:TMDPDFsb}
\tilde{\Phi}(x,\bm{b};S)=&\frac{1}{2}\Bigg[\left(\tilde{f}_{1} + {i\epsilon_{T}^{ij} b^i S_{T}^j}{M}\tilde{f}_{1 T}^{\perp(1)}\right) \slashed{n}_a+\left(\lambda_p \tilde{g}_{1L} - {i{\bm b}\cdot{\bm S}_T}{M}\tilde{g}^{(1)}_{1T}\right)\gamma_{5} \slashed{n}_a\nonumber\\
& \hspace{0.7cm} -i\sigma_{i\mu}n_a^\mu \bigg({S_{T}^i}\tilde{h}_1\gamma_5 -{b^{i}}{M}\tilde{h}_1^{\perp(1)}- i{\lambda_p b^{i}}{M}\tilde{h}_{1L}^{\perp(1)} \gamma_5\nnu
& \hspace{2.7cm}-\frac{1}{2}\left({\bm b}\cdot{\bm S}_T b^{i}-\frac{1}{2} b^2S_{T}^i\right) M^{2} \tilde{h}_{1 T}^{\perp(2)}\gamma_5\bigg)\Bigg]\,,
\eea
where $b = |\bm{b}|$ denotes the magnitude of the vector $\bm{b}$. Here, the TMDPDFs in $\bm{b}$-space are defined as
\bea\label{btilde}
\tilde{f}^{(n)}(x,b^2)=\frac{2 \pi n !}{\left({M^{2}}\right)^{n}} \int dk_T \, k_T\left(\frac{k_T}{b}\right)^{n} J_{n}\left(k_T b\right) f\!\left(x, k_T^2\right)\,,
\eea
where $n=0$ by default when denoted without a superscript. For simplicity, we have suppressed the additional scale-dependence in both $f\left(x, k_T^2\right)$ and $\tilde{f}^{(n)}(x,b^2)$, and we will specify these scale-dependence explicitly below when we present the factorization formula. 
\begin{table} 
    \centering
\begin{tabular}{ |c|c|c|c| } 
 \hline
 \diagbox[width=4em]{$H$}{$q$} & $U$ & $L$ & $T$ \\ 
  \hline
 $U$ & $f_1$ &  & $h_1^\perp$  \\ 
  \hline
$L$ &  &  $g_{1L}$& $h_{1L}^\perp$ \\ 
  \hline
$T$ & $f_{1T}^\perp$ & $g_{1T}$ & $h_1,\ h_{1T}^\perp$\\ 
  \hline
\end{tabular}
  \caption{TMDPDFs for quarks. We have quark polarizations in the row with $U=\,$unpolarized, $L=\,$longitudinal polarized, and $T=\,$transversely polarized quarks. On the other hand, the column represents polarization of the hadron $H$ (i.e. the proton in our case).}
  \label{intpdf}
\end{table}
To illustrate how structure functions are factorized, we first describe in detail the expression for $F_{UU}$, which can be factorized into different functions in $\bm{b}$-space as~\cite{Arratia:2020nxw,Buffing:2018ggv,Kang:2020xez} 
\bea
\label{eq:FUUbefore}
F_{UU} =&\hat{\sigma}_0\,H(Q, \mu)\sum_qe_q^2\, J_{q}(p_{T}R,\mu)\int\frac{d^2\bm{b}}{(2\pi)^2}e^{i\bm{q_T}\cdot\bm{b}} x\,\tilde{f}_{1}^{q, \rm unsub}(x,b^2, \mu, \zeta/\nu^2)
\nonumber\\
&\times S_{\rm global}(\bm{b},\mu,\nu)S_{cs}(\bm{b},R,\mu)\,.
\eea
where $\hat \sigma_0$ is the Born cross section for the unpolarized electron and quark scattering process, and its expression is given below in Eq.~\eqref{eq:sig0}. In the derivation of the above factorization formula we apply the narrow jet approximation with $R\ll1$. However, as shown in~\cite{Jager:2004jh,Mukherjee:2012uz,Dasgupta:2016bnd,Liu:2018ktv} this approximation works well even for fat jets with radius $R\sim \mathcal{O}(1)$, and the power corrections of $\mathcal{O}(R^{2n})$ with $n>0$ can be obtained from the perturbative matching calculation. 
Here $H(Q, \mu)$ is the hard function and $J_q(p_TR, \mu)$ is the quark jet function. They describe the partonic hard scattering and production of the outgoing jet from a hard interaction, respectively. Their renormalized expressions at the next-to-leading order (NLO) are given by~\cite{Liu:2018trl,Arratia:2020nxw,Ellis:2010rwa}
\bea
H(Q, \mu) = &\,1+\frac{\alpha_s}{2\pi}C_F\left[-\ln^2\left(\frac{\mu^2}{Q^2}\right) - 3\ln\left(\frac{\mu^2}{Q^2}\right) -8 +\frac{\pi^2}{6}\right]\,,
\\
J_q(p_TR, \mu) = &\, 1+\frac{\alpha_s}{2\pi}C_F\left[\frac{1}{2}\ln^2\left(\frac{\mu^2}{p_T^2R^2}\right) +\frac{3}{2}\ln\left(\frac{\mu^2}{p_T^2R^2}\right) + \frac{13}{2} -\frac{3\pi^2}{4}\right]\,.
\eea
On the other hand, $\tilde{f}_{1}^{q, \rm unsub}(x,b^2, \mu, \zeta/\nu^2)$ is the so-called unsubtracted unpolarized TMDPDF~\cite{Collins:2011zzd}\footnote{Compared to Eq.~\eqref{eq:TMDPDFsb}, we now include the flavor label $q$ and also write `unsub' to emphasize the role of the rapidity divergence in TMDPDFs. The basic parametrization forms given in Eqs.~\eqref{eq:TMDPDFs} and~\eqref{eq:TMDPDFsb} do not change whether we use the rapidity divergence subtracted or unsubtracted quark-quark correlation function.}, with $\mu$ and $\nu$ denoting renormalization and rapidity scales, separately, while $\zeta$ is the so-called Collins-Soper parameter~\cite{Collins:2011zzd,Ebert:2019okf}. In the rapidity regularization scheme~\cite{Chiu:2011qc,Chiu:2012ir} we implement in our calculation, we have $\zeta$ related to the large light-cone component of the quark inside the proton: $\sqrt{\zeta} = \sqrt{2}\,x\, p_A^+ = x \sqrt{s}$, which will be used in our numerical studies. Finally, the global soft function $S_{\rm global}(\bm{b},\mu,\nu)$ describes the soft radiation that has no phase space restriction and does not resolve the jet cone. The collinear-soft function $S_{cs}(\bm{b},R,\mu)$ describes the soft radiation which is only sensitive to the jet direction and resolves the jet cone. The collinear-soft function $S_{cs}(\bm{b},R,\mu)$ depends on the jet radius $R$ and is simply given by the soft radiation outside the jet cone at the NLO. The expressions of $S_{\rm global}$ and $S_{cs}$ at one-loop are given by~\cite{Buffing:2018ggv}
\bea
S_{\rm{global}}(\bm{b},\mu, \nu) &= 1+ \frac{\alpha_s}{2\pi} C_F\left[ \left( - \frac{2}{\eta} + \ln \frac{\mu^2}{\nu^2} + 2y_J + 2\ln(-2i\cos(\phi_1)) \right)\left(\frac{1}{\epsilon} + \ln\frac{\mu^2}{\mu_b^2}\right)\right. \nnu
&\hspace{2.6cm} \left.+ \frac{2}{\epsilon^2} + \frac{1}{\epsilon} \ln\frac{\mu^2}{\mu_b^2} - \frac{\pi^2}{6} \right]\,, \label{eq:Sglobalbefore}
\\
S_{cs}(\bm{b},R,\mu) &= 1- \frac{\alpha_s}{2\pi} C_F\left[\frac{1}{\epsilon^2} + \frac{2}{\epsilon}\ln \left(\frac{-2i\cos(\phi_1)\mu}{\mu_b R}\right)+2\ln^2\left(\frac{-2i\cos(\phi_1)\mu}{\mu_b R}\right) + \frac{\pi^2}{4}\right]\,,\label{eq:Scsbefore}
\eea
where we have $\mu_b = 2e^{-\gamma_E}/b$, while $\phi_1 \equiv \phi_b - \phi_J$ with $\phi_b$ and $\phi_J$ are the azimuthal angles of the vector $\bm{b}$ and jet transverse momentum $\bm{p}_T$, respectively. Note that we work in $4 - 2\epsilon$ space-time dimensions and use the rapidity regulator~\cite{Chiu:2012ir} $\eta$ and the rapidity scale parameter~$\nu$. 

It is instructive to mention that the Fourier transform given in Eq.~\eqref{eq:FUUbefore} is further complicated by the soft functions' dependence on the azimuthal angle of the jet produced, namely $\phi_1$. These non-trivial azimuthal angle dependence of the soft functions can give rise to additional azimuthal angle correlations for electron-jet production. For example, it was shown in~\cite{Hatta:2020bgy,Hatta:2021jcd} that some novel azimuthal angle correlations can arise from the soft radiation in back-to-back dijet production in unpolarized proton-proton collisions, which involve $\phi_q - \phi_J$. In general, our formalism would naturally lead to such an azimuthal dependence in the unpolarized scattering if one realizes that the azimuthal dependence in $\phi_b - \phi_J$ would translate to the $\phi_q - \phi_J$ dependence after the Fourier transform from $\bm{b}$ back to $\bm{q}_T$ space. We would expect that such $\phi_1$-dependence would lead to additional correlations in the polarized scattering, which involve $\phi_S$ in general, the azimuthal angle of the proton spin.

In the current paper, however, we concentrate on the study of the TMDPDFs and TMDFFs, and we thus ignore such type of correlations purely due to the soft radiation. To proceed and to simplify our phenomenology below, we make a few simplifications. We integrate over the jet azimuthal angle $\phi_J$ as indicated in Eq.~\eqref{eq:unpjet}, i.e., we only measure the magnitude of the jet transverse momentum $p_T$, in which case the azimuthal correlations involving $\phi_J$ would vanish. Additionally, we take $\phi_b$-averaging in both the global and collinear-soft functions, following our previous papers in~\cite{Kang:2020xez,Kang:2020xgk,Chien:2019gyf}. We therefore arrive at the following expression
\bea
\label{eq:FUU_unsub}
F_{UU}=&\hat{\sigma}_0\,H(Q,\mu)\sum_qe_q^2\, J_{q}(p_TR,\mu)\int\frac{b \,db}{2\pi}J_0(q_Tb)\, x\,\tilde{f}^{q, \rm unsub}_{1}(x,b^2, \mu, \zeta/\nu^2)
\nonumber\\
&\times \bar{S}_{\rm global}(b^2,\mu, \nu)\bar{S}_{cs}(b^2,R,\mu)\,,
\eea
where the bar in $\bar{S}$ indicates that the azimuthal angle averaged version of the soft functions are only sensitive to the magnitude of $\bm{b}$. The one-loop expressions of the azimuthal angle averaged soft functions are~\cite{Kang:2020xez,Hornig:2017pud}
\bea
\bar{S}_{\rm global}(b^2,\mu, \nu) &=1+ \frac{\alpha_s}{2\pi} C_F\left[ \left( - \frac{2}{\eta} + \ln \frac{\mu^2}{\nu^2} + 2y_J \right)\left(\frac{1}{\epsilon} + \ln\frac{\mu^2}{\mu_b^2}\right) + \frac{2}{\epsilon^2} + \frac{1}{\epsilon} \ln\frac{\mu^2}{\mu_b^2} - \frac{\pi^2}{6} \right]\,,\\
\bar{S}_{cs}(b^2,R,\mu)&= 1- \frac{\as C_F}{2\pi}\left[\frac{1}{\epsilon^2} + \frac{1}{\epsilon}\ln\frac{\mu^2}{\mu_b^2 R^2}+\frac{1}{2}\ln^2\frac{\mu^2}{\mu_b^2 R^2} - \frac{\pi^2}{12}\right]\,.
\label{eq:S_cs}
\eea
Note that in the usual SIDIS and Drell-Yan processes, the rapidity divergence and thus the rapidity scale $\nu$ cancel between the 
unsubtracted TMDPDF $\tilde{f}^{q, \rm unsub}_{1}(x,b^2, \mu, \zeta/\nu^2)$ and a square-root of the standard soft function $\sqrt{S_{ab}(b^2,\mu,\nu)}$, whose expressions are the same for SIDIS and Drell-Yan process and can be found in~\cite{Collins:2011zzd,Kang:2017glf}:
\bea
{S}_{ab}(b^2,\mu,\nu)&=1 - \frac{\alpha_s C_F}{2\pi} \Bigg[2\left(\frac{2}{\eta}+\ln\frac{\nu^2}{\mu^2}\right)\left(\frac{1}{\epsilon}+\ln\frac{\mu^2}{\mu_b^2}\right)+
\ln^2\frac{\mu^2}{\mu_b^2}
-\frac{2}{\epsilon^2}+\frac{\pi^2}{6}
\Bigg]\,.
\label{eq:S-ab}
\eea
This allows us to define the standard TMDPDFs~\cite{Collins:2011zzd} $\tilde{f}^{q}_{1}(x,b^2, \mu, \zeta)$ that are free of rapidity divergence
\bea
\tilde{f}^{q}_{1}(x,b^2, \mu,\zeta) = \tilde{f}^{q, \rm unsub}_{1}(x,b^2, \mu, \zeta/\nu^2) \sqrt{S_{ab}(b^2, \mu, \nu)}\,.
\label{eq:standard-f1}
\eea
Plugging this equation to Eq.~\eqref{eq:FUU_unsub}, we end up with the following expression
\bea
\label{eq:FUU}
F_{UU}=&\hat{\sigma}_0\,H(Q,\mu)\sum_qe_q^2\, J_{q}(p_TR,\mu)\int\frac{b \,db}{2\pi}J_0(q_Tb)\,x\,\tilde{f}^{q}_{1}(x,b^2, \mu,\zeta)
\nonumber\\
&\times \bar{S}_{\rm global}(b^2,\mu)\bar{S}_{cs}(b^2,R,\mu)\,,
\eea
where $\bar{S}_{\rm global}(b^2,\mu)$ is a rapidity divergence independent soft function defined as
\bea
\bar{S}_{\rm global}(b^2,\mu) = \frac{\bar{S}_{\rm global}(b^2,\mu, \nu)}{\sqrt{S_{ab}(b^2, \mu, \nu)}}\,,
\eea
with the following expression at the NLO
\bea
\bar{S}_{\rm global}(b^2,\mu) &=1+ \frac{\alpha_s}{2\pi} C_F\left[  2y_J \left(\frac{1}{\epsilon} + \ln\frac{\mu^2}{\mu_b}\right)+ \frac{1}{\epsilon^2} + \frac{1}{\epsilon} \ln\frac{\mu^2}{\mu_b^2}+ \frac{1}{2} \ln^2\frac{\mu^2}{\mu_b^2}  - \frac{\pi^2}{12} \right]\,.
\label{eq:S_global}
\eea
The renormalization group equations for $\bar{S}_{cs}(b^2,R,\mu)$ and $\bar{S}_{\rm global}(b^2,\mu)$ are given by
\bea
\mu\frac{d}{d\mu}\ln \bar{S}_{cs}(b^2,R,\mu) =&\, \gamma^S_{cs}\left(b, R,\mu\right)\,,\\
\mu\frac{d}{d\mu}\ln \bar{S}_{\rm global}(b^2,\mu) =&\, \gamma^S_{\rm global}\left(b, \mu\right)\,,
\eea
where the anomalous dimensions $\gamma^S_{\rm cs}$ and $\gamma^S_{\rm global}$ can be obtained from Eqs.~\eqref{eq:S_cs} and~\eqref{eq:S_global}
\bea
\gamma^S_{ cs}\left(b, R, \mu\right) &= -\frac{\alpha_sC_F}{\pi}\ln\frac{\mu^2}{\mu_b^2 R^2}\,,\\
\gamma^S_{\rm global}\left(b, \mu\right) &= \frac{\alpha_s C_F}{\pi}\left(2y_J + \ln\frac{\mu^2}{\mu_b^2}\right)\,.
\eea

To proceed, we use the usual Collins-Soper-Sterman (CSS) formalism for the unpolarized TMDPDFs $\tilde{f}^{q}_{1}(x,b^2, \mu,\zeta)$ in Eq.~\eqref{eq:FUU}, which are given by
\bea
\tilde{f}^{q}_{1}(x,b^2, \mu,\zeta) = 
\left[C_{q\leftarrow i}\otimes f_{1}^{i}\right]\left(x,\mu_{b_*}\right)
\exp\left[-S_{\rm pert}\left(\mu, \mu_{b_*} \right) - S_{\rm NP}^f\left(x, b, Q_0, \zeta\right)\right]\,.
\eea
Here we have performed an operator product expansion in terms of the unpolarized collinear PDFs $f_1^i(x, \mu_{b_*})$ with the convolution defined as follows 
\bea
\left[C_{q\leftarrow i}\otimes f_{1}^{i}\right]\left(x,\mu_{b_*}\right) = \int_x^{1} \frac{d\hat{x}}{\hat{x}} C_{q\leftarrow i}\left(\frac{x}{\hat{x}}, \mu_{b_*} \right) f_1^{i}\left(\hat{x}, \mu_{b_*}\right)\,,
\eea
where the sum over repeated indicies are understood and we follow the usual $b_*$ prescription~\cite{Collins:1984kg} with $b_* = b/\sqrt{1+b^2/b_{\rm max}^2}$ and $b_{\rm max} = 1.5$~GeV$^{-1}$. The coefficient functions $C_{q\leftarrow i}$ at the NLO can be found in~\cite{Aybat:2011zv,Kang:2015msa,Collins:2011zzd}, with even N$^3$LO results available at~\cite{Luo:2020epw,Ebert:2020yqt}. On the other hand, the perturbative Sudakov factor $S_{\rm pert}\left(\mu, \mu_{b_*} \right)$ is given by 
\bea
\label{eq:Sud-pert}
S_{\rm pert}\left(\mu, \mu_{b_*} \right)  = -\tilde K(b_*,\mu_{b_*})\ln\left(\frac{\sqrt{\zeta}}{\mu_{b_*}}\right) -\int_{\mu_{b_{*}}}^{\mu} \frac{d\mu'}{\mu'}\left[\gamma_F\left(\alpha_s(\mu'), \frac{\zeta}{\mu'^2}\right) \right]\,.
\eea
Note that at the next-to-leading logarithmic (NLL) level, $\tilde K(b_*,\mu_{b_*})=0$ and 
\bea
\gamma_F\left(\alpha_s(\mu), \frac{\zeta}{\mu^2}\right) =& \frac{\alpha_s}{\pi}C_F  \left(\ln\frac{Q^2}{{\mu}^2} - \frac{3}{2}\right)\,
\nonumber\\
& +\frac{\alpha_s^2}{\pi^2} C_F \left[C_A\left(\frac{67}{18}-\frac{\pi^2}{6}\right)-\frac{10}{9} T_R\, n_f\right]\ln\frac{\zeta}{{\mu}^2} \,.
\eea
On the other hand, for the non-perturbative Sudakov factor $S_{\rm NP}^f\left(x, b, Q_0, \zeta\right)$, we use the parametrization~\cite{Sun:2014dqm,Echevarria:2020hpy}
\bea
\label{eq:Sud-NP}
S_{\rm NP}^f(x, b, Q_0,\zeta) = \frac{g_2}{2}\ln{\frac{\sqrt{\zeta}}{Q_0}}\ln{\frac{b}{b_*}}+g_1^f b^2\,,
\eea
with $Q_0=\sqrt{2.4}$~GeV, $g_2=0.84$ and $g_1^f = 0.106$ GeV$^2$. One last ingredient for the factorization formula in Eq.~\eqref{eq:FUU} has to do with the so-called non-global logarithms~(NGLs)~\cite{Dasgupta:2001sh}. For recent developments on the NGLs, see~\cite{Becher:2015hka,Becher:2016mmh,Becher:2016omr,Becher:2017nof,Caron-Huot:2015bja,Nagy:2016pwq,Nagy:2017ggp,Larkoski:2015zka}. At the NLL accuracy, they can be included as a multiplication factor, see e.g. Refs.~\cite{Dasgupta:2001sh,Arratia:2020nxw,Liu:2020dct,Kang:2020xez,Kang:2020yqw,Gamberg:2021iat}. We find that the NGLs have only very mild effects on our numerical results below, and so we will not include them in the calculations. 

To present results for other structure functions in Eq.~\eqref{eq:unpjet} with a more compact notation, we define
\bea
\mathcal{C}_{nk}^{\text{Jet}}[A(x,b^2)]=&\hat{\sigma}_{k}H(Q, \mu)\sum_qe_q^2 J_{q}(p_TR,\mu)M^n\int\frac{b^{n+1}db}{2\pi n!}J_n(q_T b)\,x \,A(x,b^2)\,,
\eea
where $\hat \sigma_0$ and $\hat \sigma_L$ correspond to the born-level partonic cross sections of unpolarized scattering $eq\to eq$ and longitudinally polarized scattering $e_Lq_L \to eq$, respectively. They are given by 
the following expressions
\bea
\label{eq:sig0}
\hat{\sigma}_0=&\frac{\alpha_{\rm em}\alpha_s}{sQ^2}\frac{2(\hat{u}^2+\hat{s}^2)}{\hat{t}^2}\,,
\\
\hat{\sigma}_L=&\frac{\alpha_{\rm em}\alpha_s}{sQ^2}\frac{2(\hat{u}^2-\hat{s}^2)}{\hat{t}^2}\,.
\eea
Here $\hat s,~\hat t,~\hat u$ are the Mandelstam variables for the partonic $q(xp_A)+e(p_B)\to q(p_C)+e(p_D)$ process and are given by
\bea
\hat s = (xp_A + p_B)^2\,, \qquad
\hat t = (p_B - p_D)^2\,,\qquad
\hat u = (xp_A - p_D)^2\,.
\eea
With the compact notation defined, we find the following factorized expressions for the structure functions in Eq.~\eqref{eq:unpjet}
\begin{subequations}\label{eq:strjlep}
\bea
F_{UU}(q_T) &= \mathcal{C}_{00}^{\text{Jet}}[\tilde{f}_1(x,b^2, \mu, \zeta)\bar{S}_{\rm global}(b^2,\mu)\bar{S}_{cs}(b^2,R,\mu)]\,,
\label{eq:strjlep1}
\\
F_{LL}(q_T)&=\mathcal{C}_{0L}^{\text{Jet}}[\tilde{g}_{1L}(x,b^2, \mu, \zeta)\bar{S}_{\rm global}(b^2,\mu)\bar{S}_{cs}(b^2,R,\mu)]\,,
\label{eq:strjlep2}
\\
F_{TU}^{\sin({\phi}_{q}-{\phi}_{S_A})}(q_T)&= \mathcal{C}_{10}^{\text{Jet}}\left[\tilde{f}_{1T}^{\perp(1)}(x,b^2, \mu, \zeta)\bar{S}_{\rm global}(b^2,\mu)\bar{S}_{cs}(b^2,R,\mu)\right]\,,
\label{eq:strjlep3}
\\
F_{TL}^{\cos({\phi}_{q}-{\phi}_{S_A})}(q_T)&=\mathcal{C}_{1L}^{\text{Jet}}\left[\tilde{g}_{1T}^{(1)}(x,b^2, \mu, \zeta)\bar{S}_{\rm global}(b^2,\mu)\bar{S}_{cs}(b^2,R,\mu)\right]\,.
\label{eq:strjlep4}
\eea
\end{subequations}
From Eq.~\eqref{eq:strjlep}, we find that the back-to-back electron-jet production provides access to the four chiral-even TMDPDFs: $f_1,~g_{1L},~f_{1T}^{\perp},~g_{1T}$. This is what we have advocated at the beginning of the section, electron-jet production allows us to probe all chiral-even TMDPDFs, but not the chiral-odd ones which usually require the process to be coupled with another chiral-odd function. We summarize the results with their characteristic asymmetries in Table.~\ref{scn1tab}.
\begin{table}[h]
\centering
\begin{tabular}{ |c|c|c|c|c| } 
 \hline
 \diagbox[width=1.8cm]{Jet}{PDF} & $f_1$ & $g_{1L}$ & $f_{1T}^\perp$& $g_{1T}$   \\ 
  \hline
   ${J}_q$ & 1 & 1 & $\sin({\phi}_{q}-{\phi}_{S_A})$ & $\cos({\phi}_{q}-{\phi}_{S_A})$  \\ 
  \hline
\end{tabular}
  \caption{Summary of the characteristic azimuthal asymmetry with which different TMDPDFs arise for back-to-back electron-jet production. See Eqs.~\eqref{eq:unpjet} and \eqref{eq:strjlep} for details.} \label{scn1tab} 
\end{table}

\subsection{Phenomenology: the transverse-longitudinal asymmetry}
\label{subsec:g1T}
In this subsection, we study applications of back-to-back electron-jet production in $ep$ collisions. Some phenomenological applications of this framework, e.g. Sivers asymmetry $F_{TU}^{\sin({\phi}_{q}-{\phi}_{S_A})}$, have been considered previously in Refs.~\cite{Liu:2018trl,Arratia:2020nxw}, but the framework has never been completely generalized fully as in Eqs.~\eqref{eq:unpjet} and~\eqref{eq:strjlep}.

Our general framework presented in the previous subsection can bring novel insights into the study of spin and TMD effects for all chiral-even TMDPDFs, especially at the future EIC. Here, we consider an azimuthal correlation that has never been studied before in the context of back-to-back electron-jet production, specifically the double transverse-longitudinal spin asymmetry, $A_{TL}^{\cos({\phi}_{q}-\phi_{S_A})}$, defined as 
\bea\label{eq:atl}
A_{TL}^{\cos({\phi}_{q}-\phi_{S_A})}=\frac{F_{TL}^{\cos(\phi_{q}-{\phi}_{S_A})}}{F_{UU}}\,,
\eea
using the future EIC kinematics. To generate such a transverse-longitudinal spin asymmetry, we consider the situation in which a transversely polarized proton collides with a longitudinally polarized electron. As defined in Eq.~\eqref{eq:strjlep}, the denominator $F_{UU}$ is related to the unpolarized quark TMDPDFs $f_1^q(x, k_T^2)$, while the numerator $F_{TL}^{\cos(\phi_{q}-{\phi}_{S_A})}$ is related to the quark transversal helicity distributions $g_{1T}^q(x, k_T^2)$, which describe the distributions of longitudinally polarized quarks inside the transversely polarized proton as shown in Table.~\ref{intpdf}. 

To compute the unpolarized structure function $F_{UU}$, we use the unpolarized TMDPDFs $f_1^q(x, b^2, \mu, \zeta)$ extractions in ~\cite{Echevarria:2020hpy,Sun:2014dqm} based on the parametrization discussed in last section. On the other hand, for computing the polarized structure function $F_{TL}^{\cos(\phi_{q}-{\phi}_{S_A})}$, we would need the parameterization for $g_{1T}^q(x, b^2, \mu, \zeta)$ in the $b$-space. There are experimental measurements from HEREMES~\cite{Airapetian:2020zzo}, COMPASS~\cite{Parsamyan:2013fia} and Jefferson Lab~\cite{Huang:2011bc} on the double transverse-longitudinal spin asymmetry for the SIDIS process, which in principle would allow us to extract $g_{1T}^q$. However, such a global analysis for $g_{1T}^q$ is not yet available. Below, to estimate the spin asymmetry at the future EIC, we take the parametrization in~\cite{Kotzinian:2006dw}, which simply implements a Wandzura-Wilczek (WW) type approximation and has shown to describe the COMPASS data reasonably well~\cite{Bastami:2018xqd}.

A simple Gaussian form is provided in~\cite{Kotzinian:2006dw} for $g_{1T}^q$ as follows
\bea
g_{1T}^q(x, k_T^2) = g_{1T}^{q(1)}(x) \frac{2M^2}{\pi\langle k_T^2\rangle_{g_{1T}}}e^{-k_T^2/\langle k_T^2\rangle_{g_{1T}}}\,,
\eea
where $\langle k_T^2\rangle_{g_{1T}} = 0.15$~GeV$^2$ and $g_{1T}^{q(1)}(x)$ is a collinear function such that 
\bea
g_{1T}^{q(1)}(x) = \int d^2\bm{k}_T \,\frac{k_T^2}{2M^2} \,g_{1T}^q\left(x, k_T^2\right)\,.
\eea
From Eqs.~\eqref{btilde} and \eqref{eq:strjlep4}, we obtain the following expression within the Gaussian model for $\tilde{g}^{q(1)}_{1T}(x,b^2)$, the relevant quantity in the TMD factorization formula
\bea
\tilde{g}^{q(1)}_{1T}(x,b^2)&=\frac{2 \pi}{M^{2}} \int dk_T\frac{k_T^2}{b} J_{1}\left(k_T b\right) g_{1T}\left(x, k_T^2\right)=2\pi\, g_{1T}^{q(1)}(x)\,e^{-b^2\frac{\langle k_T^2\rangle_{g_{1T}}}{4}}\,.
\label{eq:g1T-gaussian}
\eea
For the numerical value of $g_{1T}^{q(1)}(x)$, one further applies a WW-type approximation to relate it to the collinear helicity distribution $g_{1L}^q(x)$ as~\cite{Metz:2008ib}
\bea
g_{1T}^{q(1)}(x)&\approx x\int_x^1\frac{dz}{z}g_{1L}^q(z)\,,
\eea
where we use the helicity PDFs $g^q_{1L}$ determined by the NNPDF collaboration~\cite{Ball:2013lla} at the scale $\mu_{b_*}$ as we will see below.

The simple Gaussian model for $\tilde{g}^{q(1)}_{1T}(x,b^2)$ in Eq.~\eqref{eq:g1T-gaussian} would then allow us to compute the double spin asymmetry $A_{TL}^{\cos({\phi}_{q}-\phi_{S_A})}$. Since we are using unpolarized TMDPDFs $\tilde{f}^{q}_{1}(x,b^2, \mu,\zeta)$ with proper TMD evolution for computing the denominator $F_{UU}$ in Eq.~\eqref{eq:atl}, we will also implement the TMD evolution to 
promote the Gaussian result for $\tilde{g}^{q(1)}_{1T}$ in Eq.~\eqref{eq:g1T-gaussian}. In this case, we include both perturbative and non-perturbative Sudakov factor, and $\tilde{g}^{q(1)}_{1T}(x,b^2, \mu,\zeta)$ can then be written as
\bea
\tilde{g}^{q(1)}_{1T}(x,b^2, \mu,\zeta) = 
2\pi g_{1T}^{q(1)}\left(x,\mu_{b_*}\right)
\exp\left[-S_{\rm pert}\left(\mu, \mu_{b_*} \right) - S_{\rm NP}^{g_{1T}}\left(x, b, Q_0, \zeta\right)\right]\,,
\eea
where the perturbative Sudakov factor $S_{\rm pert}$ is spin-independent and is given in Eq.~\eqref{eq:Sud-pert}. The non-perturbative Sudakov factor for $g_{1T}$ takes the similar form as that of the unpolarized counterpart given in Eq.~\eqref{eq:Sud-NP} as
\bea
S_{\rm NP}^{g_{1T}}\left(x, b, Q_0, \zeta\right) = \frac{g_2}{2}\ln{\frac{\sqrt{\zeta}}{Q_0}}\ln{\frac{b}{b_*}}+g_1^{g_{1T}} b^2\,.
\eea
Here $g_2$ controls the non-perturbative TMD evolution from the scale $Q_0$ to $\sqrt{\zeta}$, and it is universal for all different types of TMDs and certainly spin-independent~\cite{Collins:2011zzd}. Thus, it is the same as that for $\tilde{f}^{q}_{1}$ given below Eq.~\eqref{eq:Sud-NP}. On the other hand, the $g_1^{g_{1T}}$ parameter depends on the type of TMDs, and can be interpreted as the intrinsic transverse momentum width for the relevant TMDs at the momentum scale $Q_0=\sqrt{2.4}$ GeV ~\cite{Aybat:2011zv,Anselmino:2012aa,Echevarria:2014xaa}. To match the Gaussian model in Eq.~\eqref{eq:g1T-gaussian}, we take
\bea
g_1^{g_{1T}} = \frac{\langle k_T^2\rangle_{g_{1T}}}{4} = 0.0375~{\rm GeV}^2\,.
\eea

\begin{figure}[hbt!]
    \centering
    \includegraphics[width = 0.45\textwidth]{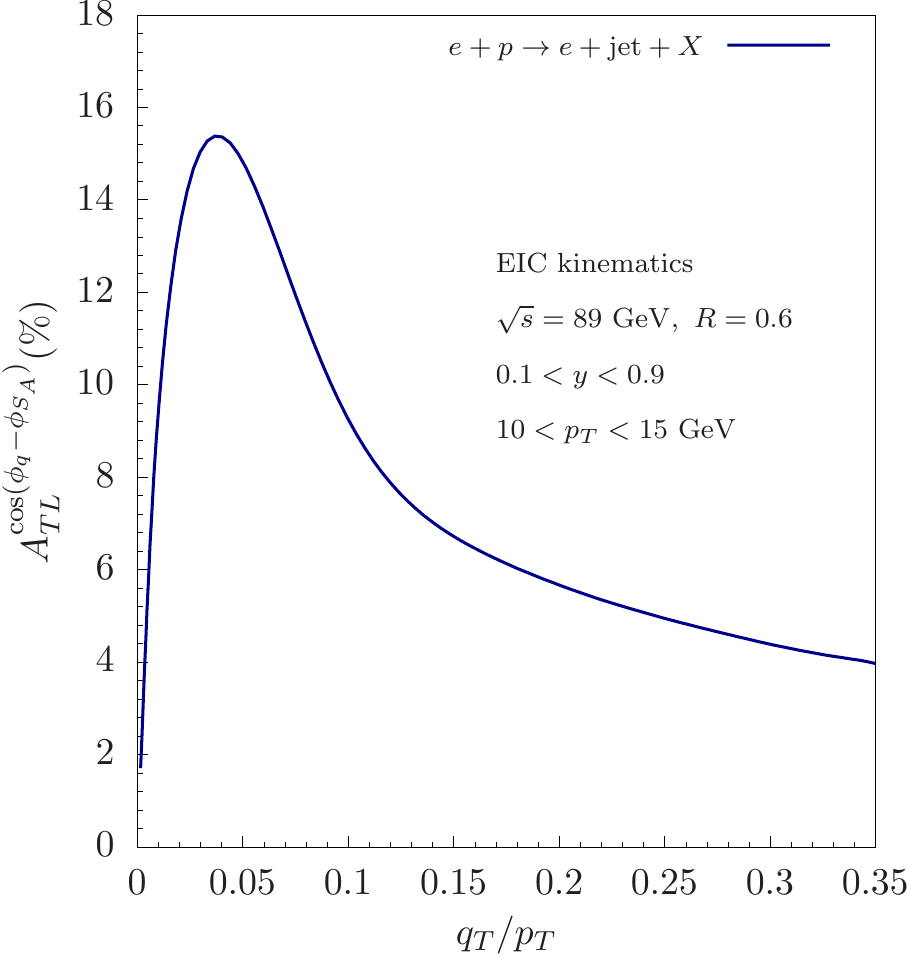}
    \caption{Spin asymmetry of back-to-back electron-jet production from the collision of transversely polarized proton and longitudinally polarized electron as a function of $q_T/p_T$, the ratio between the momentum imbalance $q_T$ and the jet transverse momentum $p_T$. Here we apply the center-of-mass energy $\sqrt{s}=89$ GeV of EIC kinematics, jet radius $R=0.6$, event inelasticity $0.1<y<0.9$ and jet transverse momentum $10 < p_T < 15$ GeV.}
    \label{fig:atl}
\end{figure}
In Fig.~\ref{fig:atl}, we present our numerical result of $A_{TL}^{\cos(\phi_{q}-\phi_{S_A})}$ as a function of $q_T/p_T$. We consider electron-proton collisions at $\sqrt{s}=89$ GeV at the future EIC, and the jets are constructed via anti-$k_T$ algorithm~\cite{Cacciari:2008gp} with jet radius $R=0.6$. We integrate over the event inelasticity $0.1<y<0.9$ and jet transverse momentum $10<p_T<15$ GeV. Note that in the back-to-back region, we have $y=1-\frac{p_T e^{y_J}}{x\sqrt{s}}$ and thus the kinematic cuts on $y$ translate into the cuts on rapidity $y_J$ and transverse momentum $p_T$ for the jet. As shown in Fig.~\ref{fig:atl}, we get a sizeable positive double spin asymmetry, demonstrating its promise at the future EIC.

\section{Unpolarized hadron inside a jet} \label{sec:un_h}
In this section, we study the back-to-back electron-jet production with unpolarized hadron observed inside jets. In particular, besides the electron-jet transverse momentum imbalance~$\bm{q}_T$, we also observe transverse momentum~$\bm{j}_\perp$ distribution of hadrons inside the jet with respect to the jet axis. Observation of a hadron inside a jet makes the process sensitive to a TMDPDF and a TMDJFF simultaneously. Unlike the counterpart process involving a hadron without observation of a jet, such as SIDIS, further dependence in $\bm{j}_\perp$ allows the two TMDs to be separately constrained. In this section, we consider only unpolarized hadron (such as pions) inside the jet and we write down the complete azimuthal modulations for the cross section. The well-known Collins asymmetry for hadrons in a jet in $ep$ collisions is one of such azimuthal modulations~\cite{Arratia:2020nxw}. In this section, to illustrate the usefulness of this process in constraining TMD functions, we carry out a new phenomenological study as an example. This azimuthal modulation is referred to as $A_{UU,U}^{\cos(\phi_{q}-\hat{\phi}_h)}$, with the dependence on azimuthal angles $\phi_q$ and $\hat \phi_h$, and it allows us to study Boer-Mulders TMDPDFs $h_{1}^\perp$ and Collins TMDFFs $H_{1}^\perp$. 

\subsection{Theoretical framework}
\label{sec3:Theoretical}
\begin{figure}
    \centering
    \includegraphics[width=3.4in]{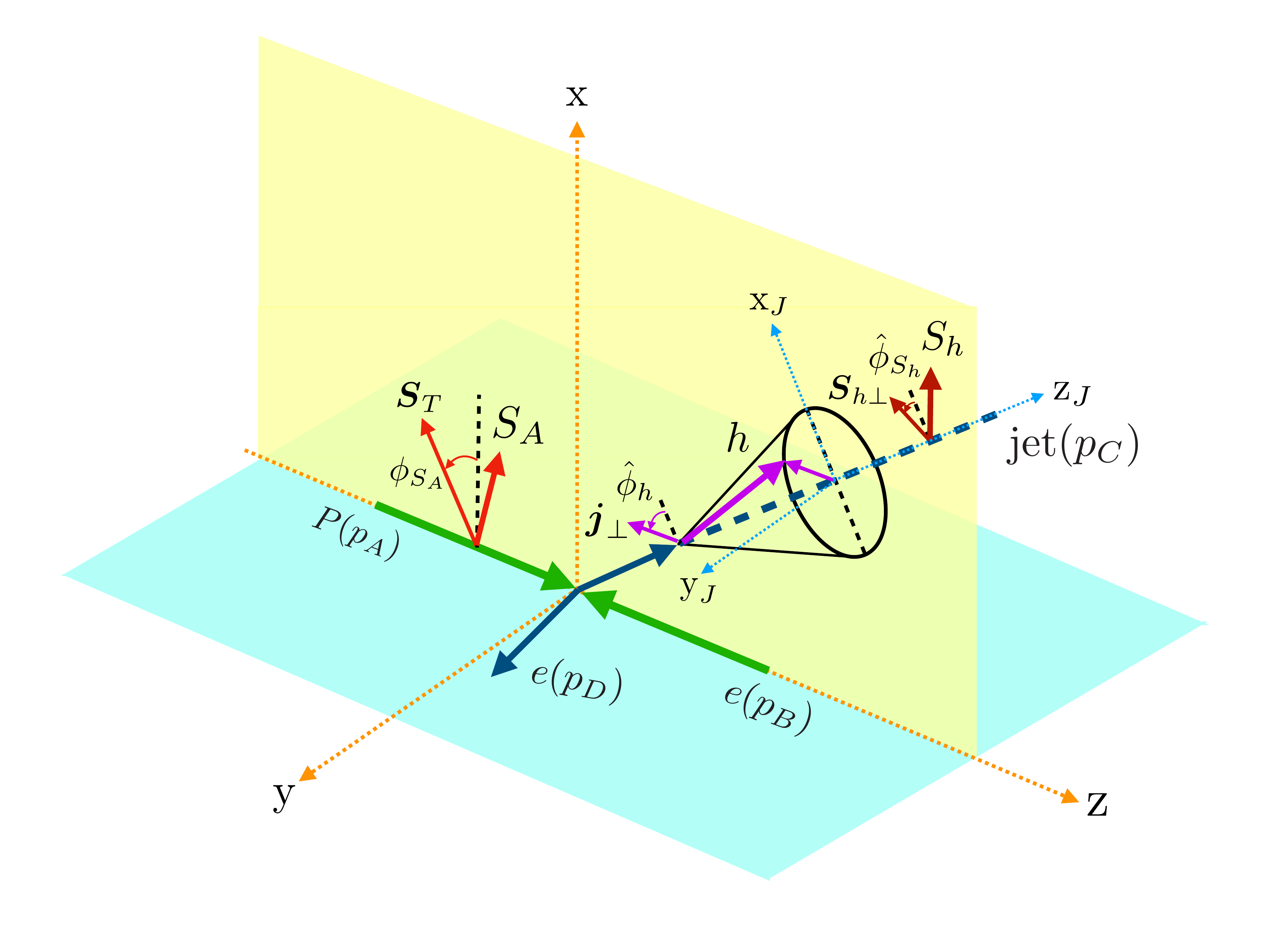} 
    \caption{Electron and hadron in jet production from back-to-back electron-proton collision, where $S_A$ indicates the spin of the incoming proton, $S_{h}$ is the spin of the produced hadron in a jet. The jet axis and colliding direction define the $xz$-plane.}
    \label{fig:jperp}
\end{figure}

We generalize the theoretical framework developed in Sec.\ \ref{sec2:theoretical} to the case where one measures distribution of unpolarized hadrons inside the jet, 
\bea
p({p}_A,{S}_{A})+e({p}_B, \lambda_e)\rightarrow \Big[\text{jet}({p}_C)\, h\left(z_h, \bm{j}_\perp\right)\Big]+e({p}_D)+X\,,
\eea
where $z_h$ is the longitudinal momentum fraction of the jet carried by the hadron $h$ and $\bm{j}_\perp$ is hadron's transverse momentum with respect to the jet axis. The details of such a scattering are illustrated in Fig.~\ref{fig:jperp}. In comparison with the subscript $T$ discussed below Eq.~\eqref{eq:sT}, which refers to the transverse momentum with respect to the incoming beam direction, we use $\perp$ to denote the transverse vector relative to the jet axis. We parametrize $\bm{j}_\perp$ in the $ep$ center-of-mass frame as
\bea
\bm{j}_{\perp}=&j_\perp(\cos\hat{\phi}_h\cos\theta_J,\sin\hat{\phi}_h,-\cos\hat{\phi}_h\sin\theta_J)\,,
\label{eq:bmj}
\eea 
where $\theta_J$ is defined in Eq.~\eqref{eq:jetmom}, and we follow the slight abuse of notation discussed in the previous section below Eq.~\eqref{eq:sT} and use $j_\perp = |\bm{j}_{\perp}|$ to denote the magnitude of the transverse vector $\bm{j}_{\perp}$. On the other hand, $\hat{\phi}_{h}$ is the azimuthal angle of the produced hadron transverse momentum $\bm{j}_\perp$ in the jet frame $x_J y_J z_J$ shown in Fig.~\ref{fig:jperp}. Recall that the scattering plane is the $xz$-plane formed by the jet momentum and the incoming electron-proton beam directions. Note also that we distinguish the azimuthal angle measured with such jet frame shown in Fig.~\ref{fig:jperp} with a hat symbol.

The differential cross section of the back-to-back electron-jet production with unpolarized hadron observed inside jets is given by 
\bea
\label{eq:unpjeth}
&\frac{d\sigma^{p(S_A)+e(\lambda_e)\to e+(\text{jet}\,h)+X}}{d{p}^2_Tdy_Jd^2{\bm q}_Tdz_h d^2{\bm j}_\perp}=F_{UU,U}+\cos(\phi_{q}-\hat{\phi}_h)F^{\cos(\phi_{q}-\hat{\phi}_h)}_{UU,U}\nnu
&\hspace{1.5cm}+\lambda_p\bigg\{\lambda_e F_{LL,U}+\sin(\phi_{q}-\hat{\phi}_h)F_{LU,U}^{\sin(\phi_{q}-\hat{\phi}_h)}\bigg\}\nnu
&\hspace{1.5cm}+S_T\bigg\{\sin(\phi_{q}-{\phi}_{S_A})F^{\sin(\phi_{q}-{\phi}_{S_A})}_{TU,U}+\sin(\phi_{S_A}-\hat{\phi}_h)F^{\sin(\phi_{S_A}-\hat{\phi}_h)}_{TU,U}\nnu
&\hspace{3cm}+\lambda_e\cos(\phi_{q}-{\phi}_{S_A})F^{\cos(\phi_{q}-{\phi}_{S_A})}_{TL,U}\nnu
&\hspace{3cm}+\sin(2\phi_{q}-\hat{\phi}_h-\phi_{S_A})F^{\sin(2\phi_{q}-\hat{\phi}_h-\phi_{S_A})}_{TU,U}\bigg\}\,,
\eea
where $F_{AB,C}$ denote the spin-dependent structure functions, with $A$, $B$, and $C$ indicating the polarization of the incoming proton, of the incoming electron, and the outgoing hadron inside the jet, respectively. Since we only consider the distribution of unpolarized hadrons inside the jet in this section, we always have $C = U$. We will consider the polarization of the hadrons in jets in the next section. 

With the hadron inside the jet observed, the factorization formula for the above structure functions $F_{AB,C}$ is very similar to that for the structure functions $F_{AB}$ discussed in~Sec.~\ref{sec2:theoretical}. In the narrow cone approximation, the jet function $J_q(p_TR, \mu)$ in $F_{AB}$ (e.g. in Eq.~\eqref{eq:FUU}) is simply replaced by appropriate TMDJFFs~\cite{Kang:2017glf,Kang:2020xyq,Kang:2019ahe}. As we have emphasized there, such a factorization formula is derived under the narrow jet approximation with $R\ll 1$. In the Appendix~\ref{appsoft}, we derive a different factorization formalism which applies to a general $R\sim 1$ and demonstrate the connection and consistency between these two formalisms. Let us start with the definition of these TMDJFFs. The correlator which describes the hadron distribution inside jet is given by~\footnote{Here for back-to-back electron-jet production where $q_T \ll p_T$, out-of-jet cone hard radiation is not allowed. The situation thus belongs to the case of exclusive jet production~\cite{Ellis:2010rwa,Kang:2019ahe} and is different from the case of inclusive jet production discussed in~\cite{Kang:2020xyq,Kang:2017glf}.}
\bea
\label{eq:jperpcorr}
\Delta_{\rm jet}^{h/q}\left(z_h,{\bm j}_\perp,S_h\right) = & \frac{1}{2N_c}\delta\left(z_h - \frac{\bar n_J\cdot p_h}{\bar n_J\cdot p_J}\right)
\langle 0| \delta\left(\bar n_J\cdot p - \bar{n}_J\cdot {\mathcal P} \right) \delta^2(\mathcal{P}_\perp/z_h+{\bm j}_\perp)\nnu
&\quad\times\chi_n(0)  |Jh\rangle\langle Jh|\bar \chi_n(0) |0\rangle,
\eea
where $h\in J$ is observed inside the jet initiated by the quark with a momentum $p$, and $p_h$ and $S_h$ are the momentum and spin vector of the final hadron described more in Sec.~\ref{sec:pol_h}. On the other hand, $\chi_n$ and $\mathcal{P}$ are the gauge invariant collinear quark field (along the jet direction) and the label momentum operator in Soft Collinear Effective Theory (SCET)~\cite{Bauer:2000ew,Bauer:2000yr,Bauer:2001ct,Bauer:2001yt}. We introduce the following two jet light-cone vectors
$n_J=\frac{1}{\sqrt{2}}(1,0,0,1)$ and $\bar n_J=\frac{1}{\sqrt{2}}(1,0,0,-1)$, which are defined by the coordinate $x_Jy_Jz_J$ found in Fig.~\ref{fig:jperp} with the jet momentum along the $n_J$ direction. For the unpolarized hadron, the correlator is parametrized by TMDJFFs at the leading twist accuracy as
\bea\label{eq:unpjffdef} 
\Delta_{\rm jet}^{h/q}\left(z_h,{\bm j}_\perp,S_h\right)=\ &\frac{1}{2}\Bigg[\mathcal{D}_1^{h/q}(z_h,j_\perp^2)\sla{n}_J+\mathcal{H}_{1}^{\perp h/q}(z_h,j_\perp^2)\frac{\sigma^{\mu\nu} n_{J,\mu} j_{\perp\nu}}{z_h\,M_h}\Bigg] \nonumber\\&+ \text{spin dependent terms}\,,
\eea
The physical interpretations of the unpolarized TMDJFFs are summarized in the first row of Table.~\ref{tabTMDff}. In other words, $\mathcal{D}_1^{h/q}$ describes an unpolarized quark initiating a jet in which an unpolarized hadron is observed, while $\mathcal{H}_{1}^{\perp h/q}$ describes a transversely polarized quark initiating a jet in which an unpolarized hadron is observed. It is thus no surprise~\cite{Kang:2020xyq,Kang:2017glf,Kang:2017btw} that TMDJFFs $\mathcal{D}_1^{h/q}$ have close relations with the unpolarized TMDFFs $D_1^{h/q}$, while TMDJFFs $\mathcal{H}_{1}^{\perp h/q}$ are closely related to the Collins TMDFFs $H_{1}^{\perp h/q}$. 

\begin{table}
    \centering
\begin{tabular}{ |c|c|c|c| } 
 \hline
 \diagbox[width=4em]{$H$}{$q$} & $U$ & $L$ & $T$ \\ 
  \hline
 $U$ & $\mathcal{D}_1^{h/q} $&  &$\mathcal{H}_1^{\perp h/q}$  \\ 
  \hline
$L$ &  &  $\mathcal{G}^{h/q}_{1L}$& $\mathcal{H}_{1L}^{h/q}$\\ 
  \hline
$T$ & $\mathcal{D}_{1\text{T}}^{\perp h/q}$ & $\mathcal{G}^{h/q}_{1\text{T}}$ & $\mathcal{H}^{h/q}_{1}$, $\mathcal{H}_{1\text{T}}^{\perp h/q}$\\ 
  \hline
\end{tabular}
  \caption{Interpretation of TMDJFFs for quarks. The rows indicate the hadron polarization — unpolarized (U), longitudinally polarized (L), transversely polarized (T). And the columns indicate the quark polarization accordingly.}
  \label{tabTMDff}
\end{table}

We now illustrate the factorization of the structure functions in the region $q_T\sim j_\perp \ll p_T R$. We replace the jet function $J_q(p_TR, \mu)$ in $F_{UU}$ of  Eq.~\eqref{eq:FUU} by the TMDJFFs $\mathcal{D}_1^{h/q}(z_h,j_\perp^2,\mu, \zeta_J)$ to obtain the factorization formula for $F_{UU,U}$ in Eq.~\eqref{eq:unpjeth}
\bea
\label{eq:FUUUbefore}
F_{UU,U} =&\hat{\sigma}_0\,H(Q,\mu)\sum_qe_q^2\, {\mathcal{D}}_{1}^{h/q}(z_h,j_\perp^2,\mu, \zeta_J)
\int\frac{b \,db}{2\pi}J_0(q_Tb)\,x\,\tilde{f}^{q}_{1}(x,b^2, \mu,\zeta)
\nonumber\\
&\times \bar{S}_{\rm global}(b^2,\mu)\bar{S}_{cs}(b^2,R,\mu)\,,
\eea
where we include renormalization scale $\mu$ and Collins-Soper parameter $\zeta_J$ for the TMDJFFs. As we will demonstrate below, $\sqrt{\zeta_J} = p_T R$. 

In the kinematic region $j_\perp\ll p_TR$, the unpolarized TMDJFF ${\mathcal{D}}_{1}^{h/q}(z_h,j_\perp^2,\mu, \zeta_J)$ can be further factorized in terms of the corresponding unpolarized TMDFF and a collinear-soft function as~\cite{Kang:2017glf,Kang:2020xyq}
\bea
{\mathcal{D}}_{1}^{h/q}(z_h,j_\perp^2,\mu, \zeta_J) &=  \int_{\bm{k}_\perp,\, \bm{\lambda}_\perp}\, D_1^{h/q,\, 
\rm unsub}(z_h,k_\perp^2,\mu,\zeta'/\nu^2) S_q(\lambda_\perp^2,\mu,\nu \mathcal{R})\nonumber\\
& =\int \frac{b\,db}{2\pi} J_0\left(\frac{j_\perp b}{z_h}\right) \tilde{D}_1^{h/q,\,\rm unsub}(z_h,b^2,\mu,\zeta'/\nu^2)S_q(b^2,\mu,\nu \mathcal{R})\,,
\label{unp_JFF_FF}
\eea
where we use the short-hand notation 
$\int_{\bm{k}_\perp,\, \bm{\lambda}_\perp}=\int d^2\bm{k}_\perp d^2\bm{\lambda}_\perp \delta^2(z_h\bm{\lambda}_\perp + \bm{k}_\perp-\bm{j}_\perp)$ in the first line, and $\sqrt{\zeta'} = \sqrt{2}n_J\cdot p_J$ is the Collins-Soper parameter for the TMDFFs. On the other hand, $S_q(b^2, \mu, \nu \mathcal{R})$ is the collinear-soft function with the following expressions~\cite{Kang:2017glf,Kang:2017mda}
\bea
S_q(b^2,\mu,\nu \mathcal{R})=1 - \frac{\alpha_s C_F}{2\pi} &\Bigg[\frac{2}{\eta}\left(\frac{1}{\epsilon}+\ln\frac{\mu^2}{\mu_b^2}\right)
-\frac{1}{\epsilon^2}+
\frac{1}{\epsilon}\ln\frac{\nu^2\mathcal{R}^2}{4\mu^2}
+\ln\frac{\mu^2}{\mu_b^2} \ln\frac{\nu^2\mathcal{R}^2}{4\mu_b^2}\nnu
&\ -\frac{1}{2}
\ln^2\frac{\mu^2}{\mu_b^2}+\frac{\pi^2}{12}
\Bigg]\,,
\label{eq:S_q}
\eea
where $\mathcal{R} = \frac{R}{\cosh{y_J}}$. To proceed, comparing Eqs.~\eqref{eq:S-ab} and \eqref{eq:S_q} we realize $S_q(b^2,\mu,\nu\mathcal{R}) = \sqrt{S_{ab}(b^2,\mu,\nu)}|_{\nu \to\nu\mathcal{R}}$ at the NLO and thus
\bea
\tilde{D}_1^{h/q,\,\rm unsub}(z_h,b^2,\mu,\zeta'/\nu^2)S_q(b^2,\mu,\nu \mathcal{R})
= & \tilde{D}_1^{h/q,\,\rm unsub}(z_h,b^2,\mu, 
\zeta'/\nu^2) \sqrt{S_{ab}(b^2,\mu,\nu\mathcal{R})} 
\nnu 
= & \tilde{D}_1^{h/q}(z_h,b^2,\mu, 
\zeta' \mathcal{R}^2)\,.
\label{eq:Dunpb}
\eea
Finally using the fact that $\sqrt{\zeta'} \mathcal{R} = \sqrt{2}n_J\cdot p_J \frac{R}{\cosh y_J} = p_T R \equiv \sqrt{\zeta_J}$, we obtain the following relation between TMDJFF ${\mathcal{D}}_{1}^{h/q}$ and TMDFF $D_{1}^{h/q}$
\bea
{\mathcal{D}}_{1}^{h/q}(z_h,j_\perp^2,\mu, \zeta_J) = \int \frac{b\,db}{2\pi} J_0\left(\frac{j_\perp b}{z_h}\right) \tilde{D}_1^{h/q}(z_h,b^2,\mu,\zeta_J) = D_1^{h/q}(z_h,j_\perp^2,\mu,\zeta_J)\,.
\label{unp_JFF_FF2}
\eea
In other words, the TMDJFF is equal to the TMDFF at the scale $\zeta_J$. Parametrization of TMDFF follows the similar form as that of the TMDPDF discussed in Sec.\ \ref{sec2:theoretical}. Using the CSS formalism, the $b$-space unpolarized TMDFF can be expressed as 
\bea
\tilde{D}_1^{h/q}(z_h,b^2,\mu, 
\zeta_J) &=\frac{1}{z_h^2}\left[\hat{C}_{i\leftarrow q}\otimes D_{1}^{h/i}\right]\left(z_h,\mu_{b_*}\right)
\exp\left[-S_{\rm pert}\left(\mu, \mu_{b_*} \right) - S_{\rm NP}^D\left(z_h, b, Q_0, \zeta_J\right)\right]\,,
\label{eq:D1param}
\eea
where we have performed an operator product expansion in terms of the unpolarized collinear FFs $D_1^{h/i}(x, \mu_{b_*})$ with the convolution defined as follows 
\bea
\left[\hat{C}_{i\leftarrow q}\otimes D_{1}^{h/i}\right]\left(z_h,\mu_{b_*}\right) = \int_{z_h}^{1} \frac{d\hat{z}_h}{\hat{z}_h} \hat{C}_{i\leftarrow q}\left(\frac{z_h}{\hat{z}_h}, \mu_{b_*} \right) D_1^{h/i}\left(\hat{z}_h, \mu_{b_*}\right)\,,
\eea
where the sum over repeated indicies are understood and we follow the same $b_*$ prescription as in TMDPDFs. The coefficient functions $\hat{C}_{i\leftarrow q}$ at the NLO can be found in~\cite{Kang:2015msa}, and the results for even-higher order are also available~\cite{Echevarria:2016scs,Luo:2019hmp,Luo:2020epw,Ebert:2020qef}. The perturbative Sudakov factor is identical to that of the TMDPDFs given in Eq.~\eqref{eq:Sud-pert}. On the other hand, for the non-perturbative Sudakov factor $S_{\rm NP}^D\left(z_h, b, Q_0, \zeta_J\right)$, we use the parametrization~\cite{Sun:2014dqm,Echevarria:2020hpy}
\bea
\label{eq:Sud-NPD}
S_{\rm NP}^D(z_h, b, Q_0,\zeta_J) = \frac{g_2}{2}\ln{\frac{\sqrt{\zeta_J}}{Q_0}}\ln{\frac{b}{b_*}}+g_1^D \frac{b^2}{z_h^2}\,,
\eea
where the values of $Q_0$ and $g_2$ are given below Eq.~\eqref{eq:Sud-NP}, and $g_1^D = 0.042$ GeV$^2$. Note that when we carry out phenomenological studies below in Secs.~\ref{sec3:pheno} and~\ref{sec4:pheno} involving $F_{UU,U}$, we parametrize the unpolarized TMDPDF $\tilde{f}_1^q$ according to Sec.~\ref{sec2:theoretical}.

Using similar arguments, one can discover similar relations between other TMDJFFs and TMDFFs, which can be found in Appendix~\ref{app1}. Explicit expressions of the rest of the structure functions in terms of TMDJFFs and TMDPDFs are given in Appendix~\ref{app2}, and we summarize the azimuthal asymmetries with which they appear in Table.~\ref{scn2tab}.  

\begin{table}
\centering
\begin{tabular}{ |c|c|c|c|c| } 
 \hline
 \diagbox[width=1.8cm]{JFF}{PDF} & $f_1$ & $g_{1L}$ & $f_{1T}^\perp$  & $g_{1T}$ \\ 
  \hline
  $\mathcal{D}_1$ & 1 & 1  & $\sin(\phi_{q}-{\phi}_{S_A})$ & $\cos(\phi_{q}-{\phi}_{S_A})$     \\ 
  \hline
  \hline
   \diagbox[width=1.8cm]{JFF}{PDF} & $h_1^\perp$ & $h_{1L}^\perp$ & $h_1$  & $h_{1T}^\perp$  \\ 
  \hline 
  $\mathcal{H}^\perp_1$ & $\cos(\phi_{q}-\hat{\phi}_h)$ & $\sin(\phi_{q}-\hat{\phi}_h)$ & $\sin({\phi}_{S_A}-\hat{\phi}_h)$ & $\sin(2\phi_{q}-\hat{\phi}_{h}-\phi_{S_A})$\\
   \hline 
\end{tabular}
  \caption{Summary of the characteristic azimuthal asymmetry with which different TMDPDFs and TMDJFFs arise for back-to-back electron-jet production, with unpolarized hadrons observed inside the jet. See Eqs.~\eqref{eq:unpjeth} and~\eqref{eq:strhlep1}-\eqref{eq:strhlep8} for parametrizations of structure functions.} \label{scn2tab}
\end{table}

\subsection{Phenomenology: Boer-Mulders correlation with Collins function}
\label{sec3:pheno}
Let us now study the phenomenology for the distribution of unpolarized hadrons inside a jet. The well-known Collins asymmetry for hadrons in a jet in the collisions of an unpolarized electron and a transversely polarized proton, manifested as a $\sin(\phi_{q}-{\phi}_{S_A})$ modulation in Eq.~\eqref{eq:unpjeth}, has been studied previously~\cite{Arratia:2020nxw}. In this section, we carry out a new phenomenological study as an example. Specifically we study the azimuthal modulation $\cos(\phi_{q}-\hat{\phi}_h)$ in Eq.~\eqref{eq:unpjeth}. This azimuthal dependence arises in the distribution of unpolarized hadrons in the unpolarized electron and proton collisions. The relevant structure function, $F_{UU,U}^{\cos(\phi_{q}-\hat{\phi}_h)}$, probes the Boer-Mulders function $h_1^\perp$ in the unpolarized proton, coupled with the Collins fragmentation function $H_1^\perp$. Since it does not require any polarization of either the beams or the final-state hadron, even the HERA experiment can measure such an asymmetry. We thus present numerical results for both HERA and EIC kinematics. 

To proceed, by normalizing the structure function $F_{UU,U}^{\cos(\phi_{q}-\hat{\phi}_h)}$ by the unpolarized and azimuthal-independent structure function $F_{UU,U}$, we define the new azimuthal asymmetry as follows
\bea
\label{eq:A-UUU}
A_{UU,U}^{\cos(\phi_{q}-\hat{\phi}_h)}=\frac{F_{UU,U}^{\cos(\phi_{q}-\hat{\phi}_h)}}{F_{UU,U}}\,,
\eea
where the denominator and the numerator are given by $F_{UU,U}$ and $F_{UU,U}^{\cos(\phi_{q}-\hat{\phi}_h)}$ in Eq.~\eqref{eq:unpjeth}, respectively. The factorization formula and the parametrization of the unpolarized TMDs for the denominator $F_{UU,U}$ has been presented in Eq.~\eqref{eq:FUUUbefore} and proceeding discussion that follows. On the other hand, the structure function $F_{UU,U}^{\cos(\phi_{q}-\hat{\phi}_h)}$ depends on the Boer-Mulders TMDPDF $h_1^\perp$ and the Collins TMDJFF $\mathcal{H}_1^\perp$. Boer-Mulders function describes the transversely polarized quarks inside an unpolarized proton, then such a transversely polarized quark scatters with the unpolarized electron. Through transverse spin transfer, we have a transversely polarized quark that initiates a jet with distribution of unpolarized hadrons measured inside the jet. Since Collins function describes a transversely polarized quark fragments into an unpolarized hadron, such a correlation function is related to the Collins function. The factorization formula is given in Eq.~\eqref{eq:strhlep2} in the Appendix and is explicitly expressed here for convenience
\bea
F_{UU,U}^{\cos(\phi_{q}-\hat{\phi}_h)} =&\hat{\sigma}_T \,H(Q,\mu)\sum_q e_q^2\, \frac{ j_\perp}{z_hM_h} H_1^{\perp\,h/q}(z_h,j_\perp^2,\mu, \zeta_J)
\nnu
&\times 
M \int\frac{b^2 \,db}{2\pi}J_1(q_Tb)\,x\,\tilde{h}_1^{\perp\, q(1)}(x,b^2, \mu,\zeta)\bar{S}_{\rm global}(b^2,\mu)\bar{S}_{cs}(b^2,R,\mu)\,,
\label{eq:FUUU-spin}
\eea
where $\hat\sigma_T$ is the transverse spin-transfer cross section given in Eq.~\eqref{eq:sigma_T}. Note that we have followed the same procedure as that for the case of the unpolarized TMDJFF $\mathcal{D}_1^{h/q}$ from Eqs.~\eqref{unp_JFF_FF} to \eqref{unp_JFF_FF2}, to derive the relationship between the TMDJFF $\mathcal{H}_1^{\perp\,h/q}$ and the Collins TMDFF $H_1^{\perp\,h/q}$. 

It is instructive to emphasize that $F_{UU,U}^{\cos(\phi_{q}-\hat{\phi}_h)}$, being differential in both $q_T$ and $j_\perp$, allows us to constrain separately TMDPDFs (i.e. Boer-Mulder function $\tilde{h}_1^{\perp}$ here) and TMDFFs (i.e., Collins function $H_1^{\perp\,h/q}$ here). This is evident in Eq.~\eqref{eq:FUUU-spin}, since all the $q_T$-dependence is contained in the Fourier transform $b$-integral while the $j_\perp$-dependence is outside such an integration. Physically this is easily understood and expected, simply because $q_T$ and $j_\perp$ are measured with respected to two directions, i.e. the beam direction and the jet direction, respectively. This is advantageous in comparison with the usual TMD measurement, e.g. in the Drell-Yan production where one measures all the transverse momenta with respect to the beam direction. 

For the phenomenology below, we use the Collins TMDFFs extracted from~\cite{Kang:2015msa}, which has proper TMD evolution. On the other hand, we still need the parametrization for the Boer-Mulders functions $h_1^{\perp}$. For the purpose of the numerical studies below,  we use the Boer-Mulder functions extracted from~\cite{Barone:2009hw}, which is based on the usual Gaussian model. Following the same method in Sec.~\ref{subsec:g1T}, we build a parametrization for $\tilde{h}_1^{\perp\, q(1)}(x,b^2, \mu,\zeta)$ with TMD evolution:
\bea
\tilde{h}_1^{\perp\, q(1)}(x,b^2, \mu,\zeta) = 
2\pi h_{1}^{\perp\, q(1)}\left(x,\mu_{b_*}\right)
\exp\left[-S_{\rm pert}\left(\mu, \mu_{b_*} \right) - S_{\rm NP}^{h_{1}^\perp}\left(x, b, Q_0, \zeta\right)\right]\,,
\eea
where the collinear function $h_{1}^{\perp\, q(1)}\left(x,\mu_{b_*}\right)$ is constructed in the Gaussian model from~\cite{Barone:2009hw} via
\bea
h_{1}^{\perp\, q(1)}(x)=\int d^2\boldsymbol{k}_T\frac{k_T^2}{2M^2}h_{1}^{\perp\,q}(x,k_T^2)\,.
\eea
On the other hand, we have the non-perturbative Sudakov factor $S_{\rm NP}^{h_{1}^\perp}$ given as
\bea
S_{\rm NP}^{h_{1}^\perp}\left(x, b, Q_0, \zeta\right) = \frac{g_2}{2}\ln{\frac{\sqrt{\zeta}}{Q_0}}\ln{\frac{b}{b_*}}+g_1^{h_{1}^\perp} b^2\,.
\eea
Here $g_1^{h_{1}^\perp}$ is again related to the intrinsic Gaussian width for the TMDPDF $h_1^\perp$ in the transverse momentum space
\bea
g_1^{h_{1}^\perp} = \frac{\langle k_T^2\rangle_{h_1^\perp}}{4} = 0.036{\rm ~GeV}^2\,,
\eea
where we used $\langle k_T^2\rangle_{h_1^\perp}=\frac{M_1^2\langle k_T^2\rangle}{M_1^2+\langle k_T^2\rangle}$  with $\langle k_T^2\rangle=0.25$ GeV$^2$ and $M_1^2=0.34$ GeV$^2$ from~\cite{Barone:2009hw}. 

\begin{figure}[hbt!]
\hspace{1.3cm}\includegraphics[width = 0.8\textwidth]{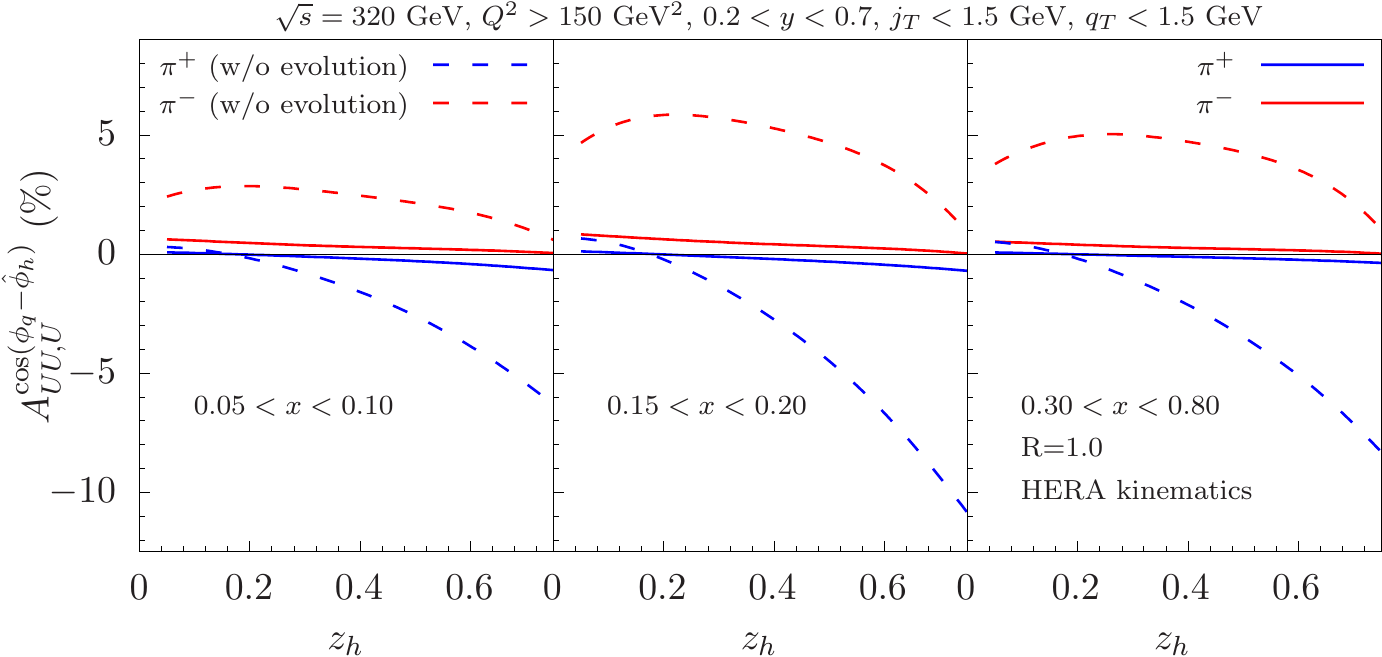}
\caption{Numerical results of $A_{UU,U}^{\cos(\phi_{q}-\hat{\phi}_{h})}$ as a function of hadron momentum fraction $z_h$ for unpolarized $\pi^\pm$ in jet production with electron in unpolarized $ep$ collision predicted for HERA using three different bins of $x$:  $[0.05,0.1],\ [0.15,0.2]$, and $[0.3,0.8]$. The solid (dashed) curves are the calculations with (without) TMD evolution. We apply the center-of-mass energy $\sqrt{s}=320$ GeV of HERA kinematics, jet radius $R=1.0$, $Q^2>150$ GeV$^2$, inelasticity $y$ in $[0.2,0.7]$ with transverse momentum imbalance $q_T$ and final hadron transverse momentum in jet are both smaller than $1.5$ GeV.}
    \label{fig:scn2_hera}
\end{figure}

Since the azimuthal asymmetry $A_{UU,U}^{\cos(\phi_{q}-\hat{\phi}_{h})}$ involves only unpolarized proton and electron beams, it can be studied in the HERA experiment at DESY. We thus present the numerical results for both HERA and the future EIC kinematics. In Fig.~\ref{fig:scn2_hera}, we plot the azimuthal asymmetry $A_{UU,U}^{\cos(\phi_{q}-\hat{\phi}_{h})}$ with unpolarized $\pi^\pm$ inside the jet with radius $R=1$ using HERA kinematics~\cite{DiStalk}. Specifically we choose electron-proton center-of-mass energy $\sqrt{s}=320$ GeV, apply the cuts $Q^2 > 150$ GeV$^2$ and $0.2< y< 0.7$. We further integrate over hadron transverse momentum $j_\perp$ and the imbalance $q_T$ with $0<j_\perp<1.5$ GeV and $0<q_T<1.5$ GeV. We conduct our analysis as a function of hadron momentum fraction $z_h$ using three different bins of $x$: $[0.05,0.1],\ [0.15,0.2]$, and $[0.3,0.8]$. Note that we select the cuts on $Q^2$, $y$, and $x$ which constrain the jet $p_T$ directly, keeping us in the TMD factorization regime discussed above. We present the numerical results with and without evolution between scales carried out, shown in solid and dashed lines in the figures, respectively. Namely, for the case without TMD evoluiton, TMDs are assumed to be pure Gaussians for the azimuthal correlations shown in dashed lines. We find that the azimuthal asymmetry is negative for $\pi^+$ production in jet and positive for $\pi^-$ production in jet, with magnitude up to around $1\%$ for HERA when the TMD evolution is turned on. On the other hand, without the TMD evolution, the size of the azimuthal asymmetry can be much larger $\sim 5\%$. This is consistent with the expectation that TMD evolution suppresses asymmetry as the radiation broadens the distribution. Thus, the azimuthal asymmetry $A_{UU,U}^{\cos(\phi_{q}-\hat{\phi}_{h})}$ could serve as a dual purpose. On one hand, it enables us to extract Boer-Mulders TMDPDFs and Collins TMDFFs. On the other hand, this asymmetry can also provide useful constraints for the TMD evolution of these TMD functions.

\begin{figure}[hbt!]
\centering
\includegraphics[width = 0.783\textwidth]{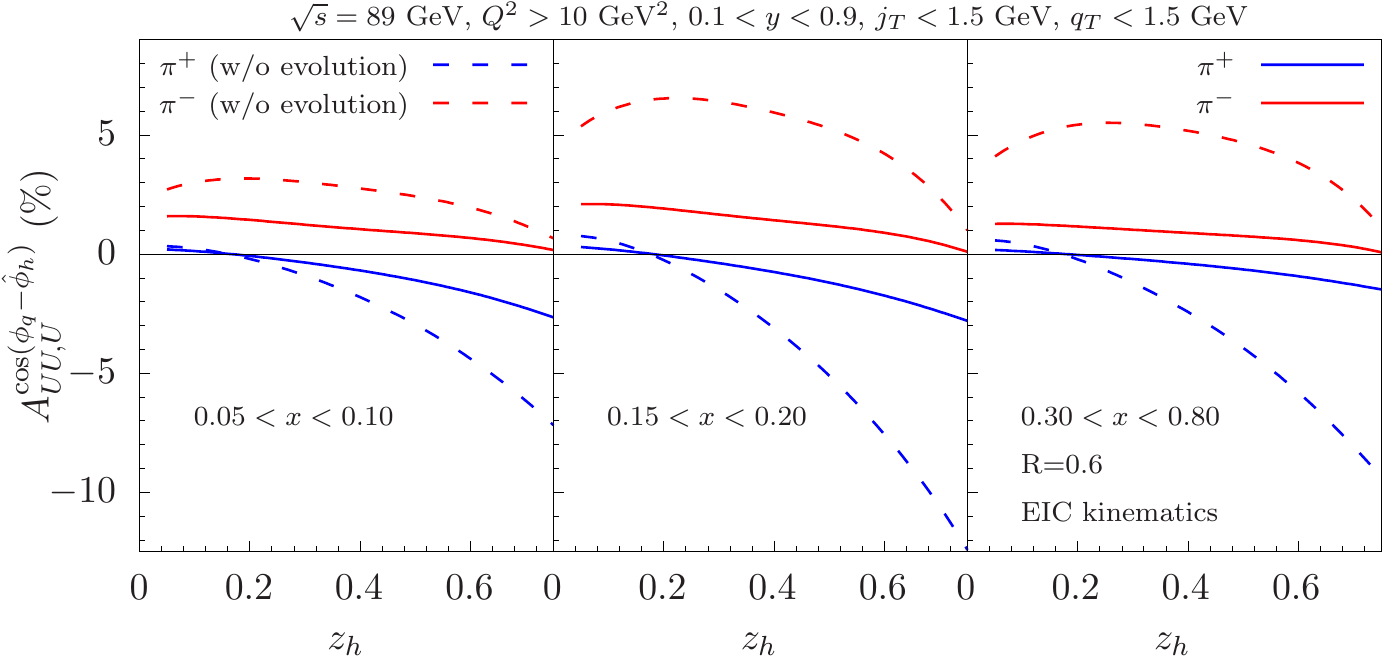}
\caption{Numerical results of $A_{UU,U}^{\cos(\phi_{q}-\hat{\phi}_{h})}$ as a function of hadron momentum fraction $z_h$ for unpolarized $\pi^\pm$ in jet production with electron in unpolarized $ep$ collision predicted for EIC  using three different bins of $x$:  $[0.05,0.1],\ [0.15,0.2]$, and $[0.3,0.8]$. The solid (dashed) curves are the calculations with (without) TMD evolution. We apply the center-of-mass energy $\sqrt{s}=89$ GeV of EIC kinematics, jet radius $R=0.6$, $Q^2>10$ GeV$^2$, inelasticity $y$ in $[0.1,0.9]$ and both transverse momentum imbalance $q_T$ and final hadron transverse momentum in jet smaller than $1.5$ GeV.}
    \label{fig:scn2_eic}
\end{figure}
We also plot the same asymmetry using the EIC kinematics with radius $R=0.6$ in Fig.~\ref{fig:scn2_eic} as a function of $z_h$ in three different bins of $x$: $[0.05,0.1],\ [0.15,0.2]$ and $[0.3,0.8]$. Specifically we calculate the asymmetry for the CM energy $\sqrt{s}=89$ GeV with the following cuts: $Q^2>10$ GeV$^2$, $0.1<y<0.9$, and $0<j_\perp, \, q_T < 1.5$ GeV. We find that the azimuthal asymmetry follows similar trends, but with larger magnitude, i.e. $\sim 2-3\%$ ($5-10\%$) with (without) TMD evolution, in comparison to the asymmetry expected with the HERA kinematics. We conclude that the experimental measurements of $A_{UU,U}^{\cos(\phi_{q}-\hat{\phi}_{h})}$ at EIC could be quite promising and can be used to constrain TMD evolution for Boer-Mulders and Collins functions. For the rest of the paper, we only present the numerical results with TMD evolution.

\begin{figure}
\includegraphics[width = 0.45\textwidth]{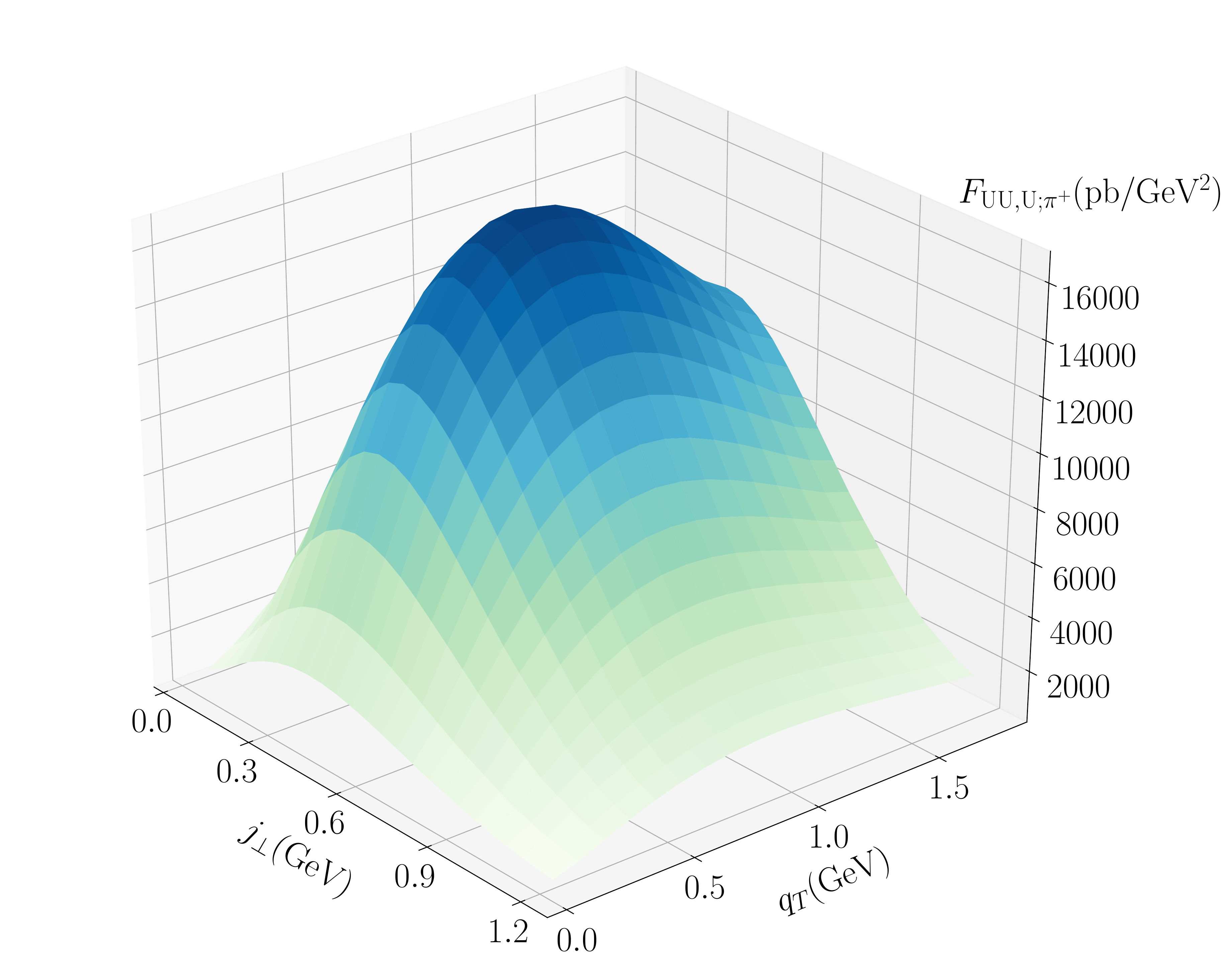}
\includegraphics[width = 0.42\textwidth]{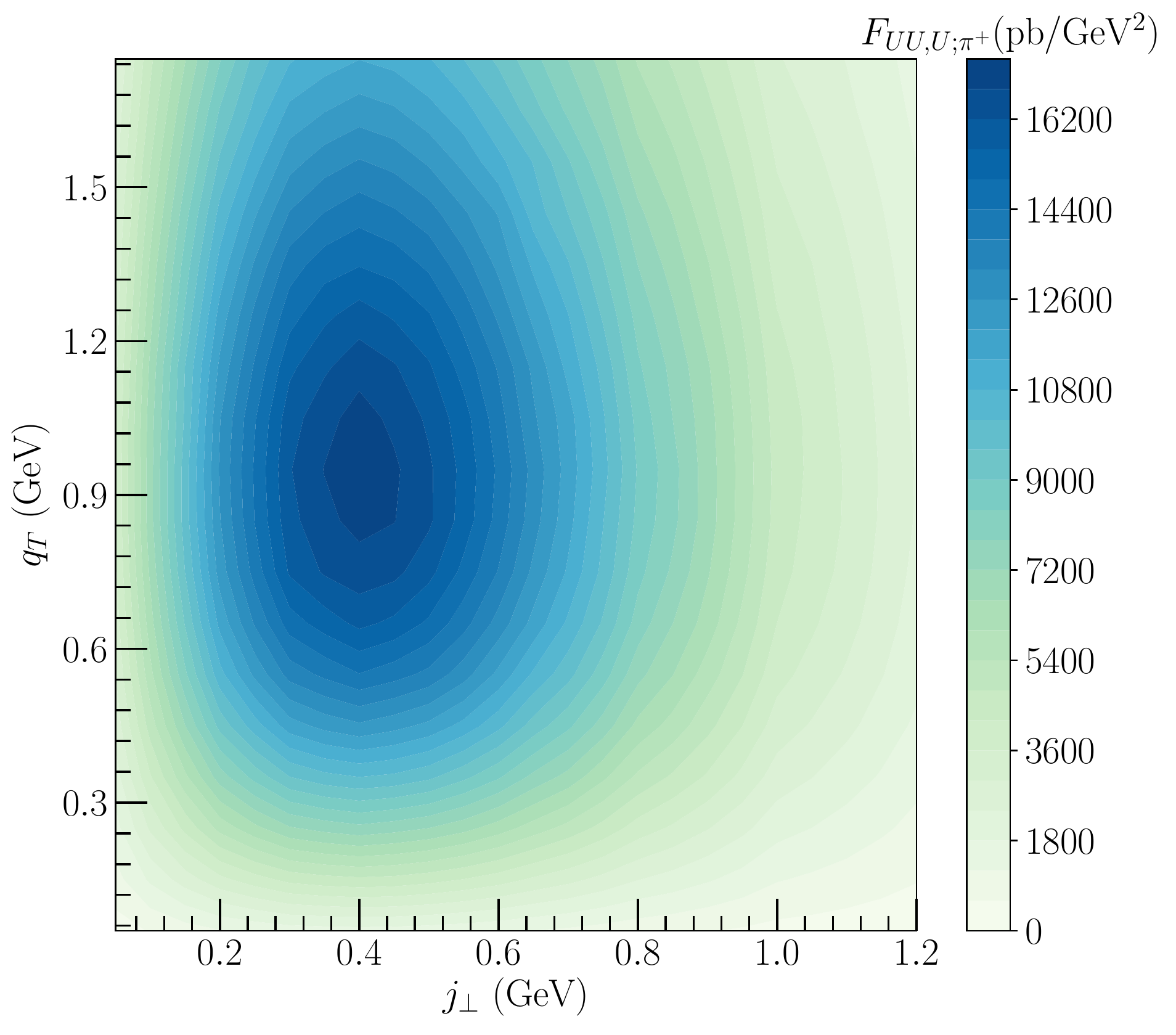}
\includegraphics[width = 0.47\textwidth]{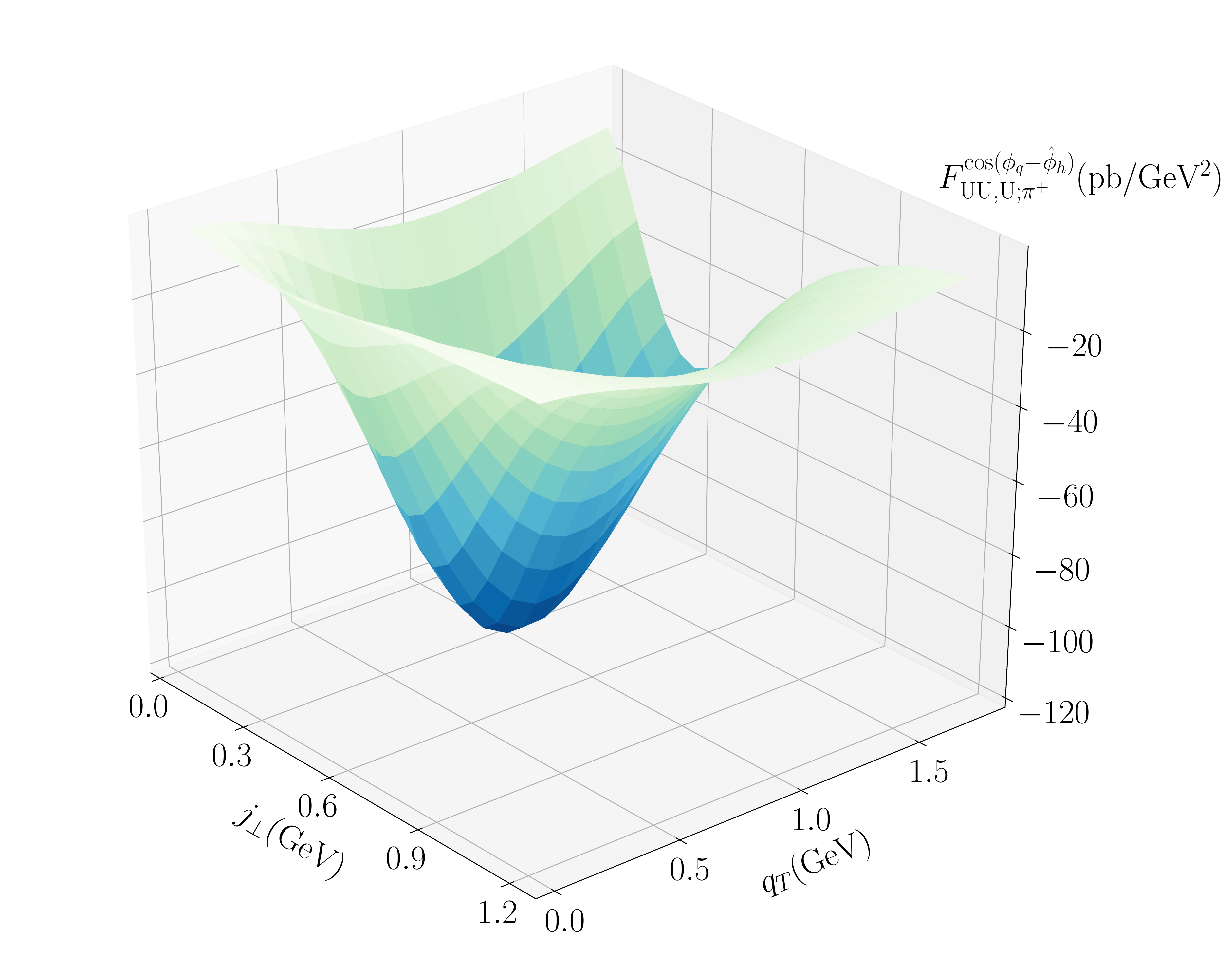}\hspace{1.5cm}
\includegraphics[width = 0.44\textwidth]{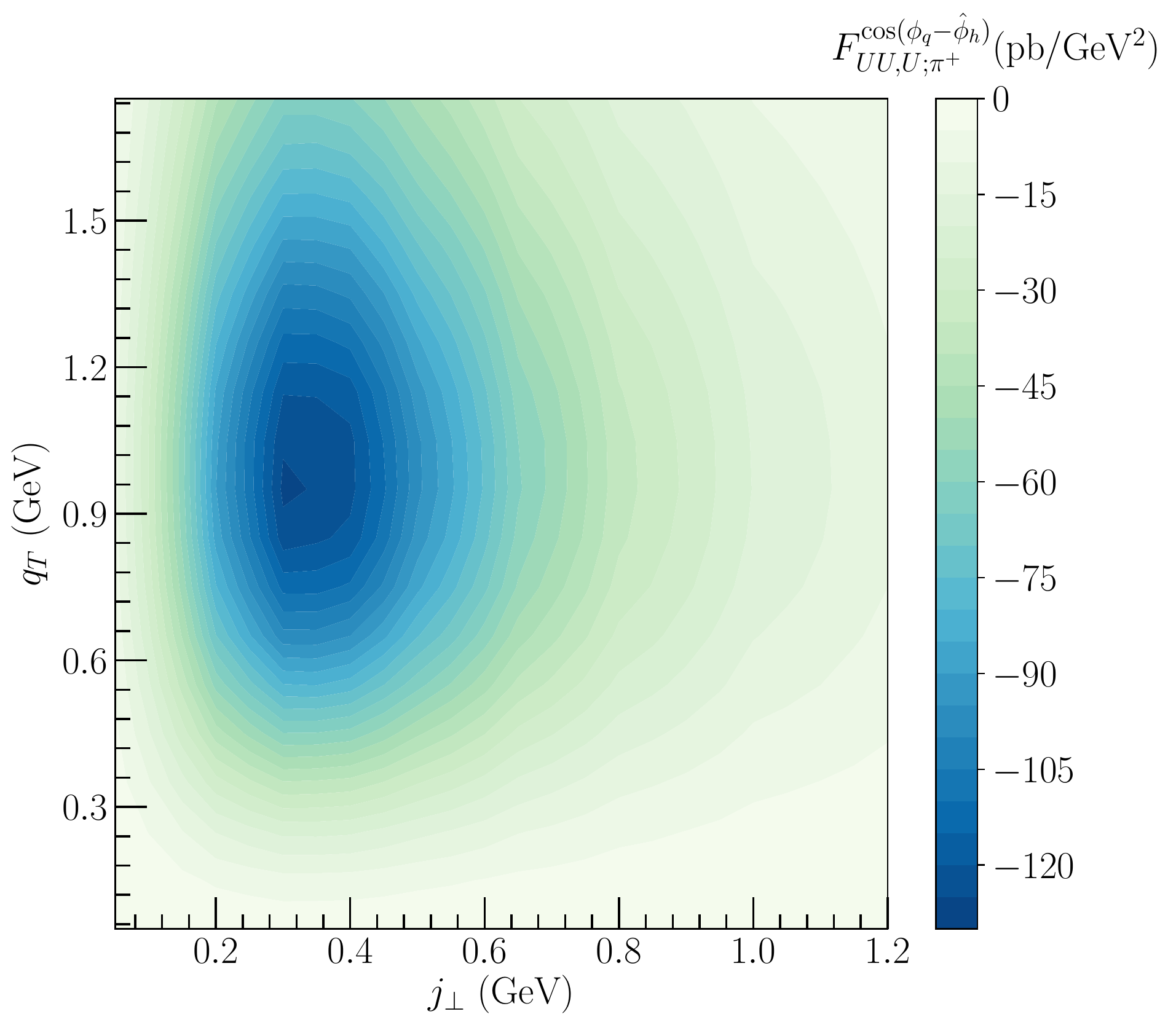}
\includegraphics[width = 0.49\textwidth]{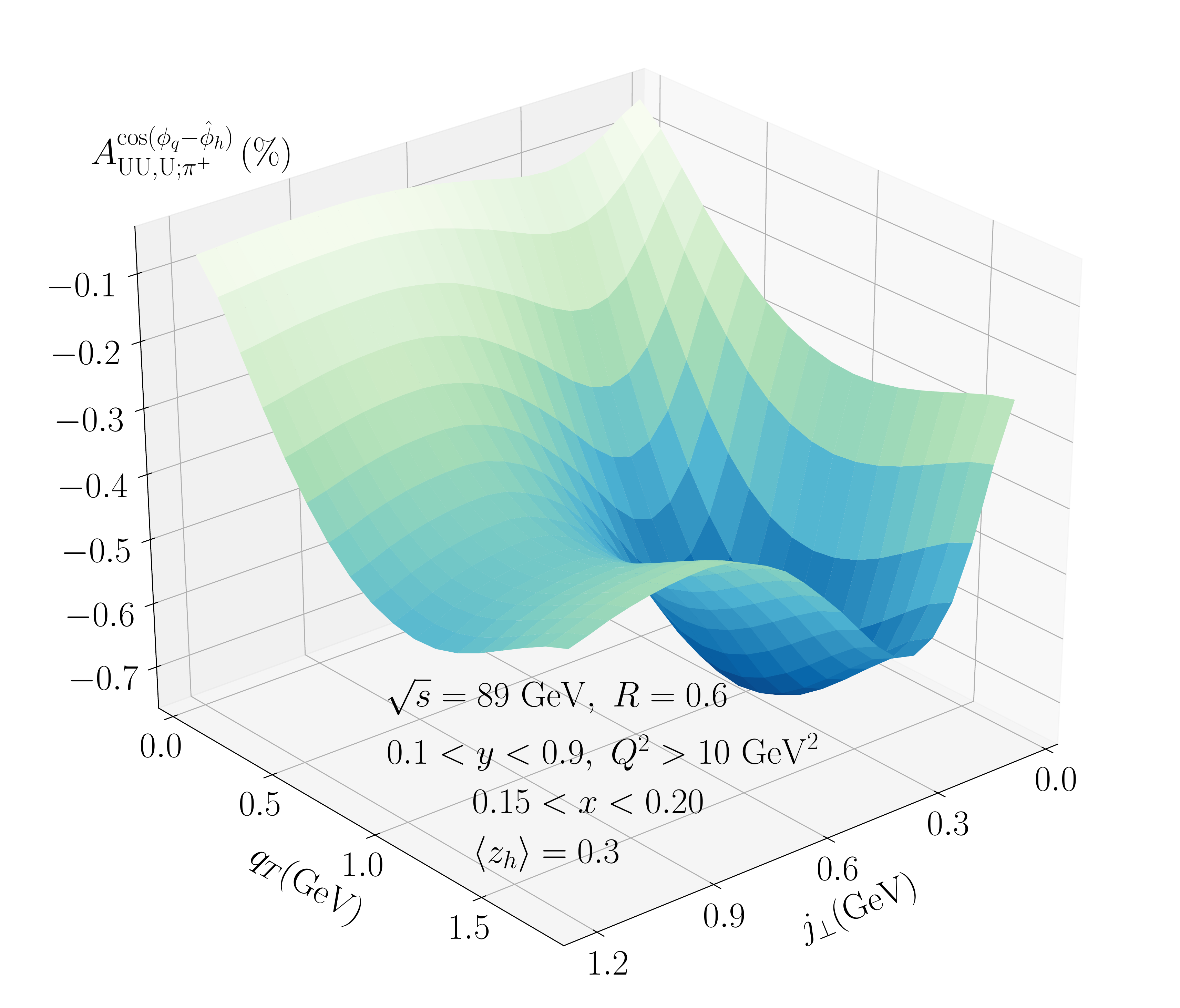}
\hspace{1.2cm}\includegraphics[width = 0.43\textwidth]{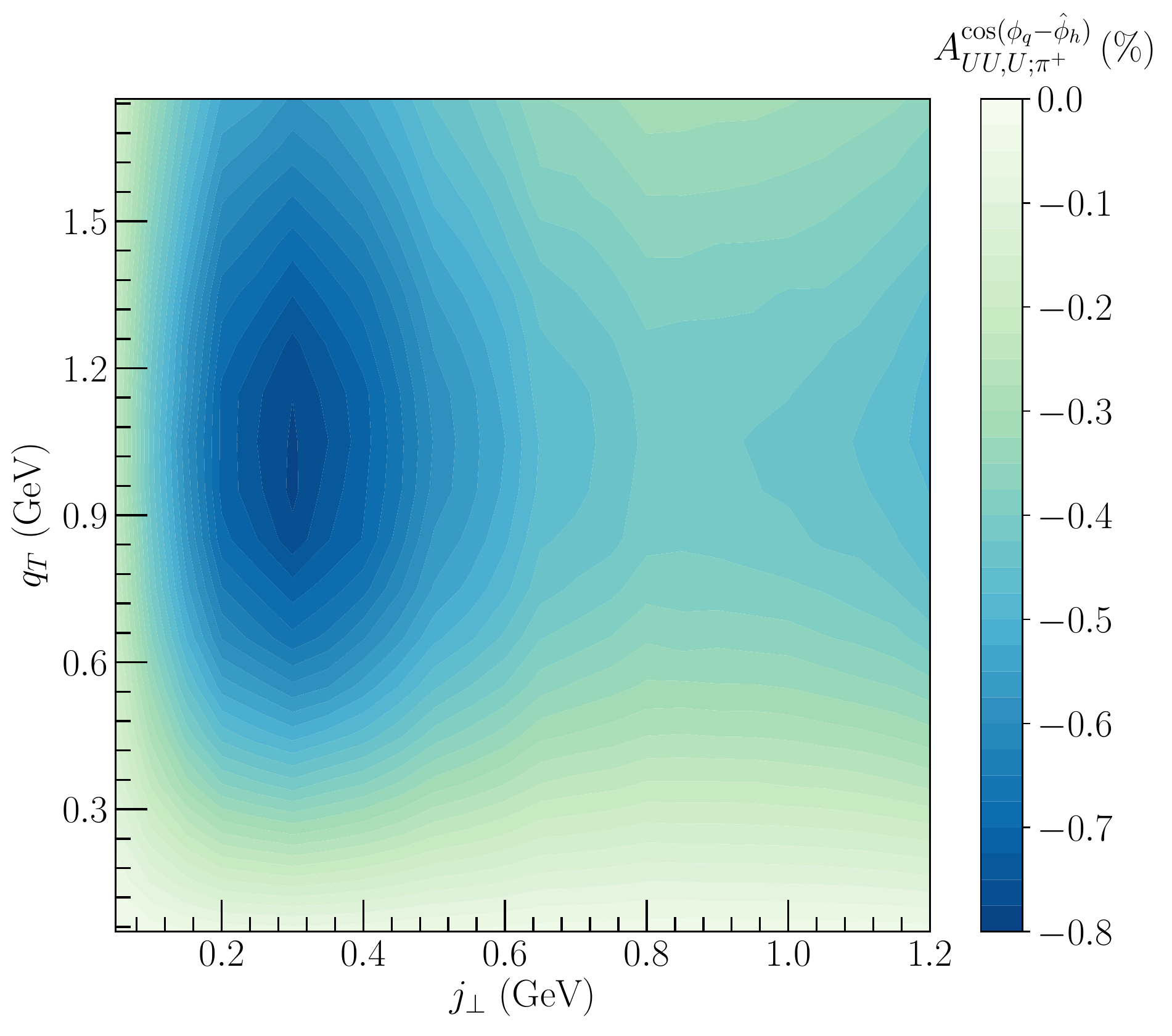}
\caption{$F_{UU,U}$ (first row), $F_{UU,U}^{\cos(\phi_{q}-\hat{\phi}_{h})}$ (second row) and $A_{UU,U}^{\cos(\phi_{q}-\hat{\phi}_{h})}$ (third row) as a function of jet imbalance $q_T$ and $j_\perp$ for unpolarized $\pi^+$ in jet production with electron in unpolarized $ep$ collision with EIC kinematics, where we have applied $\sqrt{s}=89$ GeV, jet radius $R=0.6$, inelasticity $y$ in range $[0.1,0.9]$, $Q^2>10$ GeV$^2$, Bjorken-$x$ in $[0.15,0.20]$ and average momentum fraction $\langle z_h\rangle=0.3$. Left column: Three dimensional plots of the structure functions and their ratio in $q_T$ and $j_\perp$. Right column: Contour plots of the structure functions and their ratio.}
    \label{fig:scn2_pip}
\end{figure}

\begin{figure}
\includegraphics[width = 0.45\textwidth]{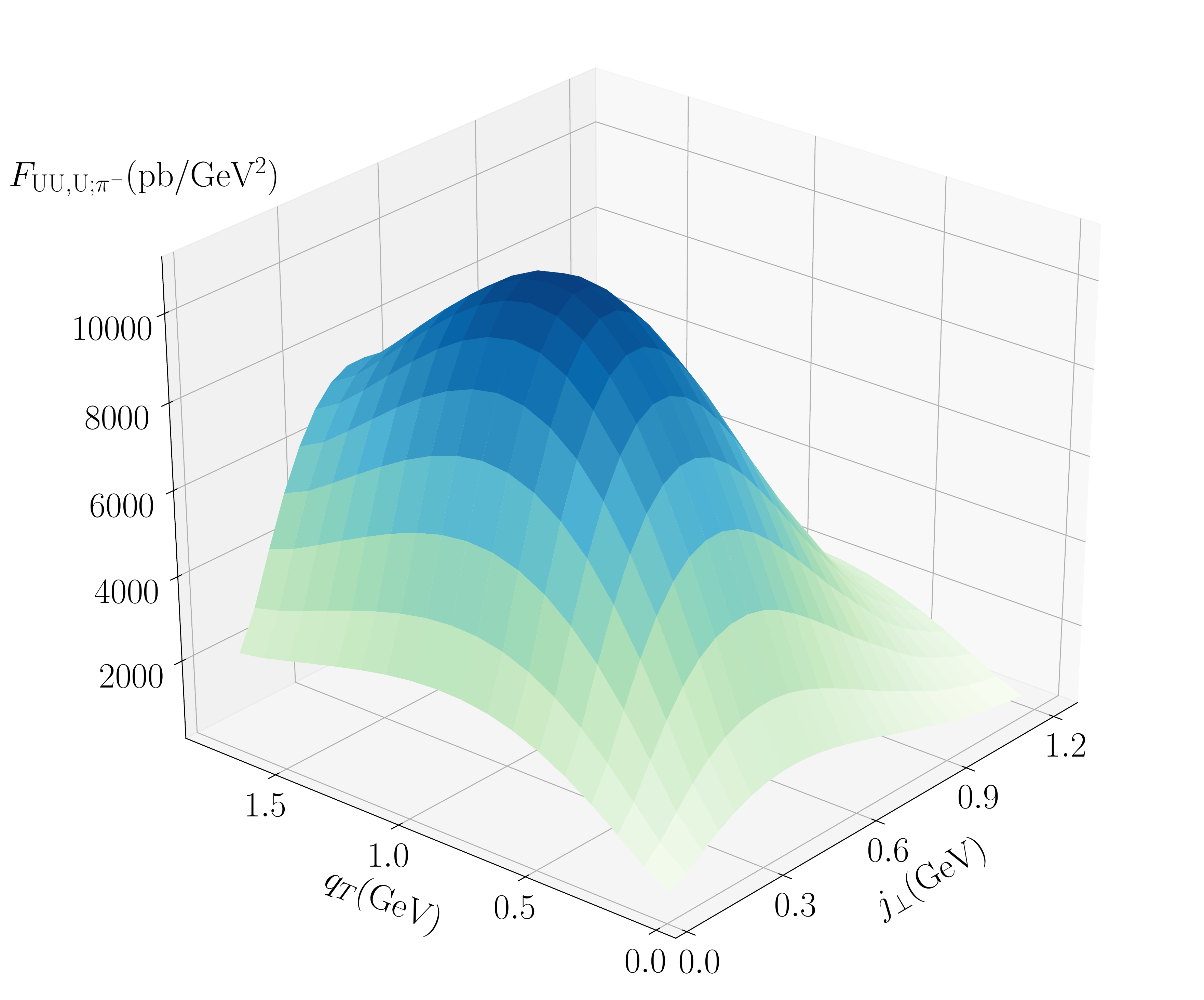}
\includegraphics[width = 0.42\textwidth]{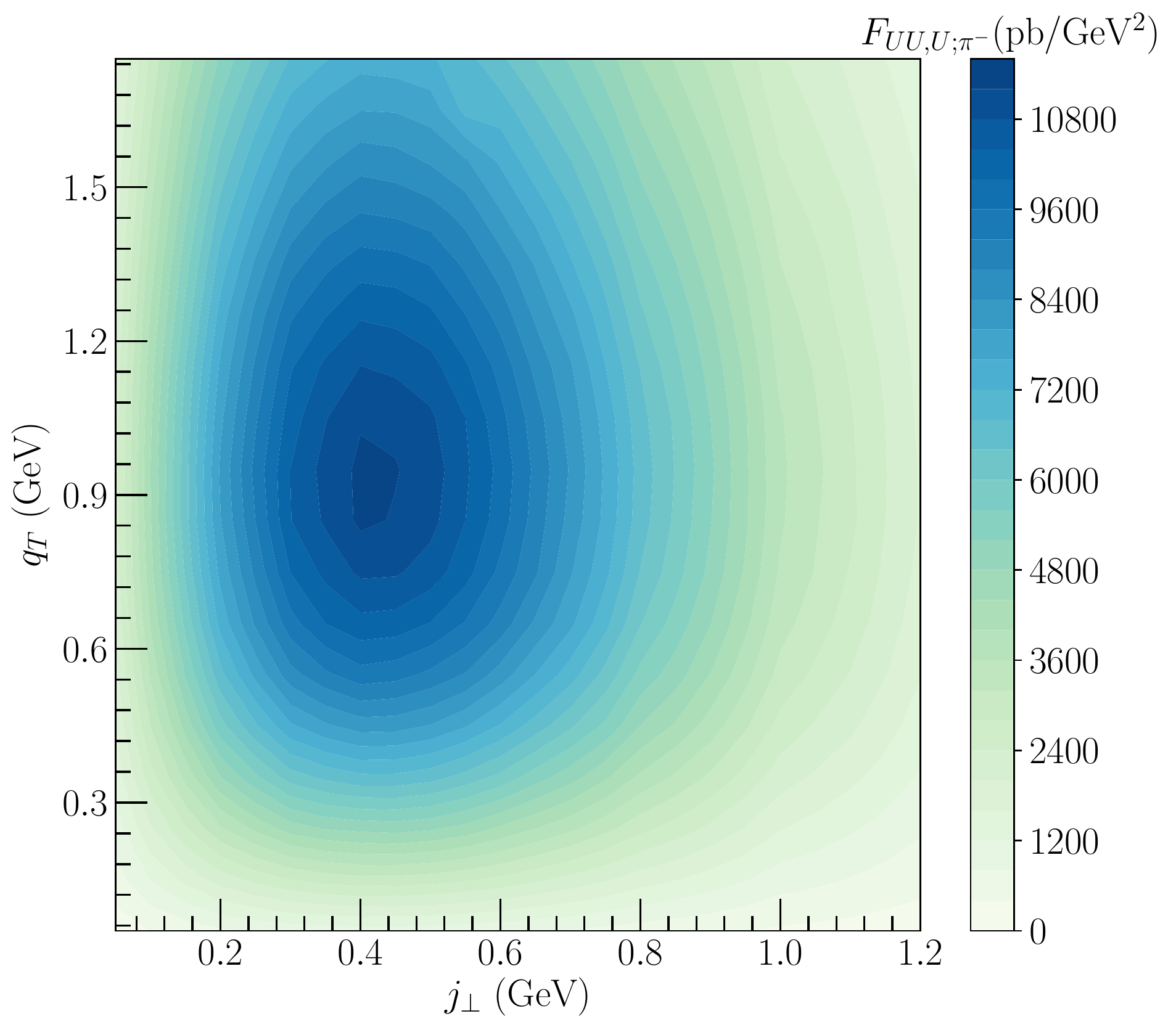}
\includegraphics[width = 0.45\textwidth]{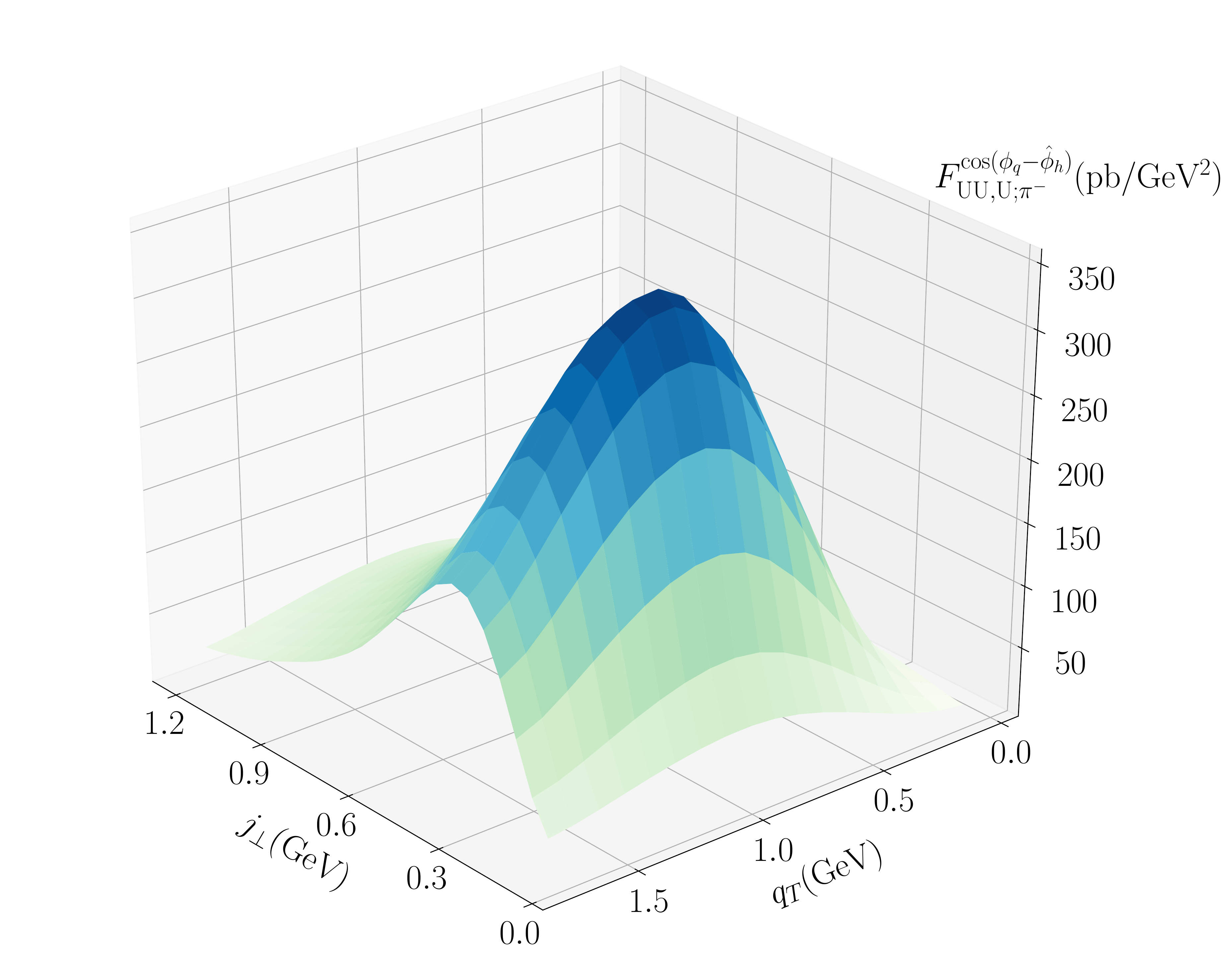}\hspace{1.8cm}
\includegraphics[width = 0.44\textwidth]{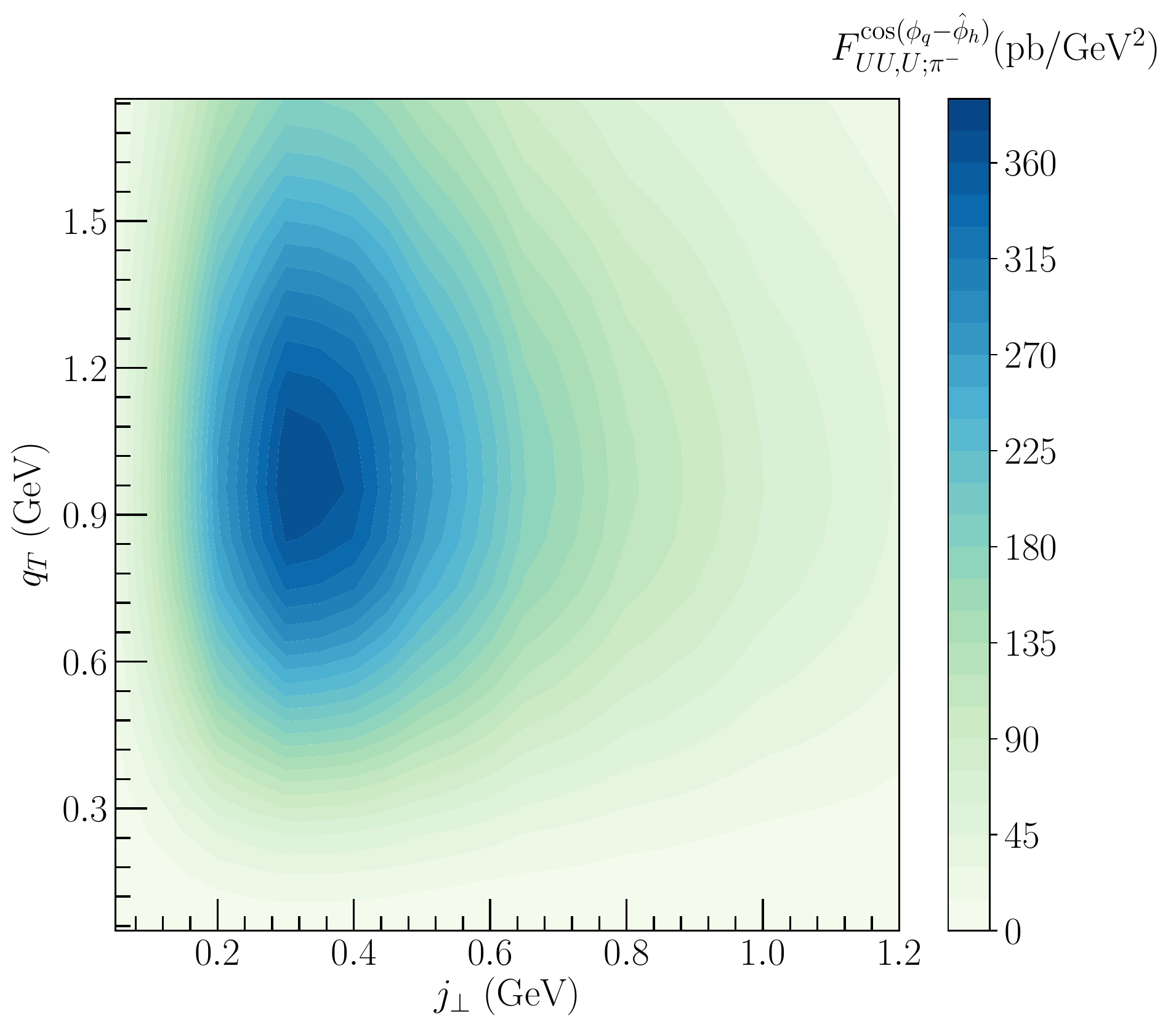}
\includegraphics[width = 0.45\textwidth]{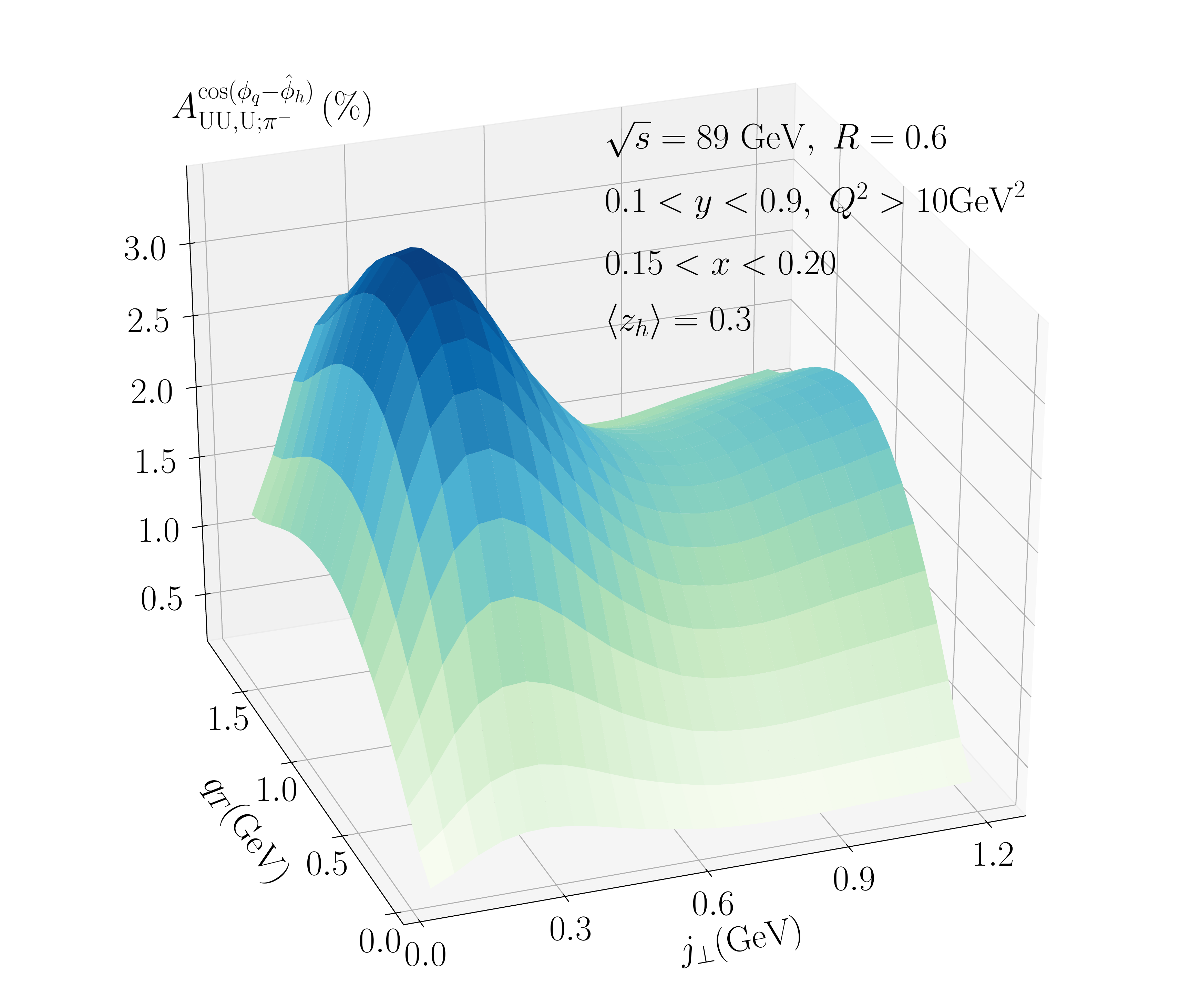}
\hspace{1.8cm}\includegraphics[width = 0.42\textwidth]{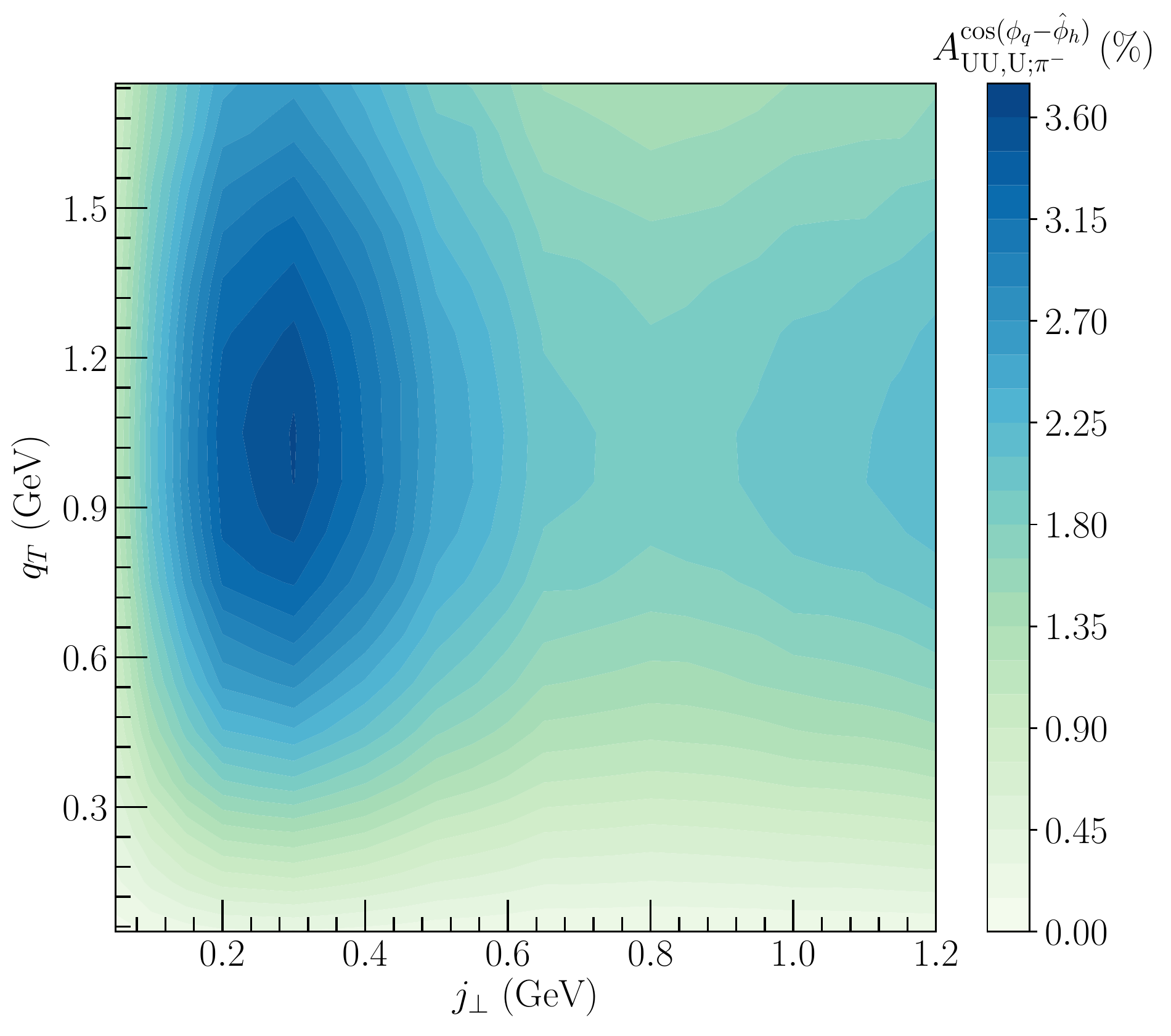}
\caption{$F_{UU,U}$ (first row), $F_{UU,U}^{\cos(\phi_{q}-\hat{\phi}_{h})}$ (second row) and $A_{UU,U}^{\cos(\phi_{q}-\hat{\phi}_{h})}$ (third row) as a function of jet imbalance $q_T$ and $j_\perp$ for unpolarized $\pi^-$ in jet production with electron in unpolarized $ep$ collision with EIC kinematics, where we have applied $\sqrt{s}=89$ GeV, jet radius $R=0.6$, inelasticity $y$ in range $[0.1,0.9]$, $Q^2>10$ GeV$^2$, Bjorken-$x$ in $[0.15,0.20]$ and average momentum fraction $\langle z_h\rangle=0.3$. Left column: Three dimensional plots of the structure functions and their ratio in $q_T$ and $j_\perp$. Right column: Contour plots of the structure functions and their ratio.}
    \label{fig:scn2_pim}
\end{figure}

Instead of integrating over $q_T$ and $j_\perp$, we can also create a plot simultaneously differential in $q_T$ and $j_\perp$. As discussed above, this is useful as TMDPDFs and TMDFFs are separately sensitive to $q_T$ and $j_T$, respectively. As the asymmetry is the largest for $0.15<x<0.2$ in Fig.~\ref{fig:scn2_eic}, we choose $0.15<x<0.2$ for the EIC kinematics with jet radius $R=0.6$, inelasticity cut $0.1<y<0.9$, momentum fraction $\langle z_h\rangle=0.3$ and $Q^2>10$ GeV$^2$ to create the three-dimensional and contour plots of the azimuthal asymmetry $A_{UU,U}^{\cos(\phi_{q}-\hat{\phi}_{h})}$ in Figs.~\ref{fig:scn2_pip} and~\ref{fig:scn2_pim}. To understand better the unpolarized and azimuthal-dependent structure function in more details, we plot the three-dimensional and contour plots of $F_{UU,U}$, $F_{UU,U}^{\cos(\phi_{q}-\hat{\phi}_{h})}$, and their ratio $A_{UU,U}^{\cos(\phi_{q}-\hat{\phi}_{h})}$, in order from the first to the third row. Fig.~\ref{fig:scn2_pip} is for $\pi^+$ inside the jet while Fig.~\ref{fig:scn2_pim} is for $\pi^-$ in the jet. From the first row of both figures, one sees the Sudakov peak from the unpolarized TMDPDF and TMDFF for constant $j_\perp$ and $q_T$ slices, respectively. From the second row, the shape of the constant $j_\perp$ slices, i.e. the $q_T$-dependence at a constant $j_\perp$ is determined by the Boer-Mulders function. On the other hand, the shape of the constant $q_T$ slices, i.e. the $j_\perp$-dependence at a constant $q_T$, is determined by the Collins function. Finally, the ratio of these plots, which define the asymmetry, is given in the third row. We find that the spin asymmetry for $\pi^+$ production in jet tends to be negative $\sim 1\%$ and for $\pi^-$ production in jet is positive with magnitude $\sim3\%$.

\begin{figure}
\centering
\includegraphics[width = 0.43\textwidth]{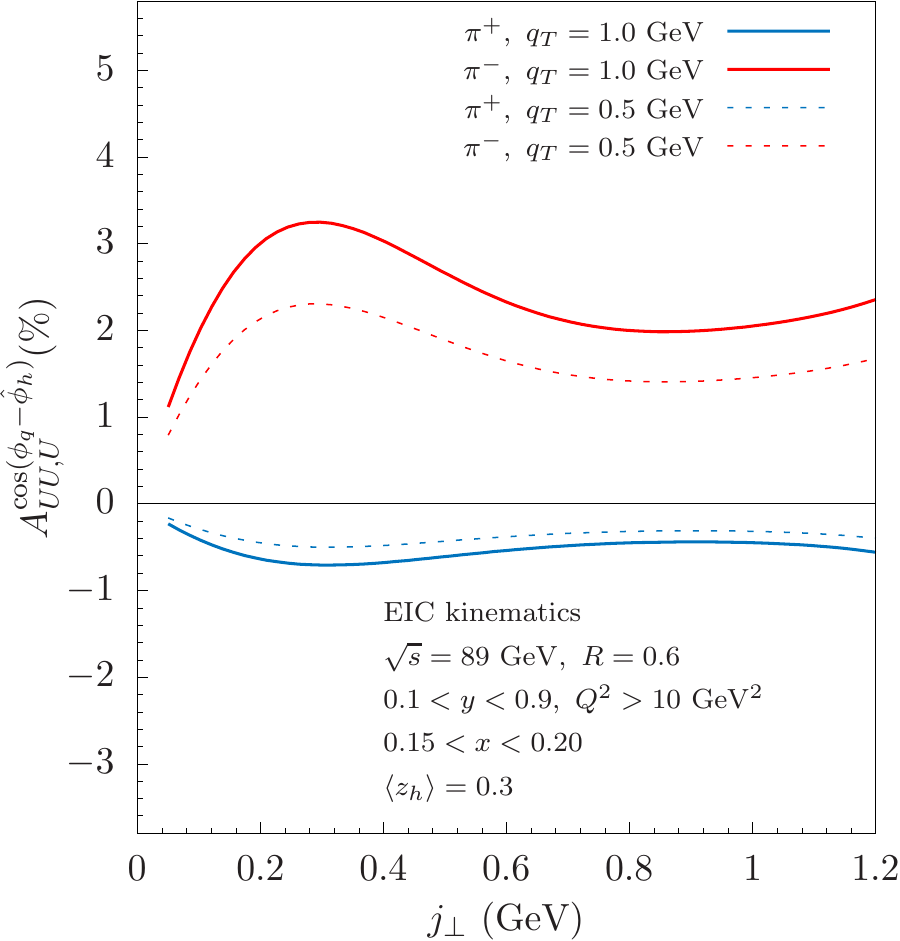}
\hspace{1cm}\includegraphics[width = 0.43\textwidth]{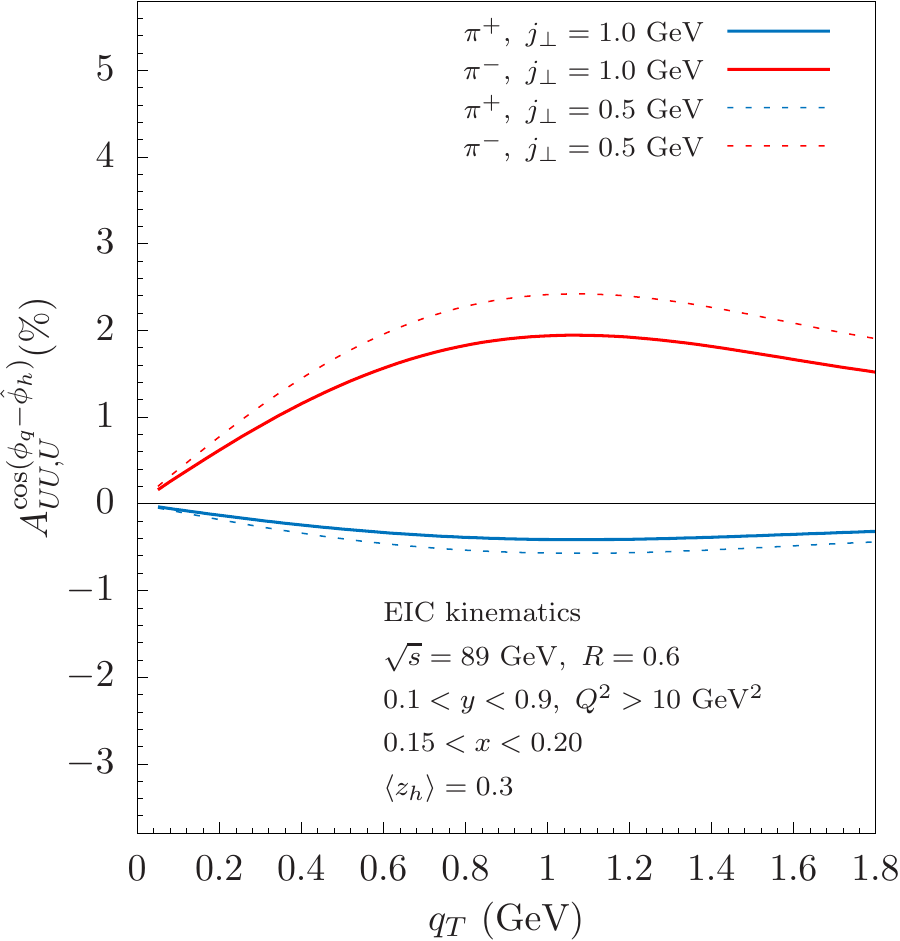}
\caption{Horizontal ($j_\perp$-dependent) and vertical ($q_T$-dependent) slices of  $A_{UU,U}^{\cos(\phi_{q}-\hat{\phi}_{h})}$ in Figs.~\ref{fig:scn2_pip} and \ref{fig:scn2_pim} for unpolarized $\pi^\pm$ in jet production with electron in unpolarized $ep$ collision as a function of transverse momentum $j_\perp$ (left panel, with $q_T=1.0,\ 0.5$ GeV for $\pi^\pm$) and jet imbalance $q_T$ (right panel, with $j_\perp=1.0,\ 0.5$ GeV for $\pi^\pm$), respectively. Here we apply the EIC kinematics with $\sqrt{s}=89$ GeV, jet radius $R=0.6$, inelasticity $y$ in range $[0.1,0.9]$, $Q^2>10$ GeV$^2$, Bjorken-$x$ in $[0.15,0.20]$ and average momentum fraction $\langle z_h\rangle=0.3$.}
    \label{fig:scn2_2d}
\end{figure}

To give a more straightforward interpretation, in Fig.~\ref{fig:scn2_2d} we show the horizontal ($j_\perp$-dependent) slices for $q_T=1.0$ GeV (solid curves) and $q_T=0.5$ GeV (dashed curves) of the third row of Figs.~\ref{fig:scn2_pip} and \ref{fig:scn2_pim} in the left plot with blue curves representing $\pi^+$ and red curves representing $\pi^-$ productions in jet. As for the right plot of Fig.~\ref{fig:scn2_2d}, we provide the vertical ($q_T$-dependent) slices for $j_\perp=1.0$ GeV (solid curves) and $j_\perp=0.5$ GeV (dashed curves) of the third row of Figs.~\ref{fig:scn2_pip} and \ref{fig:scn2_pim} with blue curves representing $\pi^+$ and red curves representing $\pi^-$ productions in jet. With the reasonable asymmetry of order negative  $\sim 1\%$ for $\pi^+$ and positive $\sim 3\%$ for $\pi^-$ with the TMD evolution turned on, this is a promising observable at the EIC to study the Boer-Mulders functions and Collins fragmentation functions.

\section{Polarized hadron inside a jet}\label{sec:pol_h}
In this section, we present the most general framework in this paper by allowing polarization for the hadron observed inside jets. In particular, in addition to the transverse momentum $\bm{j}_\perp$ with respect to the jet axis, we are now sensitive to the spin vector $S_h$, which gives arise to the additional correlations involving the polarization of the final hadron. With this most general case, we find that all of the TMDPDFs and TMDFFs make appearance in at least one of the structure functions. 

Within the context of the back-to-back electron-jet production, polarized hadron in jet is studied for the first time. See~\cite{Kang:2020xyq} for the corresponding study for the inclusive jet production case. After presenting the structure functions, we carry out a phenomenological study, namely asymmetry $A_{UU,T}^{\sin(\hat{\phi}_h-\hat{\phi}_{S_h})}$, to study the polarizing TMDFF $D_{1T}^{\perp}$ using the future EIC kinematics.

\subsection{Theoretical framework}
We continue our discussion from Sec.\ \ref{sec2:theoretical} and further generalize our study of hadron distribution inside jets by including the polarization of the final hadron. The spin vector, $S_h$, of the hadron observed inside a jet can be parametrized as
\bea
S_h = \left[\lambda_h\frac{p_h^+}{M_h},-\lambda_h\frac{M_h}{2p_h^+},\bm{S}_{h\perp}\right]_{x_Jy_Jz_J}\,,
\eea
where we use the jet coordinate system $x_Jy_Jz_J$ found in Fig.~\ref{fig:jperp}. We parametrize the transverse component of the spin $\bm{S}_{h\perp}$ in the $ep$ center-of-mass frame as
\bea
\bm{S}_{h\perp}=S_{h\perp}(\cos\hat{\phi}_{S_h}\cos\theta_J,\sin\hat{\phi}_{S_h},-\cos\hat{\phi}_{S_h}\sin\theta_J),
\eea 
where $\theta_J$ is defined in Eq.~\eqref{eq:jetmom} and $\hat{\phi}_{S_h}$ is the azimuthal angle of the transverse spin ${\bm S}_{h\perp}$ measured in the jet coordinate $x_Jy_Jz_J$ system.
\clearpage
The differential cross section of the back-to-back electron-jet production with polarized hadron observed inside jets is given by

\bea
\label{eq:poljeth}
&\frac{d\sigma^{p(S_A)+e(\lambda_e)\to e+(\text{jet}\,h(S_h))+X}}{d{p}^2_Tdy_Jd^2{\bm q}_Tdz_h d^2{\bm j}_\perp}=F_{UU,U}+\cos({\phi}_{q}-\hat{\phi}_{h})F_{UU,U}^{\cos({\phi}_{q}-\hat{\phi}_{h})}\nnu
&+\lambda_p\bigg\{\lambda_eF_{LL,U}+\sin(\phi_{q}-\hat{\phi}_h)F_{LU,U}^{\sin(\phi_{q}-\hat{\phi}_h)}\bigg\}\nnu
&+S_T\bigg\{\sin(\phi_{q}-{\phi}_{S_A})F^{\sin(\phi_{q}-{\phi}_{S_A})}_{TU,U}+\lambda_e\cos(\phi_{q}-{\phi}_{S_A})F^{\cos(\phi_{q}-{\phi}_{S_A})}_{TL,U}\nnu
&\hspace{1.3cm}+\sin(\phi_{S_A}-\hat{\phi}_h)F^{\sin(\phi_{S_A}-\hat{\phi}_h)}_{TU,U}+\sin(2\phi_{q}-\hat{\phi}_h-\phi_{S_A})F^{\sin(2\phi_{q}-\hat{\phi}_h-\phi_{S_A})}_{TU,U}\bigg\}\nnu
&+\lambda_h\bigg\{\lambda_e F_{UL,L}+\sin(\hat{\phi}_h-\phi_{q})F_{UU,L}^{\sin(\hat{\phi}_h-\phi_{q})}+\lambda_p \bigg[F_{LU,L}+{\cos(\hat{\phi}_h-\phi_{q})F_{LU,L}^{\cos(\hat{\phi}_h-\phi_{q})}}\bigg]\nnu
&\hspace{1.3cm}+S_T\bigg[\cos(\phi_{q}-{\phi}_{S_A})F^{\cos(\phi_{q}-{\phi}_{S_A})}_{TU,L}+\lambda_e\sin(\phi_{q}-{\phi}_{S_A})F^{\sin(\phi_{q}-{\phi}_{S_A})}_{TL,L}\nnu
&\hspace{2.5cm}+{\cos(\phi_{S_A}-\hat{\phi}_h)F^{\cos(\phi_{S_A}-\hat{\phi}_h)}_{TU,L}}+{\cos(2\phi_{q}-\phi_{S_A}-\hat{\phi}_h)F^{\cos(2\phi_{q}-\phi_{S_A}-\hat{\phi}_h)}_{TU,L}}\bigg]\bigg\}\nnu
&+S_{h\perp}\bigg\{\sin(\hat{\phi}_h-\hat{\phi}_{S_h})F^{\sin(\hat{\phi}_h-\hat{\phi}_{S_h})}_{UU,T}+\lambda_e{\cos(\hat{\phi}_h-\hat{\phi}_{S_h})F_{UL,T}^{\cos(\hat{\phi}_h-\hat{\phi}_{S_h})}}\nnu
&\hspace{1.3cm}+\sin(\hat{\phi}_{S_h}-\phi_{q})F_{UU,T}^{\sin(\hat{\phi}_{S_h}-\phi_{q})}+\sin(2\hat{\phi}_h-\hat{\phi}_{S_h}-\phi_{q})F_{UU,T}^{\sin(2\hat{\phi}_h-\hat{\phi}_{S_h}-\phi_{q})}\nnu
&\hspace{1.3cm}+\lambda_p\bigg[{\cos(\hat{\phi}_h-\hat{\phi}_{S_h}) F^{\cos(\hat{\phi}_h-\hat{\phi}_{S_h})}_{LU,T}}+\cos(\phi_{q}-\hat{\phi}_{S_h}) F^{\cos(\phi_{q}-\hat{\phi}_{S_h})}_{LU,T}\nnu
&\hspace{2.5cm}+\cos(2\hat{\phi}_h-\hat{\phi}_{S_h}-\phi_{q}) F^{\cos(2\hat{\phi}_h-\hat{\phi}_{S_h}-\phi_{q})}_{LU,T}+\lambda_e\sin(\hat{\phi}_h-\hat{\phi}_{S_h})F_{LL,T}^{\sin(\hat{\phi}_h-\hat{\phi}_{S_h})}\bigg]\nnu
&\hspace{1.3cm}+S_T\bigg[\cos(\phi_{S_A}-\hat{\phi}_{S_h})F^{\cos(\phi_{S_A}-\hat{\phi}_{S_h})}_{TU,T}+\cos(2\hat{\phi}_h-\hat{\phi}_{S_h}-\phi_{S_A})F^{\cos(2\hat{\phi}_h-\hat{\phi}_{S_h}-\phi_{S_A})}_{TU,T}\nnu
&\hspace{2.5cm}+\sin(\hat{\phi}_h-\hat{\phi}_{S_h})\sin(\phi_{q}-{\phi}_{S_A})F_{TU,T}^{\sin(\hat{\phi}_h-\hat{\phi}_{S_h})\sin(\phi_{q}-{\phi}_{S_A})}\nnu
&\hspace{2.5cm}+{\cos(\hat{\phi}_h-\hat{\phi}_{S_h})\cos(\phi_{q}-\phi_{S_A}))F_{TU,T}^{\cos(\hat{\phi}_h-\hat{\phi}_{S_h})\cos(\phi_{q}-\phi_{S_A})}}\nnu
&\hspace{2.5cm}+\cos(2\phi_{q}-\phi_{S_A}-\hat{\phi}_{S_h})F_{TU,T}^{\cos(2\phi_{q}-\phi_{S_A}-\hat{\phi}_{S_h})}\nnu
&\hspace{2.5cm}+\cos(2\hat{\phi}_h-\hat{\phi}_{S_h}+2\phi_{q}-\phi_{S_A})F_{TU,T}^{\cos(2\hat{\phi}_h-\hat{\phi}_{S_h}+2\phi_{q}-\phi_{S_A})}\nnu
&\hspace{2.5cm}+\lambda_e\cos(\hat{\phi}_h-\hat{\phi}_{S_h})\sin(\phi_{S_A}-\phi_{q})F_{TL,T}^{\cos(\hat{\phi}_h-\hat{\phi}_{S_h})\sin(\phi_{S_A}-\phi_{q})}\nnu
&\hspace{2.5cm}+\lambda_e\sin(\hat{\phi}_h-\hat{\phi}_{S_h})\cos(\phi_{S_A}-\phi_{q}))F_{TL,T}^{\sin(\hat{\phi}_h-\hat{\phi}_{S_h})\cos(\phi_{S_A}-\phi_{q})}\bigg]\bigg\}\,,
\eea
where the structure functions with unpolarized hadron in the final state $C=U$ already made appearances in Eq.~\eqref{eq:unpjeth} in last section.

As discussed already above in Sec.~\ref{sec3:Theoretical}, these structure functions are factorized in terms of TMDPDFs and TMDJFFs in the region $q_T\sim j_\perp \ll p_T R$. As emphasized there, TMDPDFs and TMDJFFs separately depend only on $\bm{q}_T$ and $\bm{j}_\perp$, respectively, and thus can be separately constrained. Including the spin dependence, the correlator for the TMDJFFs in Eq.~\eqref{eq:jperpcorr} is now parametrized at the leading twist accuracy as
\bea
\label{eq:TMDJFFs}
\Delta_{\rm jet}^{h/q}(z_h,\bm{j}_\perp, S_{h})=&\frac{1}{2}\Bigg\{\left(\mathcal{D}_{1} - \frac{\epsilon_{T}^{ij} j_{\perp}^i S_{h\perp}^j}{z_hM_h}\mathcal{D}_{1 T}^{\perp}\right) \slashed{n}_J+\left(\lambda_h \mathcal{G}_{1L} - \frac{{\bm j}_\perp\cdot{\bm S}_{h\perp}}{z_hM_h}\mathcal{G}_{1T}\right)\gamma_{5} \slashed{n}_J\nonumber\\
&
-i\sigma^{i\mu}n_{J,\mu} \left(\mathcal{H}_1{S_{h\perp}^i}\gamma_5 -i\mathcal{H}_1^{\perp}\frac{j_{\perp}^i}{z_hM_h} - \mathcal{H}_{1L}^\perp\frac{\lambda_h j_{\perp}^i}{z_hM_h} \gamma_5 \right.
\nonumber\\
&
\left.\hspace{15mm}+\mathcal{H}_{1T}^\perp\frac{{\bm j}_\perp\cdot{\bm S}_{h\perp} j_{\perp}^i - \frac{1}{2}{ j}_\perp^2S_{h\perp}^i}{z_h^2M_h^2}\gamma_5\right)\Bigg\}\,,
\eea
where the TMDJFFs associated with an unpolarized hadron were already given in Eq.~\eqref{eq:unpjffdef}. The physical interpretations of the TMDJFFs are summarized in Table.~\ref{tabTMDff}. Explicit expressions of the structure functions in Eq.~\eqref{eq:poljeth} in terms of these TMDJFFs in Eq.~\eqref{eq:TMDJFFs} and TMDPDFs in Eq.~\eqref{eq:TMDPDFsb} are given in Appendix~\ref{app2}. The Table.~\ref{tab5} summarizes the azimuthal asymmetries and their associated TMDPDFs and TMDJFFs. In Appendix~\ref{appTMDFFs}, we generalize the discussion given in Sec.~\ref{sec3:Theoretical} and present all of the relations between the TMDJFFs and TMDFFs in $j_\perp \ll p_T R$ region. 

\begin{table}[ht]
\centering
\begin{adjustbox}{angle=90}
\begin{tabular}[5pt]{ |c|c|c|c|c| } 
 \hline
 \diagbox[width=1.8cm]{JFF}{PDF} & $\tilde{f}_1$ &  $\tilde{f}_{1T}^{\perp(1)}$ & $\tilde{g}_{1L}$ & $\tilde{g}_{1T}^{(1)}$ \\ 
  \hline\xrowht[()]{15pt}
 $\mathcal{D}_1$ & $1$ & ${\sin(\phi_{q}-\phi_{S_A})}$ &$1$ & ${\cos(\phi_{q}-\phi_{S_A})}$   \\
  \hline\xrowht[()]{15pt}
$\mathcal{D}_{1T}^\perp$ & ${\sin(\hat{\phi}_h-\hat{\phi}_{S_h})}$ & ${\sin(\hat{\phi}_h-\hat{\phi}_{S_h})\sin(\phi_{q}-\phi_{S_A})}$ & ${\sin(\hat{\phi}_h-\hat{\phi}_{S_h})}$ & ${\sin(\hat{\phi}_h-\hat{\phi}_{S_h})\cos(\phi_{q}-\phi_{S_A})}$  \\ 
  \hline\xrowht[()]{15pt}
  $\mathcal{G}_{1L}$ &  $1$ & ${\sin(\phi_{q}-\phi_{S_A})}$ & $1$ & ${\cos(\phi_{q}-\phi_{S_A})} $ \\ 
  \hline\xrowht[()]{15pt}
$\mathcal{G}_{1T}$ & ${\cos(\hat{\phi}_h-\hat{\phi}_{S_h})}$ & 
${\cos(\hat{\phi}_h-\hat{\phi}_{S_h})\sin(\phi_{S_A}-\phi_{q})}$ & ${\cos(\hat{\phi}_h-\hat{\phi}_{S_h})}$& ${\cos(\hat{\phi}_h-\hat{\phi}_{S_h})\cos(\phi_{q}-\phi_{S_A})}$\\ 
  \hline
  \hline
 \diagbox[width=1.8cm]{JFF}{PDF}& $\tilde{h}_1^{\perp(1)}$ & $\tilde{h}_{1L}^{\perp(1)}$ & $\tilde{h}_{1}$ & $\tilde{h}_{1T}^{\perp(2)}$ \\ 
  \hline\xrowht[()]{15pt}
  $\mathcal{H}_1^\perp$ & ${\cos(\hat{\phi}_h-\phi_{q})}$ & ${\sin(\phi_{q}-\hat{\phi}_h)}$ &  ${\sin({\phi}_{S_A}-\hat{\phi}_h)}$ &  ${\sin(2\phi_{q}-\hat{\phi}_{h}-\phi_{S_A})}$  \\ 
  \hline\xrowht[()]{15pt}
$\mathcal{H}_{1L}^\perp$ & ${\sin(\hat{\phi}_h-\phi_{q})}$ & ${\cos(\phi_{q}-\hat{\phi}_h)}$ &  ${\cos({\phi}_{S_A}-\hat{\phi}_h)}$ & ${\cos(2\phi_{q}-\hat{\phi}_h-{\phi}_{S_A})}$
 \\ 
  \hline\xrowht[()]{15pt}
$\mathcal{H}_{1}$ &${\sin(\hat{\phi}_{S_h}-\phi_{q})}$ & ${\cos(\phi_{q}-\hat{\phi}_{S_h})}$ & ${\cos({\phi}_{S_A}-\hat{\phi}_{S_h})}$ & ${\cos(2{\phi}_q-\hat{\phi}_{S_h}-\phi_{S_A})}$\\ 
  \hline\xrowht[()]{15pt}
$\mathcal{H}_{1T}^\perp$ & ${\sin(2\hat{\phi}_h-\hat{\phi}_{S_h}-\phi_{q})}$ & ${\cos(2\hat{\phi}_h-\hat{\phi}_{S_h}-\phi_{q})}$ &  ${\cos(2\hat{\phi}_{h}-\hat{\phi}_{S_h}-{\phi}_{S_A})}$ & ${\cos(2\hat{\phi}_h-\hat{\phi}_{S_h}+{\phi}_{S_A}-2\phi_{q})}$\\ 
  \hline  
\end{tabular}
\end{adjustbox}
  \caption{We summarize all the azimuthal asymmetries for $e+p\rightarrow e+\text{jet}(h)+X$ process, where the initial proton and produced hadron in jet have general polarizations and the initial electron is unpolarized or carries helicity. Parametrizations of structure functions given in the table are provided in Appendix\ \ref{app2}.} \label{tab5}
  \end{table}
\clearpage

\subsection{Phenomenology: $\Lambda$ transverse polarization inside the jet}
\label{sec4:pheno}
As an example of application of studying the back-to-back electron-jet production with a polarized hadron inside the jet, we study transverse spin asymmetry of a $\Lambda$ particle inside the jet, $A^{\sin(\hat{\phi}_\Lambda-\hat{\phi}_{S_\Lambda})}_{UU,T}$, which arises from the structure function $F^{\sin(\hat{\phi}_\Lambda-\hat{\phi}_{S_\Lambda})}_{UU,T}$. The spin asymmetry is defined as
\bea\label{eq:auut}
A_{UU,T}^{\sin(\hat{\phi}_{\Lambda}-\hat{\phi}_{S_\Lambda})} = \frac{F^{\sin(\hat{\phi}_\Lambda-\hat{\phi}_{S_\Lambda})}_{UU,T}}{F_{UU,U}}\,.
\eea
The asymmetry can be measured in the unpolarized electron-proton collisions by observing the distribution of transversely polarized $\Lambda$s inside the jet. The $\Lambda$ transverse spin vector $\bm{S}_{\Lambda\perp}$ and the transverse momentum $\bm{j}_\perp$ with respect to the jet axis can correlate with each other, and leads to the $\sin(\hat{\phi}_{\Lambda}-\hat{\phi}_{S_\Lambda})$ correlation between their azimuthal angles. In practice, this is the mechanism which can describe the transverse polarization of $\Lambda$ particles inside the jet. As can be seen from Eq.~\eqref{eq:strh9}, the structure function $F^{\sin(\hat{\phi}_\Lambda-\hat{\phi}_{S_\Lambda})}_{UU,T}$ depends on the unpolarized TMDPDF $f_1^q(x,k_T^2)$ and TMDJFF $\mathcal{D}_{1T}^{\perp\ \Lambda/q}(z_\Lambda,j_\perp^2)$. The TMDJFF $\mathcal{D}_{1T}^{\perp\ \Lambda/q}(z_\Lambda,j_\perp^2)$ describes distribution of transversely polarized $\Lambda$ inside the jet initiated by an unpolarized quark. This is reminiscent of the polarizimg TMDFF $D_{1T}^{\perp\ \Lambda/q}(z_\Lambda,j_\perp^2)$, which describes distribution of transversely polarized $\Lambda$ fragmented from an unpolarized quark. As shown in the Appendix\ \ref{appTMDFFs}, these two are in fact related at $j_\perp \ll p_T R$ region. For this reason, we will also refer to the TMDJFF $\mathcal{D}_{1T}^{\perp\ \Lambda/q}(z_\Lambda,j_\perp^2)$ as polarizing TMDJFF.

The factorization formula of the denominator $F_{UU,U}$ was presented in Eq.~\eqref{eq:FUUUbefore}, which is expressed in terms of the unpolarized TMDPDF and TMDFF, and was extensively discussed there. On the other hand, the factorization formula for $F^{\sin(\hat{\phi}_\Lambda-\hat{\phi}_{S_\Lambda})}_{UU,T}$ is given in Eq.~\eqref{eq:strh9}, which is explicitly expressed as
\bea
\label{eq:FUUT-lambda}
F^{\sin(\hat{\phi}_\Lambda-\hat{\phi}_{S_\Lambda})}_{UU,T} =&\hat{\sigma}_0 \,H(Q,\mu)\sum_q e_q^2\, \frac{ j_\perp}{z_hM_h} D_{1T}^{\perp\,\Lambda/q}(z_\Lambda,j_\perp^2,\mu, \zeta_J)
\nnu
&\times 
\int\frac{b \,db}{2\pi}J_0(q_Tb)\,x\,\tilde{f}_1^{q}(x,b^2, \mu,\zeta)\bar{S}_{\rm global}(b^2,\mu)\bar{S}_{cs}(b^2,R,\mu)\,,
\eea
where we also used Eq.~\eqref{eq:TMDJFFrel} to express the polarizing TMDJFF $\mathcal{D}_{1T}^{\perp\,\Lambda/q}$ in terms of polarizing TMDFF $D_{1T}^{\perp\,\Lambda/q}$. The derivation is again similar to that for the case of the unpolarized TMDJFF $\mathcal{D}_1^{h/q}$ and the corresponding unpolarized TMDFF $D_1^{h/q}$, as shown from Eqs.~\eqref{unp_JFF_FF} to \eqref{unp_JFF_FF2}.

For the numerical analysis, we make predictions for $\Lambda$ transverse polarization inside the jet in back-to-back electron-jet production at the EIC. Specifically, we include TMD evolution to the unpolarized TMDFF $D_1^{\Lambda/q}$ and the polarizing TMDFF $D_{1T}^{\perp\,\Lambda/q}$ extracted in~\cite{Callos:2020qtu}, based on the recent measurement of back-to-back $\Lambda$ and a light hadron production in $e^+ e^-$ collisions, $e^+e^-\to \Lambda + h + X$, at the Belle Collaboration~\cite{Guan:2018ckx}. The extraction of polarizing TMDFF $D_{1T}^{\perp\,\Lambda/q}$ in~\cite{Callos:2020qtu} is again based on a Gaussian model, and we promote the parametrization to include the TMD evolution following the same method in Sec.~\ref{subsec:g1T}. At the end of day, we have the following expression for $D_{1T}^{\perp\,\Lambda/q}$ that is needed in the TMD factorization formula in Eq.~\eqref{eq:FUUT-lambda} 
\bea
D_{1T}^{\perp\,\Lambda/q}(z_\Lambda,j_\perp^2,\mu, \zeta_J)
=\int \frac{b^{2}\,db}{2\pi}\left(\frac{z_{\Lambda}^2M_{\Lambda}^2}{j_\perp}\right) J_1\left(\frac{j_\perp b}{z_{\Lambda}}\right) \tilde{D}_{1T}^{\perp(1)\,\Lambda/q}(z_\Lambda,b^2,\mu, 
\zeta_J)\,,
\eea
and $\tilde{D}_{1T}^{\perp(1)\,\Lambda/q}$ on the right-hand side takes the following form
\bea
\tilde{D}_{1T}^{\perp(1)\,\Lambda/q}(z_\Lambda,b^2,\mu, 
\zeta_J) = \,&\frac{\langle M_D^2\rangle}{2z_{\Lambda}^5M_\Lambda^2}\mathcal{N}_q(z_{\Lambda})D_1^{\Lambda/q}(z_\Lambda,\mu_{b_*})
\nnu
&\times \exp\left[-S_{\rm pert}\left(\mu, \mu_{b_*} \right) - S_{\rm NP}^{D_{1T}^\perp}\left(z_{\Lambda}, b, Q_0, \zeta_J\right)\right]\,,
\eea
where $\langle M_D^2\rangle=0.118$~GeV$^2$ and $\mathcal{N}_q(z_{\Lambda})=N_qz_{\Lambda}^{\alpha_q}(1-z_{\Lambda})^{\beta_q}\frac{(\alpha_q+\beta_q)^{(\alpha_q+\beta_q)}}{\alpha_q^{\alpha_q}\beta_q^{\beta_q}}$ with parameters $N_q,\ \alpha_q$ and $\beta_q$ determined in~\cite{Callos:2020qtu}. The non-perturbative Sudakov factor $S_{\rm NP}^{D_{1T}^\perp}$ is given by
\bea
S_{\rm NP}^{D_{1T}^\perp}(z_{\Lambda}, b, Q_0,\zeta_J) = \frac{g_2}{2}\ln{\frac{\sqrt{\zeta_J}}{Q_0}}\ln{\frac{b}{b_*}}+g_1^{D_{1T}^\perp} \frac{b^2}{z_{\Lambda}^2}\,,
\eea
with the parameter $g_1^{D_{1T}^\perp} = \langle M_D^2\rangle/4 = 0.0295$ GeV$^2$. We similarly include TMD evolution to the Gaussian model extraction of the unpolarized Lambda TMDFF of~\cite{Callos:2020qtu} to arrive at the same form as Eqs.~\eqref{eq:D1param} and~\eqref{eq:Sud-NPD}, except that we use the AKK08 parametrizations~\cite{Albino:2008fy} for the collinear $q\to \Lambda$ FFs $D_1^{\Lambda/q}(z_\Lambda, \mu_{b_*})$.

\begin{figure}[t]
\includegraphics[width = 0.48\textwidth]{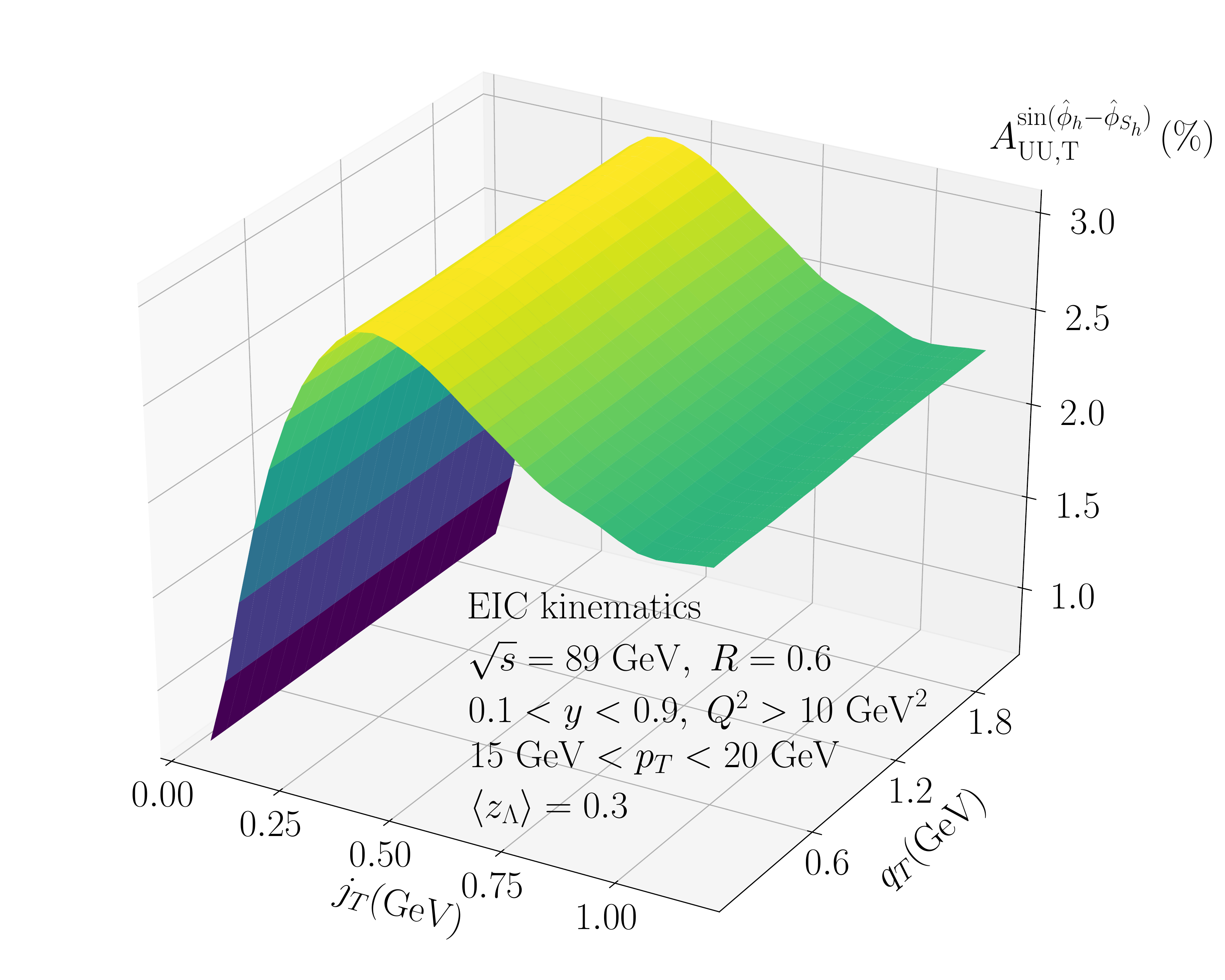}
    \includegraphics[width = 0.45\textwidth]{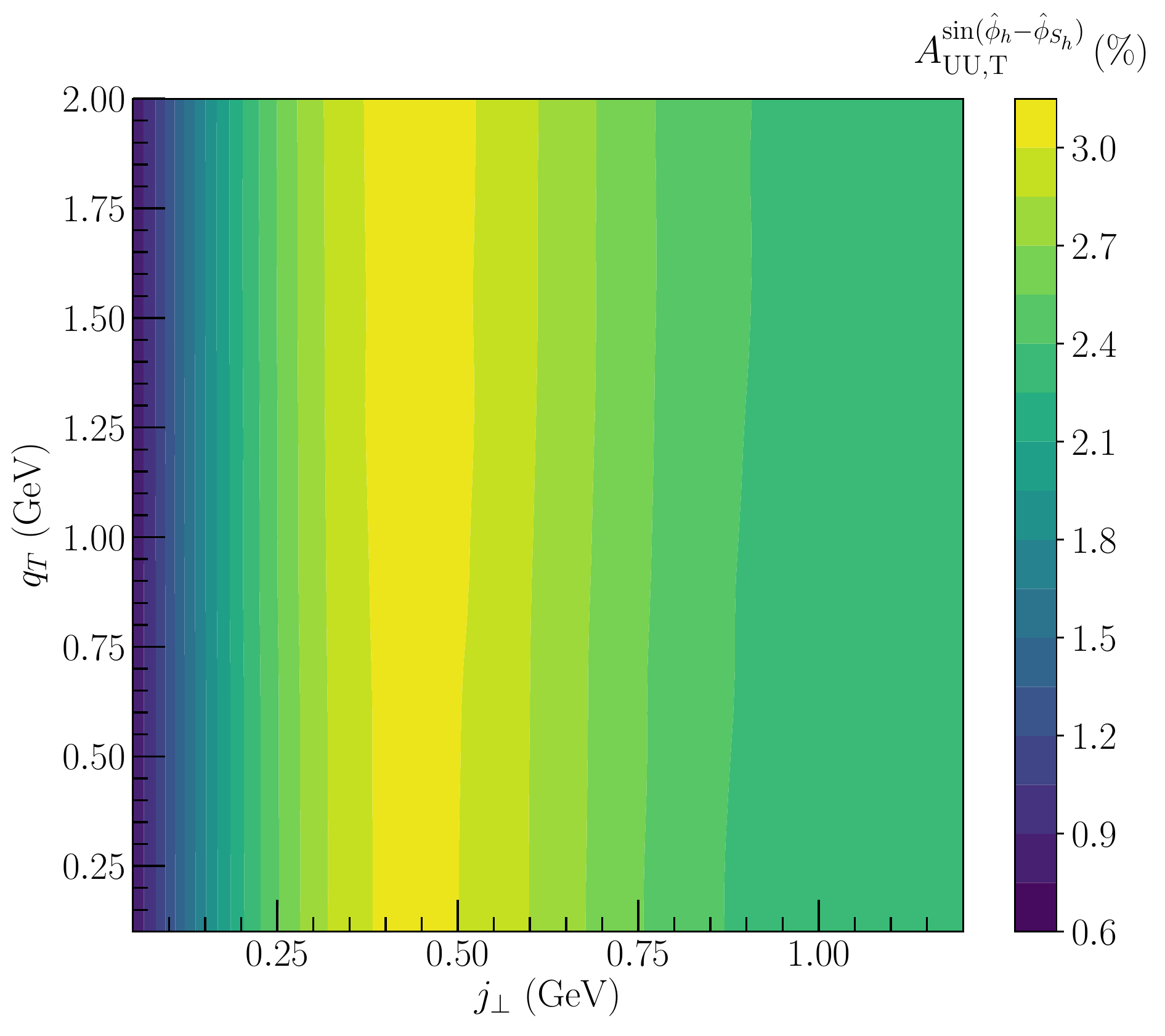}
\caption{$A_{UU,T}^{\sin(\hat{\phi}_{h}-\hat{\phi}_{S_h})}$ as a function of jet imbalance $q_T$ and $j_\perp$ for transversely polarized $\Lambda$ in jet production with electron in unpolarized $ep$ collision with EIC kinematics, where we have applied $\sqrt{s}=89$ GeV, jet radius $R=0.6$, inelasticity $y$ in range $[0.1,0.9]$, $Q^2>10$ GeV$^2$, jet transverse momentum $p_T$ in range $[15,20]$ GeV and average momentum fraction $\langle z_\Lambda\rangle=0.3$. Left: Three-dimensional plot of the spin asymmetry in $q_T$ and $j_\perp$. Right: Contour plot of the same quantity.}
    \label{fig:scn3}
\end{figure}

\begin{figure}[t]
\centering
\includegraphics[width = 0.45\textwidth]{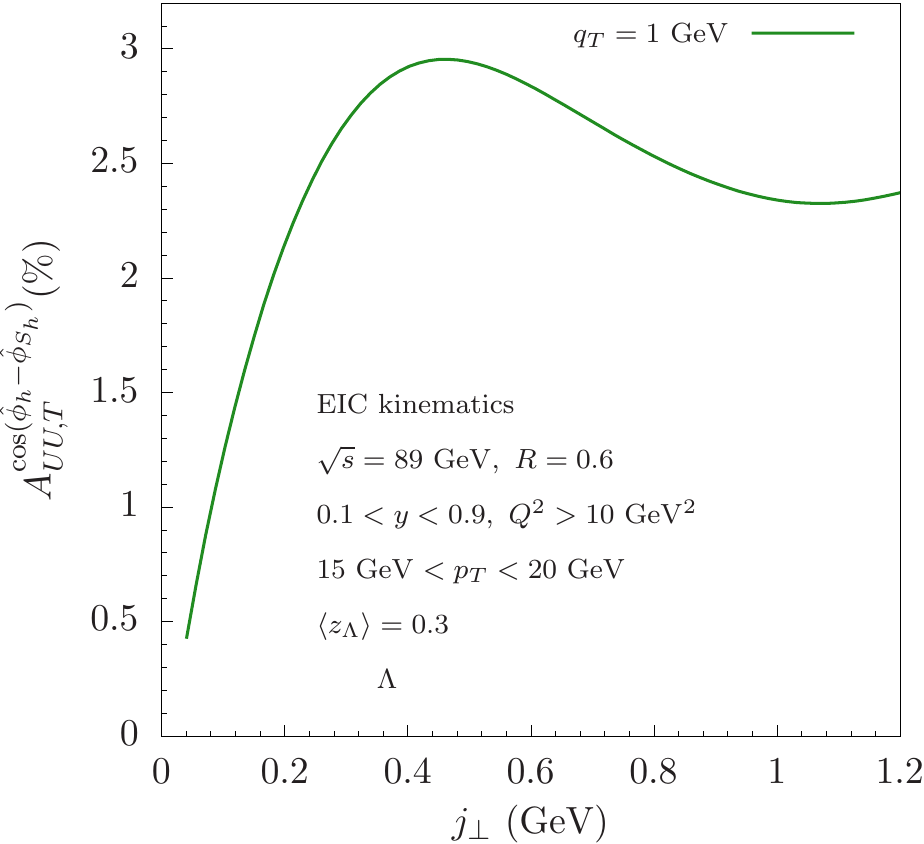}\hspace{0.5cm}
\includegraphics[width = 0.44\textwidth]{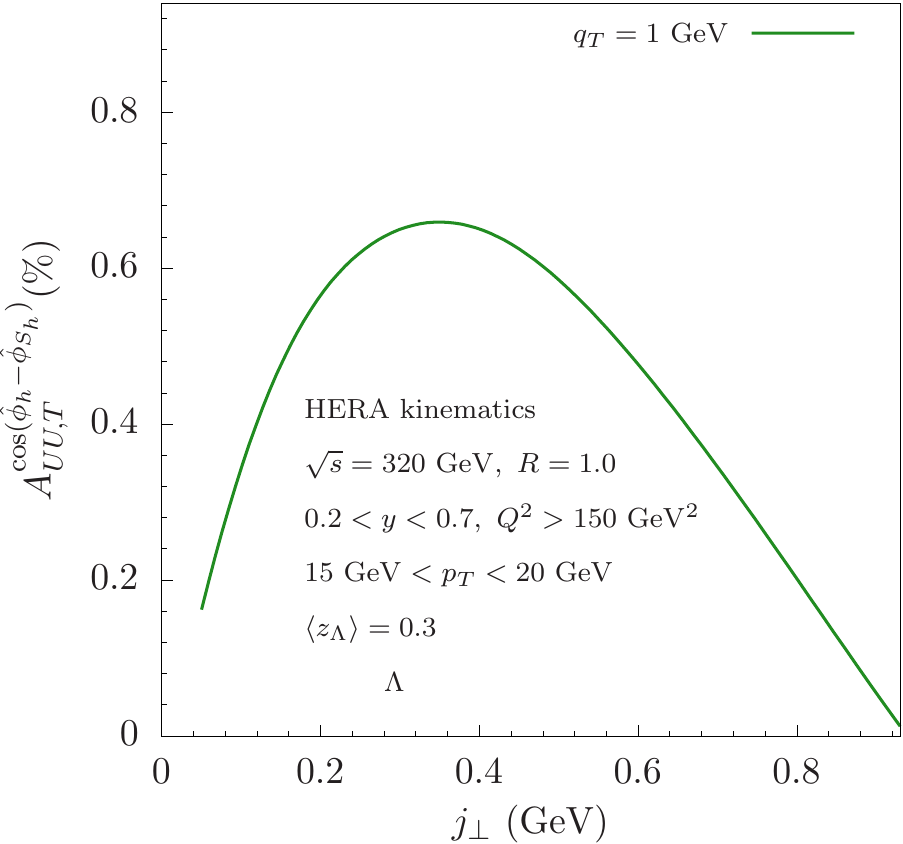}
\caption{Horizontal slices of $A_{UU,T}^{\sin(\hat{\phi}_{h}-\hat{\phi}_{S_h})}$ in Fig.~\ref{fig:scn3} as a function of transverse momentum $j_\perp$ for transversely polarized $\Lambda$ in jet production with electron in unpolarized $ep$ collision. We have jet transverse momentum $p_T$ in range $[15,20]$ GeV and average momentum fraction $\langle z_\Lambda\rangle=0.3$. Left panel: EIC kinematics, we have applied $\sqrt{s}=89$ GeV, jet radius $R=0.6$, inelasticity $y$ in range $[0.1,0.9]$, $Q^2>10$ GeV$^2$. Right panel: HERA kinematics, where we have applied $\sqrt{s}=320$ GeV, jet radius $R=1.0$, inelasticity $y$ in range $[0.2,0.7]$, $Q^2>150$ GeV$^2$.}
    \label{fig:scn3_2d}
\end{figure}

Let us make some predictions for $\Lambda$ transverse polarization at the future EIC. In Fig.~\ref{fig:scn3}, we plot the asymmetry $A_{UU,T}^{\sin(\hat{\phi}_{\Lambda}-\hat{\phi}_{\Lambda})}$ differential in both the imbalance $q_T$ and $j_\perp$ using the EIC kinematics. We choose our CM energy to be $\sqrt{s}=89$ GeV with inelasticity $y$ and Bjorken $x$ integrated between $0.1<y<0.9$ and $0.15<x<0.20$, respectively. As the $q_T$ dependence of the numerator and denominator of the asymmetry given in Eq.~\eqref{eq:auut} is both determined by the unpolarized TMDPDF $f_1$, we find that asymmetry given by the ratio is constant as expected in constant $j_\perp$ slices. As the dependence in TMDPDFs cancel in ratio, this asymmetry is particularly useful in extracting the polarizing TMDFF $D_{1T}^{\perp\,\Lambda/q}$. This is advantageous in comparison with the standard SIDIS measurements where the polarizing TMDFF $D_{1T}^{\perp\,\Lambda/q}$ would be still convolved with the  unpolarized TMDPDF $f_1$. Finally, in Fig.~\ref{fig:scn3_2d} we present horizontal slices of the contour plots $A_{UU,T}^{\sin(\hat{\phi}_{h}-\hat{\phi}_{S_h})}$ in Fig.~\ref{fig:scn3} as a function of transverse momentum $j_\perp$. In the left panel, the plot is for the EIC kinematics, while the right panel is for the HERA kinematics. For the EIC case, as $j_\perp$ increases, the asymmetry increases up to $3\%$ at $j_\perp=0.4$ GeV then slowly drops to around $2.5\%$, indicating feasibility for measurements at the future EIC. On the other hand, for the HERA kinematics, the spin asymmetry is smaller $\sim 1\%$ and hopefully it can still be measurable.

\section{Conclusion}\label{sec:summary}
In this work, we perform a comprehensive study for all possible azimuthal asymmetries that can occur in back-to-back electron-jet production with and without the hadron observed inside the jet. We develop the theoretical framework by increasing the complexity of the final state observables: electron-jet production without observation of a hadron in the jet, unpolarized hadron in jet, and polarized hadron in jet. In the back-to-back region where the transverse momentum imbalance of electron-jet 
$q_T \ll p_T $ with $p_T$ the jet transverse momentum and $R$ jet radius, we demonstrate through TMD factorization that electron-jet production is an ideal process to investigate chiral-even TMDPDFs. In particular, we study the transverse-longitudinal asymmetry $A_{TL}^{\cos(\phi_q-\phi_{S_A})}$ using the EIC kinematics to explore its sensitivity to the quark transversal helicity distribution $g_{1T}^q$. For the hadron measured inside the jet with $j_\perp$ being the hadron transverse momentum with respect to the jet axis, different TMDPDFs and TMDFFs can be sensitive to $q_T$ and $j_\perp$ distribution, respectively. When $q_T\sim j_\perp \ll p_T R$, we derive the corresponding factorization formula, where both incoming particles and outgoing hadron in the jet can have general polarizations. When the final state hadron is unpolarized, we investigate Boer-Mulders correlation with Collins function, where a new azimuthal asymmetry $A_{UU,U}^{\cos(\phi_q-\hat\phi_h)}$ is introduced. Since this asymmetry involves only unpolarized proton and electron, we present the theory results with both HERA and EIC kinematics, which all have sizable magnitude and can be promising observables for the measurements. Especially, one can measure the asymmetry $A_{UU,U}^{\cos(\phi_q-\hat\phi_h)}$ using the current HERA data, and compare with our predictions. Furthermore, we also make predictions of $A_{U U, T}^{\sin \left(\hat{\phi}_{h}-\hat{\phi}_{S_h}\right)}$ for transverse polarized $\Lambda$ production inside the jet at the EIC, which provide a new opportunity to probe $\Lambda$ polarization. Besides the examples shown in this paper, other spin asymmetries concluded in this work also worth noticing and further studying. We emphasize the advantages of this set of observables, i.e. using the simultaneous differential information on $q_T$ and $j_\perp$, one is able to separately constrain TMDPDFs and TMDFFs. As a result, we demonstrate that the electron-jet production is an excellent method for probing transverse momentum dependent parton distribution functions and fragmentation functions in both unpolarized and polarized states.

\acknowledgments
This work is supported by the National Science Foundation under award No.~PHY-1945471 (Z.K., D.Y.S. and F.Z.), the US Department of Energy, Office of Nuclear Physics (K.L.), Center for Frontiers in Nuclear Science of Stony Brook University and Brookhaven National Laboratory (D.Y.S.), and  Shanghai Natural Science Foundation through Grant No. 21ZR1406100 (D.Y.S.). This work is supported within the framework of the TMD Topical Collaboration.

\appendix
\section{TMDJFFs and TMDFFs}\label{appTMDFFs}
In Eq.~\eqref{unp_JFF_FF}, we wrote down the relation between the ${\mathcal{D}}_{1}^{h/q}(z_h,j_\perp^2,p_TR,\mu)$ and the unpolarized TMDFF in $j_\perp \ll p_T R$ region. To write down explicit relations between other TMDJFFs and TMDFFs, we start by writing the parametrization of the TMDFF correlator~\cite{Metz:2016swz} in the momentum space. 
\bea
{\Delta}\left(z_h, \boldsymbol{k}_\perp,S_h\right) =& \sum_X \int \frac{d \xi^{+}d^2{\bm \xi}_T}{(2 \pi)^3} e^{i (k^{-} \xi^{+}+{\bm k}_{\perp}\cdot{\bm \xi}_T)/z_h}\left\langle 0\left| \psi_{q}\left(\xi^{+}, 0^{-}, {\bm \xi}_{T}\right)\right| p_h, S_{h} ; X\right\rangle \nnu
& \times\left\langle p_h, S_{h} ; X\left|\bar{\psi}_{q}\left(0^{+}, 0^{-}, {\bm 0}_{T}\right) \right| 0\right\rangle \,,
\eea
where $\boldsymbol{k}_\perp$ is the transverse momentum of the final hadron $h$ with respect to the fragmenting quark $q$ and we suppress the Wilson lines that make the correlator gauge invariant. To the leading twist accuracy, the parametrization is given as
\bea
\label{eq:TMDFF}
\Delta(z_h,\bm{k}_\perp,S_{h})=&\frac{1}{2}\Bigg\{\left({{D}}_{1}- \frac{\epsilon_{T}^{ij} k_{\perp}^i S_{h\perp}^j}{z_hM_h} {{D}}_{1 T}^{\perp}\right)\slashed{n}_c+\left(\lambda_h{{G}}_{1L}- \frac{\boldsymbol{k}_\perp\cdot\boldsymbol{S}_{h\perp}}{z_hM_h} {{G}}_{1 T}\right)\slashed{n}_c\gamma_5
\nnu
&
-i\sigma_{i\mu}n_c^\mu\left(H_1S_{h\perp}^i-iH_1^\perp\frac{k_\perp^i}{z_hM_h}-H_{1L}^\perp\frac{\lambda_h k_\perp^i}{z_hM_h}\gamma_5 \right.
\nnu
&\left.\hspace{15mm}+H_{1T}^\perp\frac{\boldsymbol{k}_\perp\cdot \boldsymbol{S}_{h\perp}k_\perp^i  - \frac{1}{2} {k}_\perp^{2}S_{h\perp}^i}{z_h^2M_h^2}\gamma_5\right)\Bigg\}\,,
\eea
where $n_c$ is the light-cone vector defined by the outgoing quark direction.

Just as in Eq.~\eqref{unp_JFF_FF}, we find it more convenient to derive the relations between the TMDJFFs and TMDFFs using the Fourier space expressions of the TMDFFs. The Fourier transformation for the TMDFF correlator is defined as
\bea
\tilde{\Delta}(z_h,\bm{b},S_h)=\frac{1}{z_h^2}\int d^2\bm{k}_\perp e^{-i\bm{k}_\perp\cdot \bm{b}/z_h}\Delta(z_h,\bm{k}_\perp,S_h)\,.
\eea
The TMDFF correlator in $\bm{b}$-space is then given as
\bea
\label{eq:TMDFFb}
\tilde{\Delta}(z_h,\bm{b},S_h)=&\frac{1}{2}\Bigg\{\left({\tilde{D}}_{1}(z_h,b^2)+ i{\epsilon_{T}^{ij} b^i S_{h\perp}^j}{z_hM_h} {\tilde{D}}_{1 T}^{\perp(1)}(z_h,b^2)\right)\slashed{n}_b\nnu
&\hspace{0.5cm}+\left(\lambda_h{\tilde{G}}_{1L}(z_h,b^2) + i{\boldsymbol{b}\cdot\boldsymbol{S}_{h\perp}}{z_hM_h} {\tilde{G}}_{1 T}^{(1)}(z_h,b^2)\right)\slashed{n}_b\gamma_5\nnu
&\hspace{0.5cm} -i\sigma_{i\mu}n_b^\mu\bigg[\tilde{H}_1(z_h,b^2)S_{h\perp}^i-\tilde{H}_1^{\perp(1)}(z_h,b^2){b^i}{z_hM_h}+i\tilde{H}_{1L}^{\perp(1)}(z_h,b^2){\lambda_h b^i}{z_hM_h}\gamma_5 \nnu
&\hspace{2.3cm}-\tilde{H}_{1T}^{\perp(2)}(z_h,b^2)\frac{1}{2}\left({\boldsymbol{b}\cdot \boldsymbol{S}_{h\perp}b^i - \frac{1}{2}{b}^{2}S_{h\perp}^i}\right){z_h^2M_h^2}\gamma_5\bigg]\Bigg\}\,,
\eea
where we defined
\bea\label{btilde2}
\tilde{{F}}^{(n)}(z_h,b^2)=&\frac{1}{z_h^2}\frac{2 \pi n ! }{\left(z_h^2M_h^{2}\right)^{n}} \int d{k}_\perp {k}_\perp\left(\frac{{k}_\perp}{b}\right)^{n} J_{n}\left(\frac{{b}{k}_\perp}{z_h}\right) {F}^{h/q}\left(z_h, {k}_\perp^2\right)\,.
\eea
Note that $F$ stands generally for all TMDFFs with appropriate $n$ value and by default $n=0$. We then begin with unsubtracted TMDFFs, which follow the same parametrization, and make the scale explicit by replacing
\bea
\tilde{{F}}^{(n)}(z_h,b^2) \to \tilde{{F}}^{(n),{\rm unsub}}(z_h,b^2,\mu,\zeta'/\nu^2)\,.
\eea

Working with an assumption that soft function is independent of the polarization, we can now write down the relations between all of the TMDJFFs and TMDFFs. We find for a general TMDJFF $\mathcal{F}$ that
\bea
{\mathcal{F}}^{h/q}(z_h,j_\perp^2,\mu, \zeta_J) &= \int \frac{b^{n+1}\,db}{2\pi n!}\left(\frac{z_h^2M_h^2}{j_\perp}\right)^n J_n\left(\frac{j_\perp b}{z_h}\right) \tilde{F}^{h/q(n),{\rm unsub}}(z_h,b^2,\mu,\zeta'/\nu^2)\tilde{S}_q(b^2,\mu,\nu \mathcal{R})\nonumber\\
&=\int \frac{b^{n+1}\,db}{2\pi n!}\left(\frac{z_h^2M_h^2}{j_\perp}\right)^n J_n\left(\frac{j_\perp b}{z_h}\right) \tilde{F}^{h/q(n)}(z_h,b^2,\mu,\zeta'\mathcal{R}^2) 
\nnu
&= F^{h/q}(z_h,j_\perp^2,\mu,\zeta_J)\,,
\label{eq:TMDJFFrel}
\eea
where we generalized Eq.\ \eqref{eq:Dunpb} to define subtracted TMDFF. The values of $n$ on the right-hand-side of Eq.~\eqref{eq:TMDJFFrel} follows the $n$ values of the parametrization given in Eq.~\eqref{eq:TMDFFb}. Therefore, all of the TMDJFFs are equal to their corresponding TMDFF at the scale $\zeta_J$. As the TMD evolutions are assumed to be polarization independent, we follow the same parametrization as that of the unpolarized TMDFF presented in Sec. \ref{sec3:Theoretical} to include evolution effects for the other TMDFFs.

\section{Unpolarized hadron inside a jet with $\mathcal{O}(R)\sim 1$}\label{appsoft}

In section \ref{sec3:Theoretical} the factorization formula for the process of unpolarized hadron inside a jet has been given in the narrow jet cone limit where $R\ll 1$. As a complementary method, in this appendix, we will give the derivation of the factorization formula without the narrow cone approximation. Especially, we will show that after taking the limit of $R\ll 1$, this formula will reduce to the expressions given in the section \ref{sec3:Theoretical}.

 Since the jet radii $R$ is not a small parameter, the narrow cone approximation is not proper to construct the factorization formula. Generally speaking, the factorized cross section is expressed as the product of the hard, soft and TMD collinear functions, which reads $ d\sigma\sim H f_{1}^q\otimes\,D_{1}^{h/q}\otimes\,S$, where the soft function $S$ depends on both $\bm q_T$ and $\bm j_\perp$, and the jet algorithm dependence is also included in $S$.  Explicitly, the physics scale inside the jet is $j_\perp$, while the scale outside the jet is $q_T$.  Since we assume $q_T \sim j_\perp$, there is no large logarithms inside the soft function. If there exist scale hierachy between $q_T$ and $j_\perp$, then one needs to consider the refactorization of the soft function as shown in~\cite{Becher:2016omr}. We have the factorized cross section as 
\begin{align}\label{eq:fac_Lr}
    \frac{d\sigma^{p(S_A)+e(\lambda_e)\to e+(\text{jet}\,h)+X}}{d{p}^2_Tdy_Jd^2{\bm q}_Tdz_h d^2{\bm j}_\perp}=&\hat\sigma_0 H(Q,\mu) \int\frac{d^2\bm b}{(2\pi)^2} e^{i \bm{b} \cdot \bm{q}_T} \int\frac{d^2 \bm b'}{(2\pi)^2} e^{i \bm{b}' \cdot \bm{j}_\perp}   \\
    &\hspace{-2.5cm} \times \sum_q e_q^2 \tilde{D}_{1}^{h/q,{\rm unsub}}(z_h,b'^2,\mu,\zeta'/\nu^2) x \tilde{f}^{q,{\rm unsub}}_{1}(x,b^2,\mu,\zeta/\nu^2) S(\bm b,\bm b',y_J,R,\mu,\nu), \notag
\end{align}
where $\bm b$ and $\bm b'$ are conjugate variables of $\bm q_T$ and $\bm j_\perp$, separately. At one-loop order,  we consider only one soft gluon emission, which is either inside or outside the jet cone. Then the soft function can be factorzied as
\begin{align}\label{eq:sbpsb}
    S(\bm b,\bm b',y_J,R,\mu,\nu) = S_{\rm in}(\bm b',y_J,R,\mu,\nu) S_{\rm out}(\bm b,y_J,R,\mu,\nu). 
\end{align}
Using the above relation, we find that $q_T$ and $j_\perp$ dependence in the cross section are fully factorized. It is noted that the above factorization is an approximation, and beyond the one-loop order the expressions depending on both $b$ and $b'$ can show up. Explicitly, the one-loop soft function $S(\bm b,\bm b',y_J,R,\mu,\nu)$ is given by
\begin{align}\label{snlo}
S^{\rm NLO}(\bm b,\bm b',y_J,R,\mu,\nu) = & \,C_F \frac{\alpha_s \mu^{2\epsilon} \pi^\epsilon e^{\gamma_E\epsilon}}{\pi^2} \int d^d k \, \delta(k^2)\theta(k^0) \frac{n\cdot n_J}{n\cdot k \, k \cdot n_J} \left( \frac{\nu}{2k^0}\right)^\eta \notag \\
& \hspace{1.5cm} \times \left[ \theta(\Delta R - R) e^{i \bm k_T \cdot \bm b}  + \theta(R-\Delta R) e^{i \bm k_\perp \cdot \bm b'} \right],
\end{align}
where $\Delta R$ denotes the distance between the jet and the soft emission in rapidity and azimuthal angle plane, which is defined by $\Delta R=\sqrt{(y-y_J)^2+(\phi-\phi_J)^2}$. Therefore, $\theta(\Delta R - R)$ and $\theta(R - \Delta R)$ indicate the soft gluon with momentum $k$ is radiated outside and inside the jet, respectively. In the CM frame of incoming beams the vectors $\bm b$ and $\bm b'$ are defined as
\begin{align}
\bm b&=b(\cos\phi_1,\sin \phi_1,0), \\
\bm b'&=b'(\cos\hat\phi_2\cos\theta_J,\sin\hat\phi_2,-\cos\hat\phi_2\sin\theta_J),
\end{align}
respectively. Here, without loss of generality, we have chosen $\phi_J=0$. It is noted that the vector $\bm b'$ is the conjugate variable of $\bm k_\perp$, which describes the transverse momentum perpendicular to the jet axis.  In the jet frame, it is given as 
\begin{align}
\bm b'= b'(\cos \hat\phi_2,\sin\hat\phi_2)_J,
\end{align}
After performing the rotation transformation in the $xz$ plane, we obtain its expression in the CM frame of incoming beams.

In Eq.~\eqref{snlo} the contribution outside the jet region can be rewritten as
\begin{align}
\theta(\Delta R-R) = 1 - \theta(R-\Delta R),
\end{align}
where the first term on the right side indicates that the soft radiation is independent on the jet definition, so it is the same as the global soft function $S_{\rm global}$ introduced in Eq.~\eqref{eq:FUUbefore}. Then $S_{\rm out}(b,y_J,R,\mu,\nu)$ in Eq. \eqref{eq:sbpsb} is given by
\begin{align}
    S_{\rm out}(\bm b,y_J,R,\mu,\nu)=S_{\rm global}(\bm b,y_J,R,\mu,\nu)+S^{\rm in}_{\rm I}(\bm b,R,\mu)
\end{align}
Next we define the contribution from the second term as $S^{\rm in}_{\rm I}$, which is 
\begin{align}
S^{\rm in}_{\rm I}(\bm b,R,\mu) = -C _F \frac{\alpha_s \mu^{2\epsilon} \pi^\epsilon e^{\gamma_E\epsilon}}{\pi^2} \int d^d k \, \delta(k^2)\theta(k^0) \frac{n\cdot n_J}{n\cdot k \, k \cdot n_J}  \theta(R-\Delta R) e^{i \bm k_T \cdot \bm b},
\end{align}
where we ignore the rapidity regulator, since the integral does not contain the rapidity divergence anymore after constraining the angular integration only inside the jet. 
The phase space integration can be expressed as
\begin{align}
\int d^{d} k \delta\left(k^{2}\right) \theta(k^{0})=\frac{\pi^{\frac{1}{2}-\epsilon}}{\Gamma\left(\frac{1}{2}-\epsilon\right)} \int_{0}^{\pi} d \phi \sin ^{-2 \epsilon} \phi \int d y \int d k_{T}\, k_{T}^{1-2 \epsilon}.
\end{align}
After integrating $k_T$ with the Fourier transformation factor, we have
\begin{align}
\int d k_{T}\, k_{T}^{-1-2 \epsilon} e^{i k_{T} b \cos (\phi-\phi_1)}=\Gamma(-2 \epsilon)\left[-i b \cos \left(\phi-\phi_{1}\right)\right]^{2 \epsilon}.
\end{align}
The integration region of $y$ and $\phi$ are constrained by the jet cone as $\theta [R-\phi^2-(y-y_J)^2]$, and we express the integration variables as
\begin{align}
y=r\cos\chi+y_J, ~~~~~~\phi=r\sin\chi.
\end{align}
Then we obtain
\begin{align}
\int d y \int_{0}^{\pi} d \phi \,\theta\left[R^{2}-\left(y-y_{J}\right)^{2}-\phi^{2}\right]=\int_{0}^{R} d r \, r \int_{0}^{\pi} d \chi.
\end{align}
In the small $R$ limit, after taking the leading contribution of the integrand in the $r\ll 1$ region, we have
\begin{align}
S^{\rm in}_{\rm I}(\bm b,R,\mu) =&\, -\frac{\alpha_s}{2\pi}C_F \Bigg[  \frac{1}{\epsilon^{2}}+\frac{2}{\epsilon}\ln \left(\frac{-2i \cos \phi_{1}\mu}{\mu_{b}R} \right)+ 2\ln^2\left( \frac{-2i \cos \phi_{1}\mu}{\mu_{b}R}\right)+ \frac{\pi^2}{4}\Bigg] \,,
\end{align}  
which is exactly the same as the collinear-soft function $S_{ cs}$ given in Eq. \eqref{eq:Scsbefore}. 

Similarly, we define $S_{\rm in}(\bm b',y_J,R,\mu,\nu)$ in Eq.~\eqref{eq:sbpsb}, namely the contribution from $\theta(R-\Delta R)$ in Eq.~\eqref{snlo} as $S^{\rm in}_{\rm II}$, where $k_\perp^\mu$ is defined as
\begin{align}
    k_\perp^\mu = k^\mu - \frac{\bar n_J \cdot k}{2} n_J^\mu - \frac{n_J \cdot k}{2} \bar n_J^\mu.
\end{align}
In the small $R$ limit, we have
\begin{align}
S^{\rm in}_{\rm II}(b',\mu,\nu\mathcal{R}) = & \, -\frac{\alpha_s}{2 \pi} C_{F}\Bigg[-\frac{1}{\epsilon^{2}}+\frac{2}{\eta} \left(\frac{1}{\epsilon }+\ln\frac{\mu^2}{\mu_b'^2}\right)+\frac{1}{\epsilon} \ln \left(\frac{\nu^2 \mathcal{R}^2}{4\mu^2}\right) \notag \\
& \hspace{2cm} + \ln\frac{\mu^2}{\mu_b'^2} \ln \left(\frac{\nu^2 \mathcal{R}^2}{4\mu^2}\right) +  \frac{1}{2}\ln^2\frac{\mu^2}{\mu_b'^2} + \frac{\pi^2}{12} \Bigg],
\end{align}

which is the same as the one-loop expression of $S_q$ given in Eq.~\eqref{eq:S_q}. Therefore we show that in the small $R$ approximation, the one-loop soft function $S^{\rm NLO}$ is
\begin{align}
S^{\rm NLO} &= S_{\rm global}(\bm b,\mu,\nu)+S^{\rm in}_{\rm I}(\bm b,\mu)+S^{\rm in}_{\rm II}(b',\mu,\nu\mathcal{R}) + \mathcal{O}(R^2),\notag \\
&=S_{\rm global}(\bm b,\mu,\nu)+S_{ cs}(\bm b,\mu)+S_{q}(b',\mu,\nu\mathcal{R})+ \mathcal{O}(R^2)\,.
\end{align}
In other words the soft function $S$ in the factorization formula \eqref{eq:fac_Lr} can be expressed as
\begin{align}
S=S_{\rm global}\,S_{ cs}\,S_{q} + \mathcal{O}(R^2). 
\end{align}
Finally we obtain Eq.~\eqref{eq:FUUUbefore}, the factorization formula for the process of the unpolarized hardon production inside jet in the narrow cone approximation. 

\section{Structure functions with hadron in jets}\label{app2}
In this section, we give explicit expressions of the structure functions in Eqs.\ \eqref{eq:unpjeth} and\ \eqref{eq:poljeth}. To give a compact presentation, we define

\bea
\mathcal{C}_{mnk}[ {\mathcal{D}}_q(z_h,j_\perp^2;\mu)\tilde{A}^{(n)}(x,b^2)]=&\hat{\sigma}_{k}H(Q, \mu)\sum_qe_q^2\left(\frac{ j_\perp}{z_hM_h}\right)^m{\mathcal{D}}_q(z_h,j_\perp^2,\mu, \zeta_J)\nnu
&\times M^n\int\frac{b^{n+1}db}{2\pi n!}J_n(q_T b)x\tilde{A}^{(n)}(x,b^2)\,,
\eea
where $m$, $n$ and $k$ can be $m=0,\ 1,\ 2$, $n=0,\ 1,\ 2$ and $k=0,\ L,\ T$.

Partonic cross section $\hat{\sigma}_k$ describes scattering of electron-quark with different polarizations depending on the value of $k$. 
The $k=0$ corresponds to the partonic scattering $eq\rightarrow eq$ or $eq_L\rightarrow eq_L$, $k=L$ corresponds to the partonic scattering $e_Lq_L\rightarrow eq$ or $eq\rightarrow e_Lq_L$, and $k=T$ corresponds to the partonic scattering $eq_T\rightarrow eq_T$. Their expressions are given as
\bea
\hat{\sigma}_0=\frac{\alpha_{\rm em}\alpha_s}{sQ^2}\frac{2(\hat{u}^2+\hat{s}^2)}{\hat{t}^2}\,,\label{eq:sigma0}\\
\hat{\sigma}_L=\frac{\alpha_{\rm em}\alpha_s}{sQ^2}\frac{2(\hat{u}^2-\hat{s}^2)}{\hat{t}^2}\,,\label{eq:sigmaL}\\
\hat{\sigma}_T=\frac{\alpha_{\rm em}\alpha_s}{sQ^2}\left(\frac{-4\hat{u}\hat{s}}{\hat{t}^2}\right)\,.
\label{eq:sigma_T}
\eea
Then, we find
\bea
F_{UU,U}(q_T,j_\perp) &= \mathcal{C}_{000}[{\mathcal{D}}_{1,q}\tilde{f}_1\bar{S}_{\rm global}\bar{S}_{cs}]\,,
\label{eq:strhlep1}
\\
F_{UU,U}^{\cos(\phi_q-\hat{\phi}_h)}(q_T,j_\perp) &= \mathcal{C}_{11T}[{\mathcal{H}}_{1,q}^{\perp}\tilde{h}_1^{\perp(1)}\bar{S}_{\rm global}\bar{S}_{cs}]\,,
\label{eq:strhlep2}
\\
F_{LL,U}(q_T,j_\perp)&=\mathcal{C}_{00L}[{\mathcal{D}}_{1,q}\tilde{g}_{1L}\bar{S}_{\rm global}\bar{S}_{cs}]\,,
\label{eq:strhlep3}
\\
F_{LU,U}^{\sin(\phi_q-\hat{\phi}_h)}(q_T,j_\perp)&=\mathcal{C}_{11T}[{\mathcal{H}}_{1,q}^{\perp}\tilde{h}_{1L}^{\perp(1)}\bar{S}_{\rm global}\bar{S}_{cs}]\,,
\label{eq:strhlep4}
\\
F_{TU,U}^{\sin({\phi}_{q}-{\phi}_{S_A})}(q_T,j_\perp)&= \mathcal{C}_{010}\left[{\mathcal{D}}_{1,q}\tilde{f}_{1T}^{\perp(1)}\bar{S}_{\rm global}\bar{S}_{cs}\right]\,,
\label{eq:strhlep5}
\\
F_{TU,U}^{\sin({\phi}_{S_A}-\hat{\phi}_{h})}(q_T,j_\perp)&= \mathcal{C}_{10T}\left[{\mathcal{H}}_{1,q}^{\perp}\tilde{h}_{1}\bar{S}_{\rm global}\bar{S}_{cs}\right]\,,
\label{eq:strhlep6}
\\
F_{TL,U}^{\cos({\phi}_{q}-{\phi}_{S_A})}(q_T,j_\perp)&=\mathcal{C}_{01L}\left[{\mathcal{D}}_{1,q}\tilde{g}_{1T}^{(1)}\bar{S}_{\rm global}\bar{S}_{cs}\right]\,,
\label{eq:strhlep7}
\\
F_{TU,U}^{\cos(2{\phi}_{q}-\hat{\phi}_h-{\phi}_{S_A})}(q_T,j_\perp)&=\mathcal{C}_{12T}\left[{\mathcal{H}}_{1,q}^{\perp}\tilde{h}_{1T}^{\perp(2)}\bar{S}_{\rm global}\bar{S}_{cs}\right]\,,
\label{eq:strhlep8}
\\
F_{UL,L}(q_T,j_\perp) &= \mathcal{C}_{00L}[{\mathcal{G}}_{1L,q}\tilde{f}_1\bar{S}_{\rm global}\bar{S}_{cs}]\,,
\label{eq:strh1}
\\
F_{UU,L}^{\sin(\hat{\phi}_h-\phi_q)}(q_T,j_\perp) &= \mathcal{C}_{11T}[{\mathcal{H}}_{1L,q}^{\perp}\tilde{h}_1^{\perp(1)}\bar{S}_{\rm global}\bar{S}_{cs}]\,,
\label{eq:strh2}
\\
F_{LU,L}(q_T,j_\perp)&=\mathcal{C}_{000}[{\mathcal{G}}_{1L,q}\tilde{g}_{1L}\bar{S}_{\rm global}\bar{S}_{cs}]\,,
\label{eq:strh3}
\\
F_{LU,L}^{\cos(\hat{\phi}_h-\phi_q)}(q_T,j_\perp)&=-\mathcal{C}_{11T}[{\mathcal{H}}_{1L,q}^{\perp}\tilde{h}_{1L}^{\perp(1)}\bar{S}_{\rm global}\bar{S}_{cs}]\,,
\label{eq:strh4}
\\
F_{TU,L}^{\cos({\phi}_{q}-{\phi}_{S_A})}(q_T,j_\perp)&= \mathcal{C}_{010}\left[{\mathcal{G}}_{1L,q}\tilde{g}_{1T}^{(1)}\bar{S}_{\rm global}\bar{S}_{cs}\right]\,,
\label{eq:strh5}
\\
F_{TL,L}^{\sin({\phi}_{q}-{\phi}_{S_A})}(q_T,j_\perp)&= \mathcal{C}_{01L}\left[{\mathcal{G}}_{1L,q}\tilde{f}_{1T}^{\perp(1)}\bar{S}_{\rm global}\bar{S}_{cs}\right]\,,
\label{eq:strh6}
\\
F_{TU,L}^{\cos({\phi}_{S_A}-\hat{\phi}_{h})}(q_T,j_\perp)&=- \mathcal{C}_{10T}\left[{\mathcal{H}}_{1L,q}^{\perp}\tilde{h}_{1}\bar{S}_{\rm global}\bar{S}_{cs}\right]\,,
\label{eq:strh7}
\\
F_{TU,L}^{\cos(2{\phi}_{q}-{\phi}_{S_A}-\hat{\phi}_{h})}(q_T,j_\perp)&=- \mathcal{C}_{12T}\left[{\mathcal{H}}_{1L,q}^{\perp}\tilde{h}_{1T}^{\perp(2)}\bar{S}_{\rm global}\bar{S}_{cs}\right]\,,
\label{eq:strh8}
\\
F_{UU,T}^{\sin(\hat{\phi}_{h}-\hat{\phi}_{S_h})}(q_T,j_\perp)&= \mathcal{C}_{100}\left[{\mathcal{D}}_{1T,q}^{\perp}\tilde{f}_{1}\bar{S}_{\rm global}\bar{S}_{cs}\right]\,,
\label{eq:strh9}
\\
F_{UL,T}^{\cos(\hat{\phi}_{h}-\hat{\phi}_{S_h})}(q_T,j_\perp)&=- \mathcal{C}_{00L}\left[{\mathcal{G}}_{1T,q}\tilde{f}_{1}\bar{S}_{\rm global}\bar{S}_{cs}\right]\,,
\label{eq:strh10}
\\
F_{UU,T}^{\sin(\hat{\phi}_{S_h}-{\phi}_{q})}(q_T,j_\perp)&= \mathcal{C}_{01T}\left[{\mathcal{H}}_{1,q}\tilde{h}_{1}^{\perp(1)}\bar{S}_{\rm global}\bar{S}_{cs}\right]\,,
\label{eq:strh11}
\\
F_{UU,T}^{\cos(2\hat{\phi}_{h}-{\phi}_{q}-\hat{\phi}_{S_h})}(q_T,j_\perp)&= \mathcal{C}_{21T}\left[{\mathcal{H}}_{1T,q}^{\perp}\tilde{h}_{1}^{\perp(1)}\bar{S}_{\rm global}\bar{S}_{cs}\right]\,,
\label{eq:strh12}
\\
F_{LU,T}^{\cos(\hat{\phi}_{h}-\hat{\phi}_{S_h})}(q_T,j_\perp)&=- \mathcal{C}_{100}\left[{\mathcal{G}}_{1T,q}\tilde{g}_{1L}\bar{S}_{\rm global}\bar{S}_{cs}\right]\,,
\label{eq:strh13}
\\
F_{LU,T}^{\cos(\hat{\phi}_{S_h}-{\phi}_{q})}(q_T,j_\perp)&= \mathcal{C}_{01T}\left[{\mathcal{H}}_{1,q}\tilde{h}_{1L}^{\perp(1)}\bar{S}_{\rm global}\bar{S}_{cs}\right]\,,
\label{eq:strh14}
\\
F_{LU,T}^{\cos(2\hat{\phi}_{h}-{\phi}_{q}-\hat{\phi}_{S_h})}(q_T,j_\perp)&= \mathcal{C}_{21T}\left[{\mathcal{H}}_{1T,q}^{\perp}\tilde{h}_{1L}^{\perp(1)}\bar{S}_{\rm global}\bar{S}_{cs}\right]\,,
\label{eq:strh15}
\\
F_{LL,T}^{\sin(\hat{\phi}_{h}-\hat{\phi}_{S_h})}(q_T,j_\perp)&= \mathcal{C}_{10L}\left[{\mathcal{D}}_{1T,q}^{\perp}\tilde{g}_{1L}\bar{S}_{\rm global}\bar{S}_{cs}\right]\,,
\label{eq:strh16}
\\
F_{TU,T}^{\cos({\phi}_{S_A}-\hat{\phi}_{S_h})}(q_T,j_\perp)&= \mathcal{C}_{00T}\left[{\mathcal{H}}_{1,q}\tilde{h}_{1}\bar{S}_{\rm global}\bar{S}_{cs}\right]\,,
\label{eq:strh17}
\\
F_{TU,T}^{\cos(2\hat{\phi}_{h}-\hat{\phi}_{S_h}-{\phi}_{S_A})}(q_T,j_\perp)&= \mathcal{C}_{10T}\left[{\mathcal{H}}_{1T,q}^{\perp}\tilde{h}_{1}\bar{S}_{\rm global}\bar{S}_{cs}\right]\,,
\label{eq:strh18}
\\
F_{TU,T}^{\sin(\hat{\phi}_{h}-\hat{\phi}_{S_h})\sin({\phi}_q-{\phi}_{S_A})}(q_T,j_\perp)&= \mathcal{C}_{110}\left[{\mathcal{D}}_{1T,q}^{\perp}\tilde{f}_{1T}^{\perp(1)}\bar{S}_{\rm global}\bar{S}_{cs}\right]\,,
\label{eq:strh19}
\\
F_{TU,T}^{\cos(\hat{\phi}_{h}-\hat{\phi}_{S_h})\cos({\phi}_q-{\phi}_{S_A})}(q_T,j_\perp)&=- \mathcal{C}_{110}\left[{\mathcal{G}}_{1T,q}\tilde{g}_{1T}^{(1)}\bar{S}_{\rm global}\bar{S}_{cs}\right]\,,
\label{eq:strh20}
\\
F_{TU,T}^{\cos(2{\phi}_{q}-\hat{\phi}_{S_h}-{\phi}_{S_A})}(q_T,j_\perp)&= \mathcal{C}_{02T}\left[{\mathcal{H}}_{1,q}\tilde{h}_{1T}^{\perp(2)}\bar{S}_{\rm global}\bar{S}_{cs}\right]\,,
\label{eq:strh21}
\\
F_{TU,T}^{\cos(2{\phi}_{h}-\hat{\phi}_{S_h}+2{\phi}_{q}-{\phi}_{S_A})}(q_T,j_\perp)&= \mathcal{C}_{12T}\left[{\mathcal{H}}_{1T,q}^{\perp}\tilde{h}_{1T}^{\perp(2)}\bar{S}_{\rm global}\bar{S}_{cs}\right]\,,
\label{eq:strh22}
\\
F_{TL,T}^{\cos(\hat{\phi}_{h}-\hat{\phi}_{S_h})\sin({\phi}_{S_A}-{\phi}_q)}(q_T,j_\perp)&=\mathcal{C}_{11L}\left[{\mathcal{G}}_{1T,q}\tilde{f}_{1T}^{\perp(1)}\bar{S}_{\rm global}\bar{S}_{cs}\right]\,,
\label{eq:strh23}
\\
F_{TL,T}^{\sin(\hat{\phi}_{h}-\hat{\phi}_{S_h})\cos({\phi}_{S_A}-{\phi}_q)}(q_T,j_\perp)&=\mathcal{C}_{11L}\left[{\mathcal{D}}_{1T,q}^{\perp}\tilde{g}_{1T}^{(1)}\bar{S}_{\rm global}\bar{S}_{cs}\right]\,,
\label{eq:strh24}
\eea
where TMDJFFs found in the above equations can also be simplified in terms of TMDFFs and collinear-soft function in the region $j_\perp \ll p_T R$ as shown in the Appendix\ \ref{appTMDFFs}.

\bibliographystyle{JHEP}
\bibliography{epjet}

\end{document}